\documentclass[11pt]{article}

\usepackage[margin=1in]{geometry}
\usepackage{amsmath,amssymb,amsthm,mathtools}

\numberwithin{equation}{section}
\usepackage{booktabs}
\usepackage{enumitem}
\usepackage[linesnumbered,boxed,ruled,vlined]{algorithm2e}
\usepackage{xurl}
\usepackage{hyperref}
\usepackage{thm-restate}
\makeatletter
\@ifundefined{newcounteralias}{}{%
  \renewcommand\thmt@autorefsetup{%
    \@xa\def\csname\thmt@envname autorefname\@xa\endcsname\@xa{\thmt@thmname}%
  }%
}
\makeatother

\usepackage[nameinlink,capitalise,noabbrev]{cleveref}

\crefname{informaltheorem}{main result}{main results}
\Crefname{informaltheorem}{Main Result}{Main Results}

\usepackage{bbm}

\usepackage{complexity}

\usepackage{multirow}
\usepackage{tabularx}
\usepackage{makecell}
\usepackage{threeparttable}

\usepackage{tikz}
\usetikzlibrary{calc}
\usetikzlibrary{shapes.geometric}
\usetikzlibrary{arrows.meta}
\usetikzlibrary{decorations.pathreplacing}

\usepackage[table]{xcolor}
\definecolor{niceMagenta}{HTML}{C2185B}

\usepackage{tcolorbox}

\newtcolorbox{problembox}[1]{
  colback=gray!6,
  colbacktitle=gray!6,
  coltitle=black,
  boxrule=0pt,
  arc=3mm,
  fonttitle=\bfseries,
  title={#1},
  left=3mm,
  right=3mm,
  top=2mm,
  bottom=2mm,
  before skip=12pt,
  after skip=12pt
}

\hypersetup{
  hypertexnames=false,
  colorlinks=true,
  linkcolor=niceMagenta, 
  citecolor=niceMagenta, 
  urlcolor=niceMagenta
}

\theoremstyle{plain}
\newtheorem{theorem}{Theorem}[section]

\newtheorem{lemma}[theorem]{Lemma}
\newtheorem{claim}[theorem]{Claim}
\newtheorem{proposition}[theorem]{Proposition}

\theoremstyle{definition}
\newtheorem{definition}[theorem]{Definition}

\theoremstyle{remark}
\newtheorem{remark}[theorem]{Remark}

\usepackage{MnSymbol}
\usepackage[mathlines]{lineno}
\usepackage{bm} 

\crefname{claim}{Claim}{Claims}
\Crefname{claim}{Claim}{Claims}

\newcommand{\Xcomment}[1]{{}}

\renewcommand{\part}[2]{\frac{\partial #1}{\partial #2}}

\newlist{lemmaitems}{enumerate}{1}
\setlist[lemmaitems]{label=(\arabic*),ref=\arabic*,topsep=4pt,itemsep=5pt,leftmargin=*}
\crefname{lemmaitemsi}{Item}{Items}
\Crefname{lemmaitemsi}{Item}{Items}
\creflabelformat{lemmaitemsi}{#2(#1)#3}
\crefname{line}{Step}{Steps}
\Crefname{line}{Step}{Steps}
\makeatletter
\renewcommand{\nllabel}[1]{%
  \begingroup
  \def\@currentcounter{AlgoLine}%
  \edef\cref@currentlabel{[line][\arabic{AlgoLine}][]\theAlgoLine}%
  \label{#1}%
  \endgroup
}
\makeatother

\newcommand{\dd}{\mathrm{d}}       

\newcounter{reductionstep}
\DeclareRobustCommand{\circledstep}[1]{%
  \tikz[baseline=(stepnumber.base)]{%
    \node[draw=.,solid,circle,line width=.45pt,inner sep=0pt,
      minimum size=1.4em,font=\normalfont] (stepnumber) {#1};%
  }%
}
\crefname{reductionstep}{step}{steps}
\Crefname{reductionstep}{Step}{Steps}
\crefformat{reductionstep}{#2\circledstep{#1}#3}
\Crefformat{reductionstep}{#2\circledstep{#1}#3}
\crefmultiformat{reductionstep}{#2\circledstep{#1}#3}
  { and #2\circledstep{#1}#3}{, #2\circledstep{#1}#3}{, and #2\circledstep{#1}#3}
\Crefmultiformat{reductionstep}{#2\circledstep{#1}#3}
  { and #2\circledstep{#1}#3}{, #2\circledstep{#1}#3}{, and #2\circledstep{#1}#3}

\usepackage{natbib}
\let\oldcitet\citet
\let\oldcitep\citep
\renewcommand{\citet}[1]{\oldcitet*{#1}}
\renewcommand{\citep}[1]{\oldcitep*{#1}}

\usepackage{xcolor}
\definecolor{niceRed}{RGB}{190,38,38}
\definecolor{blueGrotto}{HTML}{059DC0}
\definecolor{royalBlue}{HTML}{057DCD}
\definecolor{navyBlue}{HTML}{0B579C}
\definecolor{limeGreen}{HTML}{81B622}
\definecolor{nicePurple}{HTML}{9c27b0}
\definecolor{lightRoyalBlue}{HTML}{def2ff} 
\definecolor{gold}{HTML}{ffa300}

\usepackage{color-edits} 
\usepackage{todonotes}

\newcommand{\ignore}[1]{}

\title{Competitive Equilibrium for Linear Chores is CLS-complete}
\author{Qian Liu\thanks{School of Computing and Data Science, The University of Hong Kong. Emails: \nolinkurl{{hecr_qian,shinuo_ma,ymjessen02}@connect.hku.hk}.}\qquad
Shinuo Ma\footnotemark[1]\qquad
Yixin Tao\thanks{Institute for Theoretical Computer Science, Shanghai University of Finance and Economics. Email: \href{mailto:taoyixin@mail.shufe.edu.cn}{\textcolor{black}{\nolinkurl{taoyixin@mail.shufe.edu.cn}}}.}\qquad
Minji Yang\footnotemark[1]\qquad
Weiqiang Zheng\thanks{Department of Computer Science, Yale University. Email: \href{mailto:weiqiang.zheng@yale.edu}{\textcolor{black}{\nolinkurl{weiqiang.zheng@yale.edu}}}.}}
\date{}
\begin{document}

\maketitle
\begin{abstract}
    The competitive equilibrium (CE) is a fundamental concept for markets, with applications to efficient and fair allocation of resources. We consider the fundamental question of computing a CE in the most basic Fisher market model where agents have linear preferences. While it is known that we can compute a CE in strongly polynomial time with linear \emph{goods}, there is no known polynomial-time algorithm for CE with linear \emph{chores}. Computing a CE for linear chores is known to be in CLS with no hardness results.
    
    In this paper, we prove that computing an exact CE in Fisher markets with divisible chores and linear disutilities is CLS-hard and thus CLS-complete. To the best of our knowledge, this is the first CLS-completeness result for a market equilibrium problem. On the technical side, we use the nonconvex dual Eisenberg-Gale (EG) program whose Karush–Kuhn–Tucker (KKT) points are CE~\citep{Chaudhury24competitive}, and the CLS-complete problem of computing a KKT point of a quadratic program~\citep{Fearnley24complexity}. We show that from any exact equilibrium of the constructed market, we can recover in polynomial time an $\varepsilon$-KKT point of the input quadratic program over the unit box.
\end{abstract}

\clearpage
\setcounter{tocdepth}{2}
\tableofcontents
\clearpage

\section{Introduction}
Competitive equilibrium (CE) for markets is a cornerstone of theoretical economics~\citep{arrow1954existence}. Through markets, a CE guarantees efficient resource allocation through prices that ensure supply meets demand. In this paper, we consider the most basic Fisher market model~\citep{brainard2005compute} with $m$ divisible items and $n$ agents, where each agent $a$ has a budget $B_a > 0$ and a linear preference. A CE is a price vector for the items such that the market clears while each agent buys their utility-maximizing budget-feasible bundle. CE is important for resource allocation since it guarantees envy-free (for equal incomes) and Pareto-optimal division~\citep{Varian74equity} and is used for real-world resource allocation~\citep{budish2011combinatorial}. Settling the  complexity of computing a CE is a fundamental question that has inspired a long line of research.

When the items are goods, the complexity of CE is now well-understood. For linear goods, the equilibrium allocations are characterized by optimal solutions to the classic Eisenberg-Gale (EG) convex program that maximizes the budget-weighted Nash social welfare~\citep{eisenberg1959consensus}, which also implies the equilibrium prices are unique. Due to the nice convex program characterization, we can compute a CE in polynomial time~\citep{devanur2008market} and even strongly polynomial time~\citep{orlin2010fisher, vegh2012strongly}. 

However, many allocation problems involve \emph{chores}, items that incur \emph{disutilities} to agents yet must be fully allocated. The Fisher market model extends naturally to the chores setting, where we can consider the prices as payments, and each agent's budget is an earning constraint that they wish to buy the disutility-minimizing bundle to meet the earning constraint. A CE is a price vector such that every chore is fully allocated. The seminal work by~\citet{bogomolnaia2017competitive} introduces CE for chores and provides characterization results. On one hand, similar to the goods setting, CE exists and implies envy-free and Pareto-optimal allocation when the agents have the same earning constraints. On the other hand, the structure of CE for chores is more nuanced. The set of CE could be non-convex and disconnected, and are critical points (in contrast to optimal solutions) to the non-convex Nash welfare program. Therefore, standard convex programming techniques can not be applied to compute a CE.

We briefly review known results for computing a CE. When the number of agents or the number of chores is fixed, an exact CE can be computed in polynomial time~\citep{garg2020computing} and even strongly polynomial time~\citep{branzei2024algorithms}. An $\epsilon$-approximate CE can be computed in $\mathrm{poly}(1/\epsilon)$ time, even for more general preferences, by the exterior point method~\citep{boodaghians2022approximate} and first-order methods on reformulations of the EG program~\citep{Chaudhury24competitive, chaudhury2026tatonnement, tao2026fisher}. On the hardness side, results are only known for richer models such exchange markets and nonlinear preferences: we know computing CE is PPAD-hard for exchange markets with linear preference, and for Fisher market with separable, piecewise-linear, concave (SPLC) preferences~\citep{chaudhury2021mixed, chaudhury2022combinatorial}.\footnote{ The class of PPAD is introduced by~\citet{papadimitriou1994parity} and captures the complexity of Nash equilibrium in normal-form games~\citep{daskalakis2009complexity, chen2009settling, rubinstein2017settling}.} For Fisher market with linear preferences,   \citet{chaudhury2022combinatorial} show that computing a CE lies in PPAD $\cap$ PLS = CLS~\citep{fearnley2023gradient}.\footnote{PLS is introduced by~\citet{johnson1988local} and CLS is introduced by~\citet{daskalakis2011continuous}. Examples of CLS-complete problems includes computing an Nash equilibrium in congestion games~\citep{babichenko2021settling}, and finding an Karush--Kuhn--Tucker (KKT) point of a quadratic program over a box \citep{Fearnley24complexity}.} However, there is no hardness result for CE with linear preferences. This motivates the following problem:

\begin{quote}
\emph{What is the complexity of finding a CE in a linear Fisher market with divisible chores?}
\end{quote}

\subsection{Our Result}
\begin{theorem}[Main Result]
\label{thm:exact-equilibrium-cls}
It is CLS-complete to compute a competitive equilibrium of Fisher market for chores with linear preferences.
\end{theorem}

Due to the CLS membership result by~\citet{chaudhury2022combinatorial}, our main contribution is to prove it is CLS-hard to compute a CE in Fisher market with linear chores. This implies CE is unlikely to have an efficient algorithm. The hardness of CLS-complete problems can also be based on several cryptographic assumptions including learning with error (LWE)~\citep{jawale2021snargs},  indistinguishability obfuscation~\citep{hubacek2020hardness}, and soundness of Fiat-Shamir~\citep{choudhuri2019finding}. 

To the best of our knowledge, \Cref{thm:exact-equilibrium-cls} provides the first CLS-completeness result for a market equilibrium problem. It adds a natural economic problem to the short list of CLS-complete problems including Nash equilibrium in congestion games~\citep{babichenko2021settling}, simultaneous contests~\citep{elkind2022simultaneous}, adversarial team games~\citep{anagnostides2026algorithms}, computing KKT points of smooth functions and quadratic programs~\citep{Fearnley24complexity, ghosh2024complexity}.

On the technical side, we use the nonconvex dual Eisenberg-Gale (EG) program whose Karush--\allowbreak{}Kuhn--\allowbreak{}Tucker (KKT) points are CE~\citep{Chaudhury24competitive}, and the CLS-complete problem of computing a KKT point of a quadratic program~\citep{Fearnley24complexity}. We show that from any exact equilibrium of the constructed market, we can recover in polynomial time an $\varepsilon$-KKT point of the input quadratic program over the unit box. We present a detailed overview of the proof in \Cref{sec:overview}.


\begin{table*}[t]
  \centering
  \caption{Computational complexity of competitive equilibrium in markets
    with divisible goods and chores.}
  \label{tab:market-complexity}
  \small
  \setlength{\tabcolsep}{6pt}
  \renewcommand{\arraystretch}{1.20}
  \begin{tabularx}{\textwidth}{
    @{}ll
    >{\raggedright\arraybackslash}X
    >{\raggedright\arraybackslash}X@{}
  }
    \toprule
    \textbf{Market} & \textbf{Preferences}
      & \textbf{Goods} & \textbf{Chores} \\
    \midrule
    Fisher & Linear
      & $\mathsf{FP}$~\citep{eisenberg1959consensus, devanur2008market, orlin2010fisher}
      & \colorbox{black!10}{\strut\textbf{CLS-complete}}
        \textbf{(this paper)}
        \\
    \addlinespace[0.3em]
    Fisher & SPLC
      & $\mathsf{PPAD}$-complete~\citep{chen2009spending, vazirani2011market}
      & $\mathsf{PPAD}$-complete~\citep{chaudhury2021mixed}. \\
    \addlinespace[0.3em]
    Exchange & Linear
      & $\mathsf{FP}$~\citep{jain2007polynomial, duan2015combinatorial, garg2019strongly}
      & $\mathsf{PPAD}$-hard~\citep{chaudhury2022combinatorial}. \\
    \bottomrule
  \end{tabularx}
  \vspace{3pt}

  \begin{minipage}{\textwidth}
    \footnotesize
    \emph{Notes.}
    SPLC = separable piecewise-linear concave.
  \end{minipage}
\end{table*}



\subsection{Further Related Work}
\label{subsec:further-related-work}

We review other algorithms and hardness results for goods markets, starting with Fisher markets and then exchange markets.

Algorithms have also been developed for Fisher markets with goods and more general preferences, or with limits on buyers' utilities and sellers' earnings. \citet{jain2005homothetic} extend the EG approach to continuous, monotone, homothetic, quasi-concave utilities. \citet{jain2010eisenberg} give strongly polynomial algorithms for several special classes of EG markets. \citet{vazirani2010spending} gives a polynomial-time algorithm for decreasing-step spending-constraint utilities, where preferences depend on the amount spent on each good. \citet{garg2024satiation} study linear utilities with limits on both buyers' total utilities and sellers' earnings. Under a money-clearing condition that guarantees a CE, they give an FPTAS and show that computing an exact CE is in $\mathrm{PPAD}\cap\mathrm{PLS}=\mathrm{CLS}$. 

On the hardness side, \citet{deligkas2026fisher} prove PPAD-hardness for Fisher markets with goods and SPLC utilities under the PCP-for-PPAD conjecture. Their result allows buyers to receive approximately optimal bundles and the market to clear approximately, with a constant relative error in utility and a constant market-clearing error. It holds even with equal budgets and linear capped utility components.

In exchange markets with goods, strongly polynomial algorithms are known for linear utilities, and polynomial-time algorithms are known for SPLC or general piecewise-linear concave (PLC) utilities when the number of goods is fixed. \citet{duan2015combinatorial} give a combinatorial polynomial-time algorithm for linear utilities. \citet{devanur2016rational} describe the equilibria using a rational convex program based on flows. Building on this approach, \citet{garg2019strongly} give a strongly polynomial algorithm. \citet{devanur2008fixed} give polynomial-time exact algorithms for SPLC utilities when either the number of goods or the number of agents is fixed, and for general PLC utilities when the number of goods is fixed. For SPLC utilities, \citet{garg2015complementary} give a complementary-pivot algorithm based on Lemke's method under assumptions that guarantee a CE. Their algorithm also gives a direct proof of PPAD membership and runs in strongly polynomial time when either the number of goods or the number of agents is fixed.

Computing a CE is PPAD-hard in exchange markets with goods and SPLC or certain constant-elasticity-of-substitution (CES) utilities, even when an approximate CE is sufficient. \citet{chen2009arrowdebreu} prove PPAD-hardness for SPLC utilities with at most two linear segments per univariate component, even when only inverse-polynomial accuracy is required. \citet{chen2017nonmonotone} give a general method for proving hardness from non-monotone demand. They show that finding an approximate CE is PPAD-complete for CES utilities with any fixed, finite rational parameter $\rho<-1$. 

A stronger result is known for exchange markets with goods and Leontief utilities: \citet{garg2017leontief} prove FIXP-hardness for exact CE computation. This strengthens the earlier PPAD-hardness result of \citet{codenotti2008complexity}, obtained by encoding bimatrix games in a special class of Leontief economies. Leontief utilities are both the $\rho\to-\infty$ limit of CES utilities and a special case of PLC utilities. Equilibria under Leontief or general PLC utilities can be irrational. On the upper-bound side, \citet{garg2016dichotomies} show that finding an exact CE is in FIXP for PLC markets under their conditions that guarantee existence.
\section{Preliminaries}

For every positive integer $k$, we write $[k]:=\{1,\ldots,k\}$.
We use $e_i$ to denote the $i$-th unit vector. The gradient of a function $f$ is denoted as $\nabla f$.
We use $x$ for points in the quadratic problem and $z$ for market allocations.

\paragraph{Fisher Market with Chores and Competitive Equilibrium.}
A Fisher market with chores consists of $m$ divisible chores $M=\{1,\ldots,m\}$ and $n$ agents $N=\{1,\ldots,n\}$.
Each chore $j\in M$ has supply $S_j>0$. Each agent $a\in N$ has an earning requirement $B_a>0$ and
a linear disutility function $d_a(z_a)=\sum_{j\in M}d_{aj}z_{aj}$, where $d_{aj}>0$ denotes the disutility incurred by agent $a$ for performing one unit
of chore $j$. The goal is to find a Fisher market equilibrium, also known as a \emph{competitive equilibrium}, consisting of a payment
vector $p\in\mathbb{R}_{\geq 0}^{m}$, where $p_j$ is the payment for performing one unit of chore $j$, and an allocation
$z\in\mathbb{R}_{\geq 0}^{n\times m}$, where $z_{aj}$ is the amount of chore $j$ assigned to agent $a$.
At equilibrium, each agent minimizes her disutility subject to meeting her earning requirement, and every chore is fully
allocated.

\begin{definition}[Fisher Market with Chores]
Consider a Fisher market instance with chores \(\mathcal{F}=(N, M, \{d_{aj}\}, \{B_a\}, \{S_j\})\),
consisting of $n$ agents and $m$ divisible chores. Each chore $j\in[m]$ has supply $S_j$.
We omit $\{S_j\}$ when all chores have unit supply.
Each agent $a\in[n]$ has an earning requirement $B_a>0$ and a linear disutility function, i.e.,
\[
d_a(z_a)=\sum_j d_{aj}z_{aj},
\qquad
d_{aj}>0.
\]

An allocation $z\in\mathbb{R}_{\geq 0}^{n\times m}$ assigns agent $a$
the bundle $z_a=(z_{a1},\ldots,z_{am})$.
Given a price vector $p\in\mathbb{R}_{\geq 0}^m$, where $p_j$ is the
payment for performing one unit of chore $j$, agent $a$'s
demand correspondence is
\[
X_a(p)
:=
\operatorname*{arg\,min}_{z_a\in\mathbb{R}_{\geq 0}^m}
\left\{
d_a(z_a):\langle p,z_a\rangle\ge B_a
\right\}.
\]
\end{definition}

\begin{definition}[Competitive Equilibrium]
A price--allocation pair
$(p^*,z^*)\in\mathbb{R}_{\geq 0}^m\times\mathbb{R}_{\geq 0}^{n\times m}$
is a competitive equilibrium of a Fisher market with chores if:
\begin{enumerate}
    \item \textbf{Individual Optimality:}
    Each agent receives a disutility-minimizing bundle
    satisfying her earning requirement $z_a^*\in X_a(p^*)$ for all  $a\in[n]$.
    \item \textbf{Market Clearing:}
    Every chore is fully allocated: $\sum_{a=1}^n z_{aj}^*=S_j$ for all $j\in[m]$.
\end{enumerate}
For linear disutilities, individual optimality is
equivalent to
$$
\sum_{j=1}^m p_j^*z_{aj}^*=B_a,
\qquad
z_{aj}^*>0
\ \Longrightarrow\
\frac{d_{aj}}{p_j^*}
=
\min_{k\in[m]}\frac{d_{ak}}{p_k^*}.
$$
Thus, each agent earns exactly $B_a$ and performs only
chores with minimum disutility per unit of payment.

Market clearing and exact earnings imply
\begin{equation}
\sum_{j=1}^m S_jp_j^*
=
\sum_{a=1}^n B_a.
\label{eq:market_clearing}
\end{equation}
We call this identity \eqref{eq:market_clearing} \emph{payment balance}. Furthermore, define the \emph{normalized price domain} $\mathcal{P}$ as
\begin{align}\label{def-normalized-price}
\mathcal{P}
:=
\left\{
p\in\mathbb{R}_{>0}^m:
\sum_{j=1}^m S_jp_j=\sum_{a=1}^n B_a
\right\}.
\end{align}
\label{def:CE}
\end{definition}
\paragraph{Equivalent Characterization of Equilibrium.}
\citet{Chaudhury24competitive} establish a correspondence
between equilibrium points in chores markets and KKT points of an
associated optimization problem they call the dual Eisenberg-Gale program. Building on their results,
we give a first-order characterization of equilibrium prices.

\begin{theorem}[First-order Characterization of Equilibrium Price]
\label{thm:first-order-character}
Consider a linear Fisher market with chores, with supplies
$S_j>0$, earning requirements $B_a>0$, and disutility
coefficients $d_{aj}>0$. Define
\begin{align}\label{def:Psi}
\mu_a(p):=\max_{j\in[m]}\frac{p_j}{d_{aj}},
\qquad
\Psi(p):=-\sum_{a=1}^n B_a\log\mu_a(p).
\end{align}
For any $p^*\in\mathcal{P}$, the following are equivalent:
\begin{enumerate}
    \item There exists an allocation
    $z^*\in\mathbb{R}_{\geq 0}^{n\times m}$ such that
    $(p^*,z^*)$ is a competitive equilibrium.

    \item The vector $p^*$ satisfies the first-order
    necessary condition for maximizing $\Psi$ over
    $\mathcal{P}$:
    \begin{align}\label{eq:first-order-condition}
        \Psi'(p^\star;v)\leq 0
    \qquad
    \text{for every }v\in\mathbb R^m
    \text{ with }\sum_{j=1}^m S_jv_j=0,
    \end{align}
    where the right directional derivative is defined by
    \[
    \Psi'(p;v)
    :=
    \lim_{h\downarrow0}
    \frac{\Psi(p+hv)-\Psi(p)}{h}.
    \]
    We use the same convention $f'(y;w)$ for other functions along
    feasible directions, whenever the right-hand limit exists.
\end{enumerate}

\end{theorem}
For completeness, the proof of \Cref{thm:first-order-character} is given in \Cref{proof-first-order-character}.

\paragraph{Approximate KKT Points of Quadratic Programs.} To prove CLS-hardness, we reduce from the problem of computing an approximate KKT point of a quadratic function over the box $[0, 1]^n$.  We define this problem formally below.
\begin{definition}[$\varepsilon$-\textsc{quadratic-KKT} Problem] Given a symmetric matrix $A\in\mathbb{Q}^{n\times n}$, a vector $c\in\mathbb{Q}^n$, and a rational parameter $\varepsilon>0$, consider the quadratic program
\begin{equation}
\begin{aligned}
    \max \quad & Q(x) :=\frac12 x^\top A x+c^\top x  \\
    \text{s.t.} \quad & 0 \leq x_i \leq 1
    \quad \forall i=1,...,n
\end{aligned}
\label{eq:quadratic-kkt}
\end{equation}
A point $x \in [0,1]^n$ is an $\varepsilon$-KKT point of
\eqref{eq:quadratic-kkt} if, for all $i \in [n]$,
\begin{itemize}
    \item if $x_i = 0$, then
    $\dfrac{\partial Q}{\partial x_i}(x) \leq \varepsilon$,
    \item if $0 < x_i < 1$, then
    $\Big|\dfrac{\partial Q}{\partial x_i}(x)\Big| \leq \varepsilon$, and
    \item if $x_i = 1$, then
    $\dfrac{\partial Q}{\partial x_i}(x)\geq -\varepsilon.$
\end{itemize}
\end{definition}
We say a direction $\tilde{v}\in\{e_i, -e_i\}$ is \textit{feasible} at $x$ if $x+h\tilde{v}\in[0,1]^n$ for all sufficiently small $h>0$.
Equivalently, the definition of $\varepsilon$-KKT point above can be rewritten as
\begin{equation}
    x \text{ is an }\varepsilon \text{-KKT point}\Longleftrightarrow \nabla Q(x)\cdot \tilde{v}\leq \varepsilon\text{, for every feasible direction }\tilde{v}.
    \label{eq:quadratic_kkt_nabla}
\end{equation}

The following complexity result provides the starting point for our reduction.

\begin{theorem}[\cite{Fearnley24complexity}, Theorem 2.1, Lemma A.1]\label{thm:kkt-cls-complete}
    The $\varepsilon$-\textup{\textsc{quadratic-KKT}} problem is $\mathrm{CLS}$-complete.
\end{theorem}

\section{Proof Overview}
\label{sec:overview}

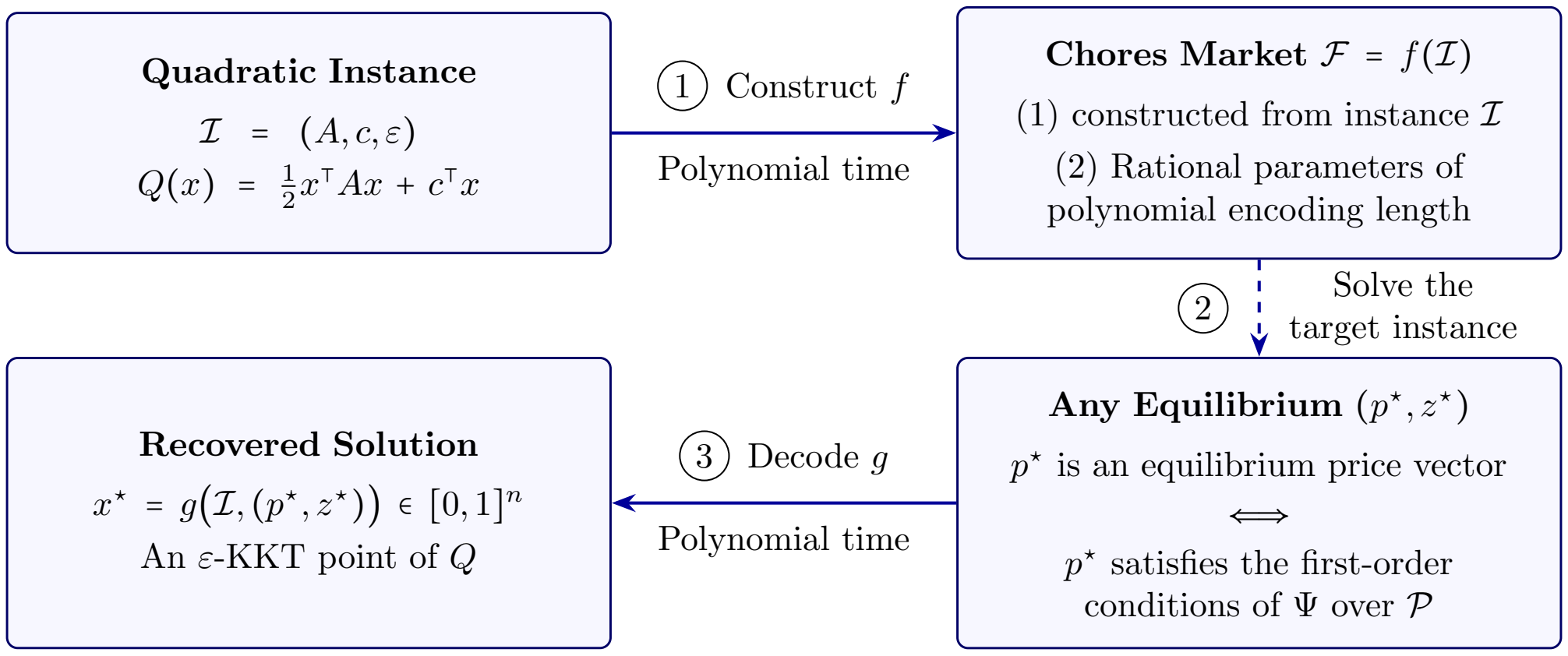
\begin{figure}[htbp]
\centering
\setcounter{reductionstep}{0}
\resizebox{0.8\linewidth}{!}{%
\begin{tikzpicture}[
  box/.style={draw=blue!40!black,fill=blue!3,rounded corners=3pt,
    line width=.65pt,align=center,text width=5.4cm,
    minimum height=2.4cm,inner sep=9pt,font=\small},
  map/.style={-{Stealth[length=2.4mm]},draw=blue!60!black,line width=.9pt},
  note/.style={font=\small,align=center,fill=white,inner sep=3pt},
  stepbadge/.style={font=\small,text=black,fill=white,inner sep=1.5pt}
]
\node[box] (input) at (0,0) {
  \textbf{Quadratic Instance}\\[5pt]
  $\mathcal{I}=(A,c,\varepsilon)$\\[3pt]
  $Q(x)=\tfrac12x^\top Ax+c^\top x$
};
\node[box] (market) at (9.5,0) {
  \textbf{Chores Market} $\mathcal F=f(\mathcal{I})$\\[5pt]
  (1) constructed from instance $\mathcal{I}$\\[3pt]
  (2) Rational parameters of polynomial encoding length
};
\node[box,minimum height=2.9cm] (answer) at (0,-3.7) {
  \textbf{Recovered Solution}\\[5pt]
  $x^\star=g\bigl(\mathcal{I},(p^\star,z^\star)\bigr)\in[0,1]^n$\\[3pt]
  An $\varepsilon$-KKT point of $Q$
};
\node[box,minimum height=2.9cm] (equilibrium) at (9.5,-3.7) {
  \textbf{Any Equilibrium $(p^\star,z^\star)$}\\[5pt]
  $p^\star$ is an equilibrium price vector\\[2pt]
  $\Longleftrightarrow$\\[2pt]
  $p^\star$ satisfies the first-order conditions of $\Psi$ over $\mathcal P$
};
\draw[map] (input.east) --
  node[note,above=3pt] {\refstepcounter{reductionstep}\label{step:construct}%
    \circledstep{\arabic{reductionstep}}\enspace Construct $f$}
  node[note,below=3pt] {Polynomial time} (market.west);
\draw[map,dashed] (market.south) --
  node[note,right=5pt] {Solve the\\target instance}
  node[stepbadge,left=7pt] {\refstepcounter{reductionstep}\label{step:equilibrium}%
    \circledstep{\arabic{reductionstep}}}
  (equilibrium.north);
\draw[map] (equilibrium.west) --
  node[note,above=3pt] {\refstepcounter{reductionstep}\label{step:recover}%
    \circledstep{\arabic{reductionstep}}\enspace Decode $g$}
  node[note,below=3pt] {Polynomial time} (answer.east);
\end{tikzpicture}%
}
\caption{Overview of our reduction. Decoder $g$ converts every equilibrium
into a solution of the original quadratic instance.
The dashed arrow represents solving the target problem.}
\label{fig:reduction-overview}
\end{figure}

Given an instance $\mathcal{I}=(A\in\mathbb{Q}^{n\times n},c\in\mathbb{Q}^n,\varepsilon\in\mathbb{Q})$ of the
$\varepsilon$-\textsc{quadratic-KKT} problem, we now outline the proof of our main result. As illustrated in \cref{fig:reduction-overview}, our reduction proceeds
in three steps.
Both the market construction and solution recovery can be carried out in polynomial time.

\paragraph{Targets of Our Construction.}
To complete the reduction, our goal is to construct
a Fisher market with chores \(\mathcal{F}\) satisfying two targets (see \cref{prop:market-target} for a formal version). Given an instance $(A,c,\varepsilon)$ of the
$\varepsilon$-\textsc{quadratic-KKT} problem, we wish to construct a market $\mathcal{F}$ that satisfies:
\begin{enumerate}[label=(\alph*),leftmargin=*]
    \item \label{target:price-encoding}
    \emph{Existence of a Mapping.}
    There exists a mapping \(p:[0,1]^n\to\mathcal P\) such that
    every equilibrium price vector \(p^\star\in\mathcal P\) has a
    preimage \(x^\star\in[0,1]^n\) satisfying $
    p(x^\star)=p^\star$. Moreover, 
    given \(p^\star\), its preimage \(x^\star\) can be computed
    in polynomial time by a decoder.

    \item \label{target:objective-approximation}
    \emph{Approximation of the Target Objective.} Recall the objective $\Psi=-\sum_{a=1}^n B_a\log\mu_a(p)$ for CE defined in~\eqref{def:Psi}. We want  \(\Psi(p(x))\) to approximate \(Q(x)=\frac12 x^\top A x+c^\top x\) as follows:
    \[
    \begin{aligned}
    \Psi(p(x))=C+\alpha Q(x)+R(x),
    \end{aligned}
    \]
    for some constant \(C\) and  \(\alpha > 0\), and a  remainder term $R(x)$. In addition, we require  \(R(x)\) to satisfy
    $
    |R(x+he_i)-R(x)|\leq\alpha\varepsilon|h|
    $
    for every $i\in[n]$ and $x,x+he_i\in[0,1]^n$.
\end{enumerate}
These targets let us transfer
the equilibrium first-order conditions of \cref{thm:first-order-character}
to approximate KKT conditions of the target quadratic program. (See \cref{fig:first-order-transfer} for an illustration).
Thus, given any quadratic instance $\mathcal{I}=(A,c,\varepsilon)$, if we can construct, in polynomial time, a Fisher market with chores that satisfies both targets, then we can also recover an $\varepsilon$-KKT point of $\mathcal{I}$ in polynomial time, which finishes the reduction.

In the following, we focus on how to construct a market satisfying
these targets.

\begin{figure}[t]
    \centering
    \resizebox{\linewidth}{!}{%
        \begin{tikzpicture}[
            condition/.style={align=center, inner sep=0pt, outer sep=0pt},
            explanation/.style={align=center, font=\small},
            implication/.style={
                double,
                double distance=1.2pt,
                line width=0.5pt,
                -{Implies[length=3.5mm,width=3.5mm]}
            }
        ]

        \node[condition, minimum width=2.7cm] (prices) at (0,0) {
            $\displaystyle \Psi'(p^\star;v)\leq 0$
        };

        \node[condition, minimum width=2.7cm] (curve) at (8.6,0) {
            $\displaystyle
            \left.\frac{\dd}{\dd h}\Psi(p(x^\star+h\tilde v))
            \right|_{h=0^+}\leq 0$
        };

        \node[condition, minimum width=3.3cm] (quadratic) at (17.2,0) {
            $\displaystyle \nabla Q(x^\star)\cdot\tilde v\leq\varepsilon$
        };

        \draw[implication]
            ([xshift=2mm]prices.east) --
            node[explanation, above=10pt] {
                Map variables to prices\\[5pt]
                $p(x^\star)=p^\star$
            }
            ([xshift=-2mm]curve.west);

        \draw[implication]
            ([xshift=2mm]curve.east) --
            node[explanation, above=10pt] {
                Approximate the quadratic\\[5pt]
                $\Psi(p(x))=C+\alpha Q(x)+R(x)$
            }
            ([xshift=-2mm]quadratic.west);

        \node[font=\small] at (0,-0.85)
            {Equilibrium prices};
        \node[font=\small] at (8.6,-0.85)
            {First-order condition for $\Psi(p(x))$};
        \node[font=\small] at (17.2,-0.85)
            {$\varepsilon$-KKT point of $Q$};

        \end{tikzpicture}%
    }

    \caption[Transferring the equilibrium first-order condition to the target quadratic.]%
    {Transferring the equilibrium first-order condition to the target quadratic.%
     \protect\footnotemark}
    \label{fig:first-order-transfer}
\end{figure}
\footnotetext{
    We define
    $\displaystyle
            \left.\frac{\dd}{\dd h}\Psi(p(x^\star+h\tilde v))
            \right|_{h=0^+}:=\displaystyle \lim_{h\downarrow 0}
    \frac{\Psi(p(x^\star+h\tilde v))-\Psi(p(x^\star))}{h}$.}

\ignore{\footnote{Here $\tilde v$ is a feasible coordinate direction at $x^\star$,
and $v$ is the induced price direction at $p^\star$. The derivative in
the middle of the figure denotes the right-hand limit
$\displaystyle \lim_{h\downarrow 0}
\frac{\Psi(p(x^\star+h\tilde v))-\Psi(p(x^\star))}{h}$.}
}

\paragraph{Core Idea of Our Construction.} The main goal of our construction is to achieve Target~\labelcref{target:objective-approximation}, which requires approximating
the quadratic objective $Q(x)=\frac12 x^\top A x+c^\top x$ using the logarithmic objective $\Psi(p(x))$.
Our starting point is the Taylor expansion: for small $t$,
\[
-\log(1+tx_j)=-tx_j+\frac{t^2x_j^2}{2}+O(t^3),
\]
which gives both linear and quadratic terms while keeping higher-order terms under control.
This suggests a plausible first route: constructing a market and an appropriate mapping \(p\) such that $\Psi(p(x))$ contains
terms of the form $-\log(1+tx_j)$. Our initial construction is a simple basic market that achieves this goal (\Cref{sec:basic-market}). However, the resulting quadratic terms obtained from this initial construction
generally do not match the target $\alpha Q(x)$. Building on the basic market, we correct the mismatch by adding agents and chores whose contributions in the optimization objective
supply the required missing terms (\Cref{sec:gadgets}), which is the most technical part in this paper.

\paragraph{Construction of the Basic Market.}
In \cref{sec:basic-market} we construct a basic market
$\mathcal{F}_0$ containing \(n+1\) agents and chores, both indexed by $0,\ldots,n$. Each agent has unit earning requirement and each chore has unit supply.
For a small positive rational $t$, set $x_0:=0$ and define mapping \(p\) by
\[
\begin{gathered}
p:[0,1]^n\longrightarrow
\mathcal{P}_{\textnormal{basic}}:=\left\{p\in\mathbb{R}_{>0}^{n+1}:
\sum_{j=0}^{n}p_j=n+1\right\},\\[3pt]
p_j(x):=\frac{1+tx_j}{1+t\bar{x}},
\qquad \bar{x}=\frac{1}{n+1}\sum_{i=0}^{n}x_i.
\end{gathered}
\]
To generate the desired logarithmic terms, we set the disutilities of agents as follows: for agent $a$ and chore $j$, the disutility $d_{aj}=1$ if $j = a$ or $j = 0$, and $d_{aj} = 1 + t$ otherwise. Every basic agent has disutility 1 for chore 0 and for her own chore, and disutility $1+t$ for every other basic chore. In particular, agent 0 has disutility $1+t$
for each basic chore other than chore 0. This implies $\mu_a(p(x))=\max_{j} p_j(x)/d_{aj} = p_a(x)$. Therefore, the optimization objective for CE is 
$$\Psi_0(p(x))=-\sum_{j=0}^{n}\log p_j(x) = -\sum_{j=0}^{n} \log \frac{1+tx_j}{1+t\bar{x}}.$$
Expanding the right-hand side in \(t\) by Taylor expansion and ignoring third-order and higher-order terms, we obtain
\[
\begin{aligned}
\Psi_0(p(x))
&\approx\frac{t^2}{2(n+1)}\sum_{0\leq j<j'\leq n}(x_j-x_{j'})^2.
\end{aligned}\]

However, the quadratic terms in $\Psi_0(p(x))$ generated from the basic market could differ from the target quadratic term \(\frac{\alpha}{2}x^\top A x\) in \(\alpha Q(x)\).
We express their differences as 
\[
-\frac{t^2}{2}
\sum_{0\leq j<j'\leq n}w_{jj'}(x_j-x_{j'})^2,
\]
with the coefficients \(w_{jj'}\) given in \eqref{eq:basic-weights}. Moreover, \(\Psi_0(p(x))\) does not provide the linear term \(\alpha c^\top x\) in \(\alpha Q(x)\). The basic market lacks the quadratic term differences and the linear term. In the following, we show how to modify the basic market to add these missing terms. 

\paragraph{Modifying the Basic Market.} 
Fix a pair $(j,j')$. In \cref{sec:gadgets}, we show how to modify the basic market that generates the missing term 
$-\frac{t^2}{2}w_{jj'}(x_j-x_{j'})^2.$ 
It is not easy to directly generate 
this quadratic term. The key observation is that we can approximate the quadratic term by the summation of a sequence of absolute-value terms, and we can generate each absolute-value term by a gadget. We make the above intuition more specific in the following. 

Recall the price mapping $p_j(x) = (1+tx_j)/(1+t\bar{x})$ and define $\Delta:=\log(p_j(x)/p_{j'}(x))$. By the Taylor expansion again, we have
\[
-\frac{w_{jj'}t^2}{2}(x_j-x_{j'})^2
=-\frac{w_{jj'}}{2}\Delta^2+O(t^3).
\]
The challenge is to generate the squared logarithm $\Delta^2$: each agent contributes
a term of the form $-B_a\log\mu_a(p)$ to $\Psi$, which is only a logarithm. 
To overcome this difficulty, we approximate $-\frac{w_{jj'}}{2}\Delta^2$ by
a summation of a sequence of absolute-value terms. We can separate an absolute-value term from the quadratic term using  the following identity
\[
-\frac{w_{jj'}}{2}\Delta^2
=-w_{jj'}s_0|\Delta|
-\frac{w_{jj'}}{2}(|\Delta|-s_0)^2
+\frac{w_{jj'}}{2}s_0^2.
\]
Notice that we can apply the identity again on $-\frac{w_{jj'}}{2}(|\Delta|-s_0)^2$ with $s_1 > 0$. More generally, we can continue this process with positive constants \(s_1,\dots,s_{L-1}\).  Define $\Delta_0:=\Delta$ and
$\Delta_{k+1}:=|\Delta_k|-s_k$ for $k  \ge 0$. After $L$ steps we obtain
\[
-\frac{w_{jj'}}{2}\Delta^2
=
\underbrace{-w_{jj'}\sum_{k=0}^{L-1}s_k|\Delta_k|}
_{\text{absolute-value terms}}
\underbrace{-\frac{w_{jj'}}{2}\Delta_L^2}_{\text{remainder}}
+
\underbrace{\frac{w_{jj'}}{2}\sum_{k=0}^{L-1}s_k^2}
_{\text{constant}}.
\]
The upshot is, with a suitable choice of \(\{s_k\}\), we can make the remainder $\Delta_L^2$ sufficiently small. Hence, we can express \(-\frac{w_{jj'}}{2}\Delta^2\) by a summation of a sequence of absolute-value terms, a small remainder and a constant.

It remains to generate the absolute-value terms. In \cref{subsec:max-min-gadget} we introduce the \emph{Maximum-Minimum gadget}. Given any two chores $j, j'$ as input, the gadget adds 3 agents and 2 chores to market and generates the desired $|\Delta|=|\log(p_j(x)/p_{j'}(x))|$ term. Moreover, we can apply this gadget recursively on the new chores and generate the required absolute-value terms recursively as follows (details in \Cref{subsec:gadget-telescoping}):
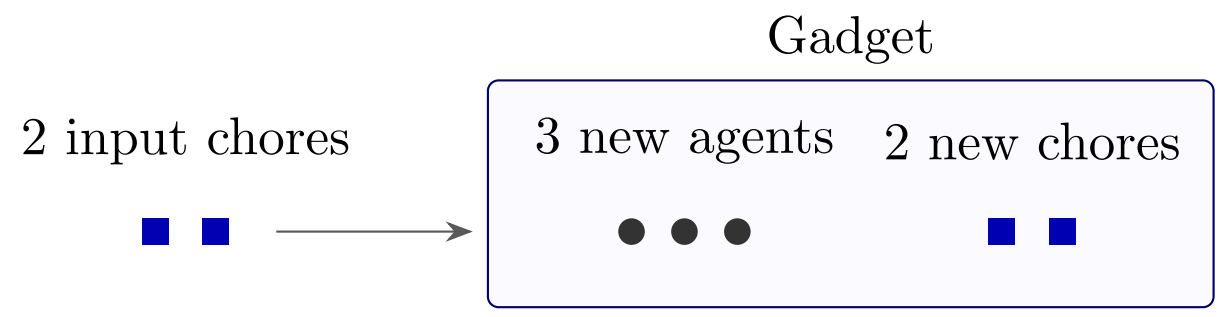
\begin{center}
\begin{tikzpicture}[
  x=1cm,y=1cm,font=\small,
  chore/.style={rectangle,fill=blue!70!black,inner sep=0pt,minimum size=5pt},
  agent/.style={circle,fill=black!80,inner sep=0pt,minimum size=5pt}
]
  \node at (0,.65) {2 input chores};
  \node[chore] at (-.2,.05) {};
  \node[chore] at (.2,.05) {};
  \draw[-{Stealth[length=1.8mm]},black!65] (.6,.05) -- (1.9,.05);
  \draw[rounded corners=2pt,draw=blue!40!black,fill=blue!2,line width=.4pt]
    (2,-.45) rectangle (6.8,1.05);
  \node at (4.4,1.32) {Gadget};
  \node at (3.3,.65) {3 new agents};
  \node[agent] at (2.95,.05) {};
  \node[agent] at (3.3,.05) {};
  \node[agent] at (3.65,.05) {};
  \node at (5.6,.65) {2 new chores};
  \node[chore] at (5.4,.05) {};
  \node[chore] at (5.8,.05) {};
\end{tikzpicture}
\end{center}


\begin{enumerate}[label=\textbf{Step \arabic*.},leftmargin=*,
                  topsep=3pt,itemsep=3pt,parsep=0pt]
\item The first gadget, takes 2 chores \(j,j^{\prime}\) from the basic market as inputs, 
adds 3 agents and 2 chores to the basic market. These 3 new agents generate the $-w_{jj'}s_0|\Delta_0|$ approximately and some extra terms.
\item The second gadget, takes 2 chores added by the first gadget as inputs,
adds 3 agents and 2 chores to the market. These 3 new agents generate the $-w_{jj'}s_1|\Delta_1|$ approximately and some extra terms.
\item[{\makebox[\labelwidth][c]{\raisebox{2pt}[0pt][0pt]{\(\vdots\;\vdots\)}}}] \mbox{}
\item[\textbf{Step $L$.}] The \(L\)-th gadget takes 2 chores added by the \((L-1)\)-th gadget as inputs, adds 3 agents and 2 chores to the market. 
These 3 new agents generate the $-w_{jj'}s_{L-1}|\Delta_{L-1}|$ approximately and some extra terms.
\end{enumerate}

These steps form a \emph{chain of gadgets}. We choose the constants $s_k$ explicitly in \cref{clm:final-parameter-choice}
to ensure the final quadratic remainder satisfying target \labelcref{target:objective-approximation}.

There are a few remarks on the construction. First, the gadget may introduce unwanted extra terms at each step. In \cref{subsec:extra-terms}, we adjust the basic agents' earning
requirements to cancel these terms and generate the desired linear term
$\alpha c^\top x$ in $\alpha Q(x)$. Second, the changes introduced in the basic market $\mathcal F_0$ could break the payment balance condition \eqref{eq:market_clearing}.
In \Cref{subsec:normalization-final-program}, we adjust the supplies of chores and renormalize
the prices so that the final \(p\) satisfies payment balance.

\paragraph{Completing the Market and Verifying the Targets.}
In \Cref{sec:completed-market}, we complete the market
construction and specify all parameters in our new market.
We then verify the two targets.
For Target~\labelcref{target:price-encoding},
\Cref{sec:equilibrium-recovery} shows that every equilibrium
price vector lies in the constructed price family.
The corresponding point can be recovered in polynomial time from equilibrium.
For Target~\labelcref{target:objective-approximation},
\Cref{subsec:market-error-bound} combines the earlier error
bounds to obtain the required objective approximation.
\ignore{We also verify the required right derivatives in
\Cref{subsec:market-right-derivative}.
Together, these results establish all properties in
\Cref{prop:market-target}.}

\section{Sufficient Conditions for the Reduction}

In this section, we formalize
Targets~\labelcref{target:price-encoding}
and~\labelcref{target:objective-approximation}
from \Cref{sec:overview} as \cref{prop:market-target}
and give a proof.
In the remaining sections, we
construct a market that satisfies these targets, which gives the reduction needed to establish \cref{thm:exact-equilibrium-cls}.

\begin{proposition}
\label{prop:market-target}
Given an instance $\mathcal{I}=(A,c,\varepsilon)$ of
$\varepsilon$-\textup{\textsc{quadratic-KKT}}, let
$\mathcal{F}=f(\mathcal{I})
=(N,M,\allowbreak {}\{d_{aj}\},\{B_a\},\{S_j\})$
be a linear Fisher market constructed from $\mathcal{I}$
in \Cref{fig:reduction-overview},
with strictly positive rational data.
Suppose that $\mathcal{F}$ satisfies the following properties:

\begin{enumerate}[label=(\alph*)]
    \item \label{item:market-price-mapping}
    \textbf{Price Mapping.}
    There exists a mapping $p:[0,1]^n\to\mathcal P$ with $\mathcal P$ defined in~\eqref{def-normalized-price}, such that:
    \begin{enumerate}[label=(\roman*)]
                \item \label{item:market-decoding}
        There exists a polynomial-time decoding mapping $g$
        that returns a rational
        point $x^\star\in[0,1]^n$ satisfying
        $$
        x^\star=g(\mathcal{I},(p^\star,z^\star))
        \qquad\text{and}\qquad
        p(x^\star)=p^\star,
        $$
        for any exact rational equilibrium
        $(p^\star,z^\star)$ of $\mathcal{F}$.

      \item \label{item:market-right-derivative}
        Given any $x\in[0,1]^n$, 
        $p(x+h\tilde v)$ has a finite right derivative
        at $h=0$, where $\tilde v$ is a feasible
        coordinate direction at $x$.
        
    \end{enumerate}

    \item \label{item:market-error-bound}
    \textbf{Error Bound.}
    There exist constants $\alpha>0$ and $C\in\mathbb R$,
    independent of $x$, and a function
    $R:[0,1]^n\to\mathbb R$ such that
    $$
    \Psi(p(x))=C+\alpha Q(x)+R(x)
    \qquad\text{for every }x\in[0,1]^n.
    $$
    Moreover, there exists a constant
    $0\leq\beta\leq\alpha\varepsilon$ such that, for every
    $i\in[n]$ and every coordinate segment from $x$ to
    $x+he_i$ contained in $[0,1]^n$,
    $$
    |R(x+he_i)-R(x)|\leq\beta|h|.
    $$
\end{enumerate}

Then, we can recover an
$\varepsilon$-\textnormal{KKT} point of $\mathcal{I}$
in polynomial time from any exact rational equilibrium
of $\mathcal{F}$.
\end{proposition}
\begin{proof}
Let $(p^\star,z^\star)$ be an exact rational equilibrium of
$\mathcal F$. By \Cref{item:market-decoding} we can compute
$x^\star=g(\mathcal I,(p^\star,z^\star))\in [0,1]^n$ satisfying $p(x^\star)=p^\star$.
Fix any feasible coordinate direction
$\tilde v\in\{\pm e_i:i\in[n]\}$ at $x^\star$, and define
$\gamma(h):=p(x^\star+h\tilde v)$ for sufficiently small $h\geq 0$.
The right-derivative assumption (\cref{item:market-right-derivative}) gives a vector $v$ such that
$\gamma(h)=p^\star+hv+o(h)$ as $h\downarrow0$.
Since $\gamma(h)\in\mathcal P$ and $\gamma(0)=p^\star$,
\[
\sum_{j\in M}S_jv_j
=\lim_{h\downarrow0}
\frac{\sum_{j\in M}S_j\gamma_j(h)
-\sum_{j\in M}S_jp_j^\star}{h}
=0.
\]
Together with $p^\star>0$, this implies that $v$ is a feasible direction at $p^\star$. Hence
\cref{thm:first-order-character} gives
$\Psi'(p^\star;v)\leq0$.

Each $\mu_a(p)=\max_{j\in M}p_j/d_{aj}$ is the maximum of
finitely many linear functions and is positive near $p^\star$.
Thus $\Psi(p)=-\sum_{a\in N}B_a\log\mu_a(p)$ is locally
Lipschitz there. For some constant $K>0$ and all sufficiently
small $h>0$, we therefore have
\[
\left|\Psi(\gamma(h))-\Psi(p^\star+hv)\right|
\leq K\left\|\gamma(h)-p^\star-hv\right\|
=o(h).
\]
Consequently, the right directional derivative of $\Psi(p(x))$
at $x^\star$ in direction $\tilde v$ exists and satisfies
\begin{equation}
\label{eq:composition-derivative-1}
\begin{aligned}
\left.\frac{\dd}{\dd h}\Psi(p(x^\star+h\tilde v))\right|_{h=0^+}
&=\lim_{h\downarrow0}
\frac{\Psi(\gamma(h))-\Psi(p^\star)}{h}\\
&=\lim_{h\downarrow0}
\frac{\Psi(p^\star+hv)-\Psi(p^\star)}{h}\\
&=\Psi'(p^\star;v)\leq0.
\end{aligned}
\end{equation}
On the other hand, by \Cref{item:market-error-bound} we have
\begin{equation}
\label{eq:composition-derivative-2}
\begin{aligned}
\left.\frac{\dd}{\dd h}\Psi(p(x^\star+h\tilde v))\right|_{h=0^+}
&=\lim_{h\downarrow0} \frac{\Psi(p(x^\star+h\tilde v))-\Psi(p(x^\star))}{h}\\
&=\lim_{h\downarrow0}\left ( \alpha\frac{Q(x^\star+h\tilde v)-Q(x^\star)}{h}
+\frac{R(x^\star+h\tilde v)-R(x^\star)}{h}\right )\\
&\geq\alpha\cdot\lim_{h\downarrow0}\frac{Q(x^\star+h\tilde v)-Q(x^\star)}{h}
-\beta\\
&=\alpha\nabla Q(x^\star)\cdot\tilde v-\beta,
\end{aligned}
\end{equation}
where the second equality follows from $\Psi(p(x))=C+\alpha Q(x)+R(x)$ for every $x\in[0,1]^n$, and the inequality holds because $|R(x+he_i)-R(x)|\leq\beta|h|$ for all $x,x+he_i\in[0,1]^n$. Combining \eqref{eq:composition-derivative-1} and \eqref{eq:composition-derivative-2}, we get
\[
\nabla Q(x^\star)\cdot\tilde v\leq\frac{\beta}{\alpha}\leq\varepsilon.
\]
Since $\tilde v$ was arbitrary feasible direction at $x^\star$, then $x^\star$ is an
$\varepsilon$-KKT point by~\eqref{eq:quadratic_kkt_nabla}.
The decoder $g$ computes this rational point in polynomial
time by assumption.
\end{proof}

\ignore{
\begin{remark}\label{rem:injectivity}
For $\varepsilon>0$, a fixed price mapping $p$ satisfying \cref{prop:market-target}
for every target quadratic is injective.
The formal statement and proof appear in
\cref{lem:universal-encoding-injective} in \cref{app:injectivity}.
\end{remark}
}

\section{Basic Market and Quadratic Approximation}
\label{sec:basic-market}

As outlined in \Cref{prop:market-target}, our goal is to use market
equilibrium prices to encode an $\varepsilon$-KKT point $x^\star$
of the target quadratic program.
\Cref{thm:first-order-character} characterizes equilibrium prices
through the first-order conditions of a logarithmic program with
objective
$$
\Psi(p)=-\sum_{a=1}^n B_a\log\mu_a(p).
$$
This motivates us to relate the logarithmic objective to the
target quadratic function.

The core intuition relies on the Taylor expansion of the logarithmic terms.
Suppose $\mu_a(p)=1+\tau_a$ for small $\tau_a$. By Taylor expansion, we have
\begin{align}
\Psi(p)=-\sum_{a=1}^n B_a \log(1+\tau_a)
&=-\sum_{a=1}^n B_a \left(\tau_a-\frac{\tau_a^2}{2}+O(\tau_a^3)\right). \label{eq::log::expansion}
\end{align}
Under a suitable normalization condition, the linear term $\sum_{a=1}^n B_a \tau_a$ vanishes identically. The remaining quadratic term  $\frac{1}{2} \sum_{a=1}^n B_a \tau_a^2$ provides the structural quadratic term $Q(x)$, while the higher-order terms $O(\tau_a^3)$ are tightly bounded and absorbed into an error term $R_0(x)$. 


In this section, we construct a \emph{basic
market} whose logarithmic objective has a quadratic leading term
proportional to a sum of squared coordinate differences.
The basic market's quadratic contribution provides the
starting point for approximating a positive multiple of
$\frac12x^\top Ax$, the quadratic part of $Q(x)$.
The remaining quadratic corrections identified at the end of this section are implemented
by the market gadgets in \Cref{sec:gadgets}. The linear term $c^\top x$ of $Q(x)$ is treated separately in \Cref{subsec:extra-terms}.

The construction and analysis proceed as follows.
We first specify the basic market's chores, agents, and
disutilities, together with a family of unnormalized prices encoding $x$.
We then rescale the unnormalized prices to satisfy payment
balance while preserving the price ratios that encode $x$.
Next, we derive a quadratic approximation of the basic
market's logarithmic objective evaluated at the normalized prices
and bound changes in the remainder when a single coordinate of $x$ varies.
Finally, we compare the basic market's quadratic contribution
with the scaled target
$\frac{\alpha}{2}x^\top Ax$ to determine the remaining quadratic corrections.

\paragraph{Construction of the Basic Market.}
To apply the expansion \eqref{eq::log::expansion} to the variables of the quadratic program,
we introduce a positive rational parameter $t\in(0,1]$, to be
chosen sufficiently small. Then
$$
\log(1+tx_j)
=
tx_j-\frac{t^2x_j^2}{2}+O(t^3),
$$
uniformly over $x_j\in[0,1]$.

We therefore construct the basic market to generate terms of the
form $-\log(1+tx_j)$. For each coordinate $x_j$, $j\in[n]$, introduce a corresponding
chore $j$.
We then add a reference chore $0$ with the fixed auxiliary
coordinate $x_0:=0$.
The resulting $n+1$ chores form the set $M=\{0,\ldots,n\}$, each with unit supply
$S_j=1$.
For $x\in[0,1]^n$, define the unnormalized prices
\[
u_j(x):=1+tx_j,\qquad j\in\{0,\ldots,n\}.
\]
In particular, $u_0(x)=1$.

Introduce $n+1$ agents, one associated with each chore and
having $B_a=1$.
Let $N$ denote the agent set. For the agent $a$ associated with chore $j$, set
$$
d_{aj'}
=
\begin{cases}
1, & j'\in\{0,j\},\\
1+t, & \text{otherwise}.
\end{cases}
$$
At prices $u(x)$, her associated chore $j$ offers ratio $u_j(x)/d_{aj}=u_j(x) = 1 + t x_j \geq1$,
while the reference chore 0 offers ratio $1$. For every $j' \notin \{0, j\}$, the price-to-disutility ratio is
$$
\frac{u_{j'}(x)}{1+t}
=
\frac{1+tx_{j'}}{1+t}
\leq1.
$$
Thus, her associated chore attains her maximum
price-to-disutility ratio:
$$
\mu_a(u(x))
=
\max_{j'\in M}\frac{u_{j'}(x)}{d_{aj'}}
=
u_j(x).
$$
Since there is one agent for each chore, summing these contributions gives the basic market's logarithmic objective at the unnormalized prices $u(x)$:
$$
-\sum_{a\in N}\log\mu_a(u(x))
=
-\sum_{j=0}^{n}\log(1+tx_j).
$$

\paragraph{Normalization of the Prices.}\label{para:normalization}
The prices $u(x)$ yield the desired logarithmic terms, but they need not satisfy payment balance. Recall from \Cref{def:CE} that every equilibrium satisfies
payment balance:
$
\sum_{j\in M}S_jp_j=\sum_{a\in N}B_a.
$
In the basic market, all supplies are one and the total earning requirement is $n+1$. For
$x\neq\mathbf 0$, the unnormalized prices satisfy
$$
\sum_{j=0}^{n}u_j(x)
=
n+1+t\sum_{j=1}^{n}x_j
>
n+1
=
\sum_{a\in N}B_a.
$$
We therefore rescale the prices to satisfy payment balance by multiplying every price by the same positive factor..
Let
$
\bar{x}:=\frac{1}{n+1}\sum_{j=0}^{n}x_j,
$
and set
\begin{equation}
p_j(x)
:=
\frac{(n+1)u_j(x)}{\sum_{j'=0}^{n}u_{j'}(x)}
=
\frac{1+tx_j}{1+t\bar{x}},
\qquad j\in\{0,\ldots,n\}.
\label{eq:naive-price}
\end{equation}
These prices are positive and satisfy
$
\sum_{j=0}^{n}p_j(x)
=
n+1
=
\sum_{a\in N}B_a.
$

This common rescaling preserves the ordering of each agent's price-to-disutility ratios. Hence, for the agent $a$ associated
with chore $j$, chore $j$ remains optimal and
$$
\mu_a(p(x))
=
\frac{\mu_a(u(x))}{1+t\bar{x}}
=
p_j(x).
$$
Writing
$
\Psi_0(p):=-\sum_{a\in N}\log\mu_a(p)
$
for the basic-market logarithmic objective, we therefore obtain
$$
\begin{aligned}
\Psi_0(p(x))
&=-\sum_{j=0}^{n}\log p_j(x)\\
&=(n+1)\log(1+t\bar{x})
-\sum_{j=0}^{n}\log(1+tx_j).
\end{aligned}
$$
The term $(n+1)\log(1+t\bar{x})$ comes from rescaling all $n+1$ prices.

\ignore{
\begin{remark}
\label{rem:basic-reference-chore}
Encoding $x\in[0,1]^n$ with $n$ agents and $n$ chores generates
the desired logarithmic terms at unnormalized prices $u_j(x)=1+tx_j$,
but payment balance requires
$$
\widehat p_j(x):=\frac{n(1+tx_j)}{n+t\sum_{j'=1}^{n}x_{j'}},\qquad j\in[n].
$$
Since $\widehat p(r\mathbf 1)=\mathbf 1$ for every $r\in[0,1]$,
this violates the injectivity requirement in \Cref{rem:injectivity}.
The reference chore fixes $u_0(x)=1$, allowing recovery by
$x_j=\frac1t\left(\frac{p_j(x)}{p_0(x)}-1\right)$ for $j\in[n]$;
thus $p(x)=p(x')$ implies $x=x'$.
\end{remark}}

\paragraph{Quadratic Approximation and Error Bound.}
We now apply Taylor expansion to $\Psi_0(p(x))$ using the exact logarithmic expression obtained by normalization. Since $x_j,\bar{x}\in[0,1]$, the uniform Taylor expansion used in the construction applies to every logarithm in this expression.
The first-order terms sum to
\[
t\left((n+1)\bar{x}-\sum_{j=0}^{n}x_j\right)=0.
\]
The second-order contribution is
\[
\frac{t^2}{2}\left(\sum_{j=0}^{n}x_j^2-(n+1)\bar{x}^2\right)
=\frac{t^2}{2}\sum_{j=0}^{n}(x_j-\bar{x})^2.
\]
Thus the common normalization cancels the linear terms
and leaves a quadratic term measuring squared deviations from the mean.

Summing the Taylor remainders gives an error
$O((n+1)t^3)$, uniformly over $x\in[0,1]^n$ as $t\downarrow0$.
A bound on the remainder's value alone does not control
its derivatives, which are what the first-order conditions depend on.
The following theorem identifies this term and bounds the error
in the form needed for \Cref{prop:market-target}.\footnote{%
Throughout the market construction, we maintain \emph{third-order
coordinate error bounds}. The Taylor expansion
$\log(1+tx_j)=tx_j-t^2x_j^2/2+O(t^3)$ gives the quadratic term of
order $t^2$ and third-order remainder in \Cref{thm:basic-quadratic-approximation},
so we choose $\alpha$ proportional to $t^2$.
The total coordinate error bound $\beta=O(t^3)$ then gives
$\beta/\alpha=O(t)$, so sufficiently small $t$ ensures
$\beta\leq\alpha\varepsilon$, as required by
\Cref{item:market-error-bound} of \Cref{prop:market-target}.
See \eqref{eq:reduction-parameter-choice} for the precise choice.}

\begin{theorem}[Quadratic Approximation]
\label{thm:basic-quadratic-approximation}
For every $x\in[0,1]^n$,
\begin{equation}
\label{eq:basic-quadratic-approximation}
\begin{aligned}
\Psi_0(p(x))
&=
\frac{t^2}{2(n+1)}
\sum_{0\leq j<j'\leq n}(x_j-x_{j'})^2+R_0(x).
\end{aligned}
\end{equation}

The remainder satisfies
\begin{equation*}
|R_0(x+he_j)-R_0(x)|\leq2t^3|h|
\end{equation*}
for every $j\in[n]$ and every coordinate segment from $x$
to $x+he_j$ contained in $[0,1]^n$.
Moreover, for every feasible coordinate direction
$\widetilde v\in\{\pm e_j:j\in[n]\}$ at $x$, the right
directional derivative exists and satisfies
\[
\left|
R_0'(x;\widetilde v)
\right|
\leq2t^3.
\]
\end{theorem}
We defer the proof to \cref{app:error-basic}.

\ignore{
\begin{proof}[\textcolor{red}{to be polished}]
Using
$$
\sum_{0\leq j<j'\leq n}(x_j-x_{j'})^2
=
(n+1)\sum_{j=0}^{n}x_j^2
-
\left(\sum_{j=0}^{n}x_j\right)^2,
$$
we define the remainder directly by
$$
R_0(x):=\Psi_0(p(x))
-\frac{t^2}{2}
\left(
\sum_{j=0}^{n}x_j^2-(n+1)\bar{x}^2
\right).
$$
This gives the claimed quadratic decomposition.

The explicit logarithmic expression for $\Psi_0(p(x))$ shows that
$R_0$ has a smooth extension to a neighborhood of $[0,1]^n$.
Since
$$
\Psi_0(p(x))
=
(n+1)\log(1+t\bar{x})
-\sum_{j=0}^{n}\log(1+tx_j),
$$
differentiation gives, for every $j\in[n]$,
$$
\begin{aligned}
\frac{\partial R_0}{\partial x_j}(x)
&=
\frac{t}{1+t\bar{x}}
-\frac{t}{1+tx_j}
-t^2(x_j-\bar{x})\\
&=
t^3\left(
\frac{\bar{x}^2}{1+t\bar{x}}
-
\frac{x_j^2}{1+tx_j}
\right).
\end{aligned}
$$
Because $0\leq\bar{x},x_j\leq1$,
$$
\left|\frac{\partial R_0}{\partial x_j}(x)\right|
\leq
t^3\left(
\frac{\bar{x}^2}{1+t\bar{x}}
+
\frac{x_j^2}{1+tx_j}
\right)
\leq2t^3.
$$
The mean value theorem along any coordinate segment contained
in the box therefore yields
$$
|R_0(x+he_j)-R_0(x)|\leq2t^3|h|.
$$

Finally, smoothness ensures that the right directional derivative
along every feasible direction $\widetilde v=\pm e_j$ exists
and equals
$$
R_0'(x;\widetilde v)
=
\pm\frac{\partial R_0}{\partial x_j}(x).
$$
Its absolute value is thus at most $2t^3$.
\end{proof}
}

\paragraph{Correction for the Target Quadratic of $Q(x)$.}

We now determine how to adjust the basic market's quadratic
contribution identified in \Cref{thm:basic-quadratic-approximation},
$$
\frac{t^2}{2(n+1)}
\sum_{0\leq j<j'\leq n}(x_j-x_{j'})^2,
$$
to match the target $\frac{\alpha}{2}x^\intercal Ax$, where
$\alpha>0$ is the scaling parameter
introduced in \Cref{prop:market-target}.

Since $A$ is symmetric and $x_0=0$, the discrepancy can be written as
$$
\begin{aligned}
&\frac{t^2}{2(n+1)}
\sum_{0\leq j<j'\leq n}(x_j-x_{j'})^2
-\frac{\alpha}{2}x^\intercal Ax \\
&~~~~~~~~~~~~= \frac{t^2}{2}
\sum_{1\leq j<j'\leq n}
\left(
\frac{1}{n+1}+\frac{\alpha}{t^2}A_{jj'}
\right)
(x_j-x_{j'})^2
+
\frac{t^2}{2}
\sum_{j=1}^{n}
\left(
\frac{1}{n+1}
-\frac{\alpha}{t^2}\sum_{j'=1}^{n}A_{jj'}
\right)
(x_j-x_0)^2.
\end{aligned}
$$
Define the correction weights by the coefficients in parentheses:
\begin{equation}
\label{eq:basic-weights}
\begin{aligned}
w_{jj'}
&:=
\frac{1}{n+1}+\frac{\alpha}{t^2}A_{jj'},
&&1\leq j<j'\leq n,\\
w_{0j}
&:=
\frac{1}{n+1}-\frac{\alpha}{t^2}
\sum_{j'=1}^{n}A_{jj'},
&&1\leq j\leq n,
\end{aligned}
\end{equation}
and extend the notation symmetrically by $w_{j'j}=w_{jj'}$.

To ensure that all correction weights are positive, we set
\begin{align}\label{eq:choice-of-alpha}
   \alpha:= \frac{t^2}{2(n+1)\left(
1+\sum_{i,j=1}^{n}|A_{ij}|+\sum_{i=1}^{n}|c_i|
\right)}.
\end{align}

We choose $\alpha$ in this way to facilitate the subsequent analysis and $t>0$ will be specified later.

The required correction is therefore
$$
-\frac{t^2}{2}
\sum_{0\leq j<j'\leq n}w_{jj'}(x_j-x_{j'})^2.
$$
In \Cref{sec:gadgets}, we realize this second-order contribution
by adding a market gadget for each pair $0\leq j<j'\leq n$,
while introducing only third-order error along feasible
coordinate changes.

\section{Approximating the Target Objective}
\label{sec:gadgets}
In this section, we introduce auxiliary agents and chores to implement the
quadratic corrections identified in \cref{sec:basic-market}.
For each pair of base chores $0\leq j<j'\leq n$, the required
contribution is
\begin{equation}
    \label{eq:required-contribution}
    -\frac{w_{jj'}t^2}{2}(x_j-x_{j'})^2,
\end{equation}
where $w_{jj'}>0$ is defined in~\eqref{eq:basic-weights} and $x_0=0$.
These negative squared differences modify the fixed quadratic of
the basic market to represent the target matrix $A$.

\subsection{Observation and Motivation}
\label{subsec:gadget-motivation}

To obtain the quadratic correction~\eqref{eq:required-contribution}, we first relate squared variable
differences to logarithmic price ratios. Fix a pair of base chores
$(j,j')$ with $0\leq j<j'\leq n$. The price mapping of
\cref{sec:basic-market} satisfies
\[
\frac{p_j(x)}{p_{j'}(x)}
=\frac{1+tx_j}{1+tx_{j'}}.
\]
A Taylor expansion gives
\begin{equation}
\label{eq:gadget-log-square-approximation}
\begin{aligned}
\log\frac{p_j(x)}{p_{j'}(x)}
&=t(x_j-x_{j'})+O(t^2),\\
\left(\log\frac{p_j(x)}{p_{j'}(x)}\right)^2
&=t^2(x_j-x_{j'})^2+O(t^3),
\end{aligned}
\end{equation}
over $x\in[0,1]^n$. Thus, the required
correction can be approximated by a squared logarithm,
\begin{equation}
\label{eq:desired-contribution}
-\frac{w_{jj'}t^2}{2}(x_j-x_{j'})^2 = -\frac{w_{jj'}}{2}\left(\log\frac{p_j(x)}{p_{j'}(x)}\right)^2 + O(t^3),
\end{equation}

However, generating this squared logarithm using auxiliary agents and chores is not immediate:
each auxiliary agent contributes
$$-B_a\log \mu_a(p)$$
to $\Psi$, and this contribution contains no explicit squared term.
Hence, we introduce an algebraic identity:

\begin{equation}
  -\frac12 a^2
=
-b|a|
\;-\;
\frac12(|a|-b)^2
\;+\;\frac12 b^2,\quad\text{for all }a,b>0.
\label{eq:algebraic-identity}
\end{equation}
Given arbitrary constant \(s_0>0\), this identity does a transfer and approximates the desired square as the sum of an
absolute-value term plus a remainder term. In \cref{subsec:max-min-gadget}, we use a technical tool to add auxiliary agents producing the absolute-value term. 
\begin{equation}
\label{eq:gadget-square-identity-1}
-\frac{w_{jj'}}{2}
\left(\log\frac{p_j(x)}{p_{j'}(x)}\right)^2
=
\underbrace{
  -w_{jj'}s_0\left|\log\frac{p_j(x)}{p_{j'}(x)}\right|
}_{\text{absolute-value term}}
\underbrace{
  -
  \frac{w_{jj'}}{2}
  \left(\left|\log\frac{p_j(x)}{p_{j'}(x)}\right|-s_0\right)^2
  +
  \frac{w_{jj'}}{2}s_0^2
}_{\text{a remainder term}}
\end{equation}
This identity suggests approximating the desired contribution in \eqref{eq:desired-contribution} by the absolute-value term.
To do so, it would suffice to introduce auxiliary agents that generate this term and choose $s_0$ so that the remainder has Lipschitz constant $O(t^3)$.

However, choosing $s_0$ alone does not provide the required Lipschitz bound.
Since the remainder is again a quadratic term plus a constant, we can apply the same decomposition to it and seek a new remainder with the required bound.

Specifically, we choose $s_1>0$ and apply \eqref{eq:algebraic-identity} to the quadratic part of the remainder in \eqref{eq:gadget-square-identity-1}, obtaining a new absolute-value term and a new remainder.
Iterating this step yields a sum of absolute-value terms and a final remainder.
With a suitable choice of the shifts $s_k$, the final remainder satisfies the required Lipschitz bound, as established in \cref{subsec:gadget-telescoping}.

We write the above procedure formally as follows. Suppose \(\{s_k\}\) has been determined in advance. For each input pair $(j,j')$, define
\[
\begin{aligned}
\Delta_{(j,j'),0}:=\log\frac{p_j(x)}{p_{j'}(x)},
\quad \Delta_{(j,j'),1}&:=|\Delta_{(j,j'),0}|-s_0,\quad \Delta_{(j,j'),k+1}:=|\Delta_{(j,j'),k}|-s_k.
\end{aligned}
\]
Then, we have
\begin{equation}
\label{eq:gadget-square-identity}
\begin{aligned}
-\frac{w_{jj'}}{2}\Delta_{(j,j'),0}^2
&=\underbrace{-w_{jj'}s_0|\Delta_{(j,j'),0}|}_{\text{old absolute-value term}}
  \underbrace{-\frac{w_{jj'}}{2}\Delta_{(j,j'),1}^2
  +\frac{w_{jj'}}{2}s_0^2}_{\text{old remainder term}}\\[1.2ex]
&=-w_{jj'}\sum_{k=0}^{1}s_k|\Delta_{(j,j'),k}|
-\frac{w_{jj'}}{2}\Delta_{(j,j'),2}^2
+\frac{w_{jj'}}{2}(s_0^2+s_1^2)\\[0.6ex]
&\mathrel{\phantom{=}}\vdots\\[0.6ex]
&=\underbrace{-w_{jj'}\sum_{k=0}^{L-1}s_k|\Delta_{(j,j'),k}|}_{\text{Final absolute-value terms}}
  \underbrace{-\frac{w_{jj'}}{2}\Delta_{(j,j'),L}^2
  +\frac{w_{jj'}}{2}\sum_{k=0}^{L-1}s_k^2}_{\text{Final remainder term}}.
\end{aligned}
\end{equation}

The remaining subsections turn this decomposition into a market construction. From discussion above, we need to add auxiliary agent who can contribute the absolute-value terms.
We formally introduce our technical tool, \emph{Maximum-Minimum gadget}
term to achieve it in \cref{subsec:max-min-gadget}. 
In \cref{subsec:gadget-telescoping}, we connect
these gadgets into chains, derive the telescoping identity \eqref{eq:gadget-square-identity}, and choose
parameters so that the resulting remainder term satisfies the required
Lipschitz bound. 
The gadgets also introduce additional terms. \cref{subsec:extra-terms} cancels these terms by adjusting the basic
agents' earning requirements and recovers the target linear term.
Finally, \cref{subsec:normalization-final-program} adjusts supplies and
normalizes prices to enforce payment balance, controls the resulting
normalization error, and specifies the complete market.

\subsection{The Maximum-Minimum Gadget}
\label{subsec:max-min-gadget}
We use the same notation from \cref{subsec:gadget-motivation}:
\[
\Delta_{(j,j'),0}:=\log\frac{p_j(x)}{p_{j'}(x)}.
\]

\paragraph{Goal of This Section.}
Motivated by \cref{subsec:gadget-motivation}, we seek to introduce auxiliary agents that contribute
absolute-value terms to \(\Psi(p)\). 
In this section, we focus on generating a term involving
\(|\Delta_{(j,j'),0}|\).

At first glance, introducing auxiliary agents and chores still appears to be difficult:
each auxiliary agent still contributes \(-B_a \log \mu_a(p)\) to \(\Psi(p)\), and this contribution contains no explicit absolute-value term.
However, the following identity relates the absolute-value term to \(\log\max\):
\begin{equation*}
\log\max\{p_j(x),p_{j'}(x)\}
=\frac12\left(
\log p_j(x)+\log p_{j'}(x)+
|\Delta_{(j,j'),0}|
\right).
\end{equation*}
This motivates our method.
\paragraph{Construction of the Maximum-Minimum Gadget.}
Suppose we add an agent with earning requirement $B>0$ and a
new chore whose price is proportional to $\max\{p_j(x),p_{j'}(x)\}$. If the agent's optimal ratio \(\mu_a(p)\) is attained on this chore, then the agent's contribution to
$\Psi$ is
\[
\begin{aligned}
-B\log\max\{p_j(x),p_{j'}(x)\}
={}&
-\underbrace{\frac{B}{2}|\Delta_{(j,j'),0}|}_{\text{desired absolute-value term}}
-\underbrace{\frac{B}{2}\bigl(\log p_j(x)+\log p_{j'}(x)\bigr)}_{\text{extra terms}}
\end{aligned}
\]
up to an additive constant. This construction produces the desired absolute-value term together with additional logarithmic terms, motivating the formal construction below.
We will cancel the additional logarithmic terms in \cref{subsec:extra-terms}.

We now formalize the \emph{Maximum-Minimum gadget} motivated by the preceding discussion. Intuitively, a gadget takes two existing chores as inputs and adds three agents
and two output chores to the market. 

To simplify notation, we omit the dependence on \(x\) and write \(p_j\) and \(p_{j^{\prime}}\) for \(p_j(x)\) and \(p_{j^{\prime}}(x)\), respectively.
By a slight abuse of notation, we use $p^+_{(j,j')},p^-_{(j,j')}$ for both the output chores and their
prices.\footnote{The superscripts do not imply a price ordering.}

\begin{definition}[Maximum-Minimum gadget]
\label{def:max-min-gadget}
Fix base chores $(j,j')$, a
\emph{Maximum-Minimum gadget on (\(j,j^{\prime}\))} outputs 
\begin{itemize}
  \item three new agents $a^{\max},a^{(1)},a^{(2)}$ with the same positive earning requirement \(B\),
  \item two new chores $p^+_{(j,j')},p^-_{(j,j')}$ with the same positive supply \(S\),
  \item a disutility matrix with parameter \(q\).
\end{itemize}
Their disutility matrix is defined as
\[
\begin{array}{c|ccccc}
 & j & j' & p^+_{(j,j')} & p^-_{(j,j')} & \text{Other base chores}\\ \hline
a^{\max} & q & q & 1 & \text{--} & \text{--}\\
a^{(1)} & 1/q & \text{--} & \text{--} & 1 & \text{--}\\
a^{(2)} & \text{--} & 1/q & \text{--} & 1 & \text{--}
\end{array}
\]
The entries other than dashes specify each agent's
\emph{designated choices}.
The last column applies to every base chore other than the input chores
$j,j'$.
\end{definition}

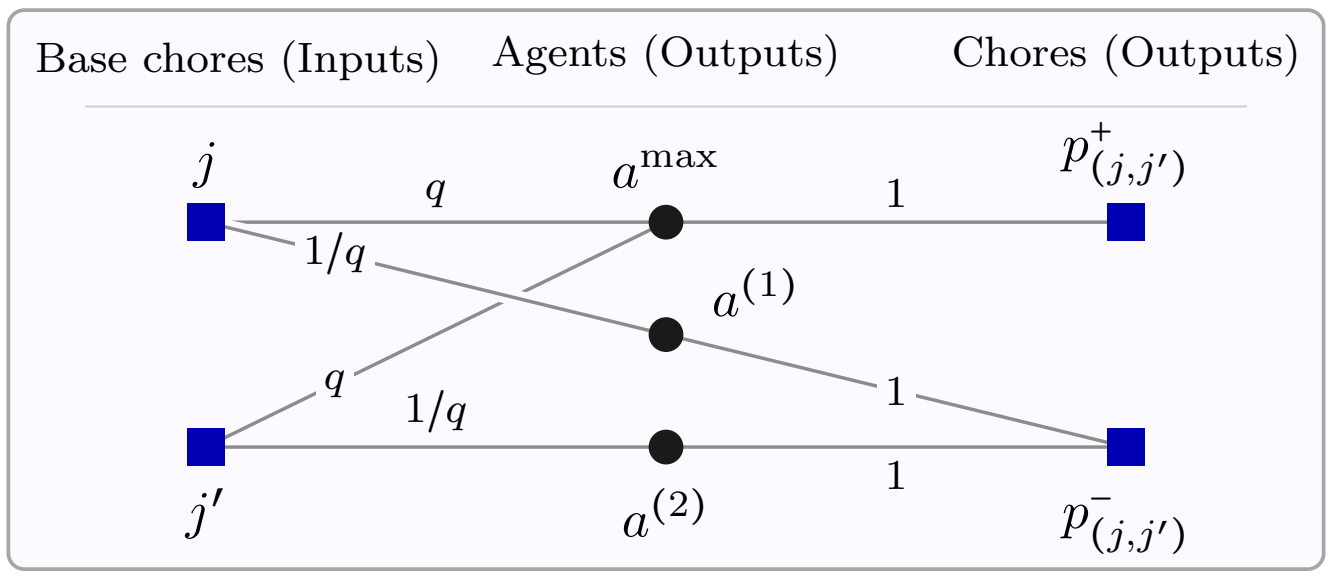
\begin{figure}[htbp]
\centering
\begin{tikzpicture}[
  x=1.45cm,y=.85cm,
  chore/.style={rectangle,fill=blue!70!black,
    minimum size=2.5mm,inner sep=0pt},
  agent/.style={circle,fill=black!90,
    minimum size=2.3mm,inner sep=0pt},
  choice/.style={draw=black!45,line width=.65pt},
  price/.style={font=\small,inner sep=1pt},
  head/.style={font=\scriptsize\rmfamily,align=center},
  cost/.style={font=\scriptsize,fill=blue!2,inner sep=1.5pt}
]
\draw[draw=black!35,fill=blue!2,line width=.7pt,
  rounded corners=3pt] (0,0) rectangle (6,4.35);
\node[head] at (1.05,3.95) {Base chores (Inputs)};
\node[head] at (3,4.0) {Agents (Outputs)};
\node[head] at (5.1,4.0) {Chores (Outputs)};
\draw[black!20,line width=.35pt] (.35,3.6)--(5.65,3.6);

\draw[choice] (.9,2.7)--node[cost,above=1pt] {$q$} (3,2.7);
\draw[choice] (.9,.95)--node[cost,pos=.28] {$q$} (3,2.7);
\draw[choice,preaction={draw=blue!2,line width=3pt}]
  (.9,2.7)--node[cost,pos=.28] {$1/q$} (3,1.825);
\draw[choice] (.9,.95)--node[cost,above=1pt] {$1/q$} (3,.95);

\draw[choice] (3,2.7)--node[cost,above=1pt] {$1$} (5.1,2.7);
\draw[choice] (3,1.825)--node[cost] {$1$} (5.1,.95);
\draw[choice] (3,.95)--node[cost,below=1pt] {$1$} (5.1,.95);

\node[chore] at (.9,2.7) {};
\node[chore] at (.9,.95) {};
\node[price,anchor=south] at (.9,2.92) {$j$};
\node[price,anchor=north] at (.9,.68) {$j^{\prime}$};

\node[agent] at (3,2.7) {};
\node[agent] at (3,1.825) {};
\node[agent] at (3,.95) {};
\node[price,anchor=south] at (3,2.92) {$a^{\max}$};
\node[price,anchor=south west] at (3.18,1.92) {$a^{(1)}$};
\node[price,anchor=north] at (3,.68) {$a^{(2)}$};

\node[chore] at (5.1,2.7) {};
\node[chore] at (5.1,.95) {};
\node[price,anchor=south] at (5.1,2.92) {$p^+_{(j,j')}$};
\node[price,anchor=north] at (5.1,.68) {$p^-_{(j,j')}$};
\end{tikzpicture}
\caption{A Maximum-Minimum gadget on $(j,j')$ with parameter $q>1$.
}
\label{fig:max-min-gadget}
\end{figure}

\begin{remark}
\label{rem:symbol-dash-gadget}
Each dash in the disutility matrix denotes a finite positive
value to be chosen later. We will choose these values sufficiently large, so that
at every equilibrium of the completed market, the gadget
agents use only their designated chores and the base agents use
only base chores. \Cref{fig:max-min-gadget} illustrates the gadget.  
\end{remark}

\paragraph{Price Condition.}
Applying the gadget to base chores \((j,j^{\prime})\), we can add three agents and two output chores.
We next address two questions: 
\begin{enumerate}
  \item How do we force the price of one of the output chores to be proportional to \(\max\{p_j,p_{j^{\prime}}\}\)?
  \item How do we guarantee that the corresponding agent's optimal ratio is attained on this chore?
\end{enumerate} 
These requirements lead to additional restrictions on prices, which we call the \emph{price condition}.

We first consider \(p^+_{(j,j')}\). Let \(N_{\text{out}}^+\) denote the set of agents outside the gadget who choose this chore.
If we have
\[
\sum_{a\in N_{\text{out}}^+} B_a < Sp^+_{(j,j')} < B,
\]
then the price of the output chore \(p^+_{(j,j')}\) is proportional to \(\max\{p_j,p_{j^{\prime}}\}\), and the optimal ratio of agent \(a^\text{max}\) is attained on this chore.

One may wonder what \(N_\text{out}^+\) really is.
Although we currently consider a single gadget that generates one
absolute-value term, implementing the recursive decomposition
in \eqref{eq:gadget-square-identity} requires a chain of gadgets,
as illustrated in \cref{fig:gadget-chain-vector}.
Agents in subsequent gadgets may choose chores produced by the current gadget.
The set $N_{\text{out}}^+$ accounts for those agents outside the current
gadget.
We describe this set explicitly in \cref{rem:elements-N}.

Similarly, let \(N_{\text{out}}^-\) denote the set of agents outside the gadget who choose \(p^-_{(j,j')}\).
If we have
\[
\sum_{a\in N_{\text{out}}^-} B_a < Sp^-_{(j,j')} < B,
\]
then this also causes restriction to the price of chore \(p^-_{(j,j')}\) and the optimal ratios of
\(a^{(1)}, a^{(2)}\). 
At this stage, the output chore $p^-_{(j,j')}$ and the two agents
$a^{(1)},a^{(2)}$ may appear peripheral to our immediate goal of
producing an absolute-value term. Their role becomes clear in
\cref{subsec:gadget-telescoping}, where they enable the recursive
construction. We defer the explanation
of this role until then.

We summarize these requirements in the following \emph{price condition}:
\begin{equation}
\tag{Price Condition}
\label{eq:gadget-payment-window}
\sum_{a\in N_{\text{out}}^+} B_a < Sp^+_{(j,j')} < B,
\qquad
\sum_{a\in N_{\text{out}}^-} B_a < Sp^-_{(j,j')} < B.
\end{equation}
Under these conditions, agent \(a^\text{max}\) generates the desired absolute-value term,
while the other agents introduce additional contributions that we discuss later.
The following lemma states the resulting output prices and optimal ratios.

\begin{lemma}
\label{lem:max-min-gadget}
Consider a market containing a gadget on chores 
\((j,j^{\prime})\). Suppose at the equilibrium,
\begin{enumerate}
    \item The three agents added by the gadget only use their designated chores,
    \item \eqref{eq:gadget-payment-window} holds.
\end{enumerate} 
The output prices then satisfy
\begin{equation}
\label{eq:gadget-output-prices}
\begin{gathered}
p^+_{(j,j')}=\frac{\max\{p_j,p_{j'}\}}{q},
\quad
p^-_{(j,j')}=q\min\{p_j,p_{j'}\},\quad\log\frac{p^+_{(j,j')}}{p^-_{(j,j')}}
=|\Delta_{(j,j'),0}|-2\log q.
\end{gathered}
\end{equation}
The three agents' optimal ratios are
\begin{equation}
\label{eq:gadget-optimal-ratios}
\mu_{a^{\max}}(p)=\frac{\max\{p_j,p_{j'}\}}{q},
\qquad
\mu_{a^{(1)}}(p)=qp_j,
\qquad
\mu_{a^{(2)}}(p)=qp_{j'}.
\end{equation}
\end{lemma}

\begin{proof}
First we have \(\mu_{a^{\max}}(p)=p^+_{(j,j^{\prime})}\). By assumption,
the new agents use only their designated chores.
Since $Sp^+>\sum_{a\in N_{\text{out}}^+}B_a$, market clearing forces
$a^{\max}$ to choose $p^+$. This gives
\[
\mu_{a^{\max}}(p)\geq \frac{\max\{p_j,p_{j'}\}}{q}.
\]
Since $Sp^+<B$, this agent must also choose an input chore, which rules out the strict inequality.
\[
\mu_{a^{\max}}(p)=\frac{\max\{p_j,p_{j'}\}}{q}.
\]

Similarly, $Sp^-<B$ forces each of $a^{(1)}$ and $a^{(2)}$
to choose its designated input chore. Hence
\[
\mu_{a^{(1)}}(p)=qp_j,\qquad
\mu_{a^{(2)}}(p)=qp_{j'},\qquad
p^-\le q\min\{p_j,p_{j'}\}.
\]
The lower bound on $Sp^-$ forces at least one of these two agents
to perform $p^-$, so $p^-$ equals that agent's optimal ratio.
Together with the preceding inequality, this yields
$p^-=q\min\{p_j,p_{j'}\}$.
Dividing the two output prices and taking logarithms gives the stated log ratio.
\end{proof}

It remains to ensure that \eqref{eq:gadget-payment-window} holds at every equilibrium of our constructed market.
We claim it here and defer the verification later. From \cref{lem:max-min-gadget}, the three new agents together contribute
\begin{equation*}
\begin{aligned}
&-B\left(
\log\frac{\max\{p_j,p_{j'}\}}{q}
+\log(qp_j)+\log(qp_{j'})
\right)\\
&\qquad=
\underbrace{-\frac{B}{2}|\Delta_{(j,j'),0}|}_{\text{desired absolute-value term}}
-\frac{3B}{2}(\log p_j+\log p_{j'})
-B\log q
\end{aligned}
\end{equation*}
to $\Psi$. The first term is the contribution we will
use to form a negative square. The remaining terms are treated in \cref{subsec:extra-terms}.

\paragraph{The Gadget Contribution.}
We have constructed a \emph{Maximum-Minimum gadget} on base chores \((j,j^{\prime})\).
Under the price condition, its agents contribute an absolute-value term involving \(|\Delta_{(j,j^{\prime}),0}|\) to \(\Psi\).
In the next subsection, we use this gadget to construct the desired market.

\subsection{Chains of Maximum-minimum Gadgets}
\label{subsec:gadget-telescoping}

\paragraph{Goal of This Section.}
In this section, we will connect Maximum-Minimum gadgets into a chain so that their
absolute-value contributions approximate the desired negative quadratic term.
We choose the chain length and parameters to ensure that the remainder
satisfies the required bound.

\paragraph{Layer Notation.}
From now, we index gadgets by layer, starting at $0$.
The gadget on $(j,j')$ introduced in \cref{subsec:max-min-gadget} is
called the layer-$0$ gadget. Its parameters $B$, $S$, and $q$ are
henceforth denoted by $B_{(j,j'),0}$, $S_{(j,j'),0}$, and $q_0$,
respectively, and output chore and its price
$p^+_{(j,j')}$ are denoted by $p^+_{(j,j'),0}$.\footnote{The discussion holds for \(p^-_{(j,j^{\prime})}\).}

\paragraph{Construction of a Gadget Chain.}
Recall the first formula in \eqref{eq:gadget-square-identity}. We need to use auxiliary agents to produce \(|\Delta_{(j,j^{\prime}),0}|\) part and choose \(s_0\) to make sure the decomposition works,
\[
-\frac{w_{jj'}}{2}\Delta_{(j,j'),0}^2
=
-w_{jj'}s_0|\Delta_{(j,j'),0}| 
-\frac{w_{jj'}}{2}\Delta_{(j,j'),1}^2
+\frac{w_{jj'}}{2}s_0^2,
\quad
\Delta_{(j,j'),1} = |\Delta_{(j,j'),0}|-s_0.
\]
Through \cref{subsec:max-min-gadget}, we have added the \emph{first} gadget taking \(p_j\) and \(p_{j^{\prime}}\) as input, output agents who can produce absolute-value part, so the current goal is to choose suitable \(s_0\) to define appropriate \(\Delta_{(j,j'),1}\), 
then implement the decomposition to \(\Delta_{(j,j^{\prime}),1}\). Surprisingly, the price relations from \eqref{eq:gadget-output-prices} 
\[
\log\frac{p^+_{(j,j'),0}}{p^-_{(j,j'),0}}
=|\Delta_{(j,j'),0}|-2\log q_0
\]
gives this structure. This motivates us to set \(s_0=2\log q_0\)\footnote{Note that we still haven't determined the specific value of \(s_0\).}
and define 
\[\Delta_{(j,j'),1}:=\log\frac{p^+_{(j,j'),0}}{p^-_{(j,j'),0}}.\]
Then, the task reduces to producing \(|\Delta_{(j,j^{\prime}),1}|\) part. To achieve this, we imitate the execution of \emph{first} maximum-minimum gadget in \cref{subsec:max-min-gadget}, take chores \(p^+_{(j,j^{\prime})},p^-_{(j,j^{\prime})}\) as the input, and continue constructing the \emph{second} maximum-minimum gadget.

\begin{figure}[htbp]
\centering
\resizebox{\linewidth}{!}{%
\begin{tikzpicture}[
  x=1cm,y=1cm,
  chore/.style={rectangle,fill=blue!70!black,
    minimum size=2.5mm,inner sep=0pt},
  agent/.style={circle,fill=black!90,
    minimum size=2.3mm,inner sep=0pt},
  choice/.style={draw=black!45,line width=.65pt},
  reuse/.style={draw=blue!70!black,dashed,line width=.75pt},
  price/.style={font=\small,inner sep=1pt},
  cost/.style={font=\scriptsize,fill=blue!2,inner sep=1.5pt},
  head/.style={
    font=\scriptsize\rmfamily,text=black,align=center,
    text width=1.75cm,inner sep=1pt
  }
]

\foreach \id/\layer/\ip/\im/\dx in {
  0/0/{p_j}/{p_{j'}}/0,
  1/1/{p^+_{(j,j'),0}}/{p^-_{(j,j'),0}}/7,
  2/{L-1}/{p^+_{(j,j'),L-2}}/{p^-_{(j,j'),L-2}}/15.4
}{
  \begin{scope}[shift={(\dx,0)}]
    \draw[
      draw=black!35,fill=blue!2,
      line width=.7pt,rounded corners=3pt
    ] (0,0) rectangle (6,4.5);

    \node[font=\small] at (3,4.95) {Layer $\layer$};

    \ifnum\id=0
      \node[head] at (.9,4.0) {Base chores\\(Inputs)};
    \else
      \node[head] at (.9,4.0) {Inputs};
    \fi
    \node[head] at (3,4.0) {Agents\\(Outputs)};
    \node[head] at (5.1,4.0) {Chores\\(Outputs)};
    \draw[black!20,line width=.35pt]
      (.35,3.6)--(5.65,3.6);

    \tikzset{
      uppercost/.style={above=1pt},
      diagcost/.style={pos=.28},
      lowercost/.style={above=1pt}
    }
    \ifnum\id=2
      \tikzset{
        uppercost/.style={pos=.6,above=1pt},
        diagcost/.style={pos=.34},
        lowercost/.style={pos=.57,below=1pt}
      }
    \fi
    \draw[choice] (.9,2.7)--node[cost,uppercost] {$q_{\layer}$} (3,2.7);
    \draw[choice] (.9,.95)--node[cost,diagcost] {$q_{\layer}$} (3,2.7);
    \draw[choice,preaction={draw=blue!2,line width=3pt}]
      (.9,2.7)--node[cost,diagcost] {$1/q_{\layer}$} (3,1.825);
    \draw[choice] (.9,.95)--node[cost,lowercost] {$1/q_{\layer}$} (3,.95);

    \draw[choice] (3,2.7)--node[cost,above=1pt] {$1$} (5.1,2.7);
    \draw[choice] (3,1.825)--node[cost] {$1$} (5.1,.95);
    \draw[choice] (3,.95)--node[cost,below=1pt] {$1$} (5.1,.95);

    \node[chore] (inplus-\id) at (.9,2.7) {};
    \node[chore] (inminus-\id) at (.9,.95) {};
    \node[price,anchor=south] at (.9,2.92) {$\ip$};
    \node[price,anchor=north] at (.9,.68) {$\im$};

    \node[agent] at (3,2.7) {};
    \node[agent] at (3,1.825) {};
    \node[agent] at (3,.95) {};
    \node[price,anchor=south]
      at (3,2.92) {$a^{\max}$};
    \node[price,anchor=south west]
      at (3.18,1.92) {$a^{(1)}$};
    \node[price,anchor=north]
      at (3,.68) {$a^{(2)}$};

    \node[chore] (outplus-\id) at (5.1,2.7) {};
    \node[chore] (outminus-\id) at (5.1,.95) {};
    \node[price,anchor=south]
      at (5.1,2.92) {$p^+_{(j,j'),\layer}$};
    \node[price,anchor=north]
      at (5.1,.68) {$p^-_{(j,j'),\layer}$};
  \end{scope}
}

\foreach \sign/\y in {plus/2.7,minus/.95}{
  \draw[reuse] (out\sign-0.east)--(in\sign-1.west);
  \draw[reuse] (out\sign-1.east)--(13.7,\y);
  \node[text=blue!70!black] at (14.2,\y) {$\cdots$};
  \draw[reuse] (14.7,\y)--(in\sign-2.west);
}

\node[chore] at (2.2,-.65) {};
\node[font=\small,anchor=west] at (2.45,-.65) {Chore};

\node[agent] at (5.1,-.65) {};
\node[font=\small,anchor=west] at (5.35,-.65) {Agent};

\draw[choice] (8,-.65)--(8.8,-.65);
\node[font=\small,anchor=west]
  at (9,-.65) {Designated choice};

\draw[reuse] (14,-.65)--(14.8,-.65);
\node[font=\small,anchor=west]
  at (15,-.65) {Same chore, reused};

\end{tikzpicture}%
}
\caption{A chain of Maximum-Minimum gadgets on $(j,j')$.
}
\label{fig:gadget-chain-vector}
\end{figure}
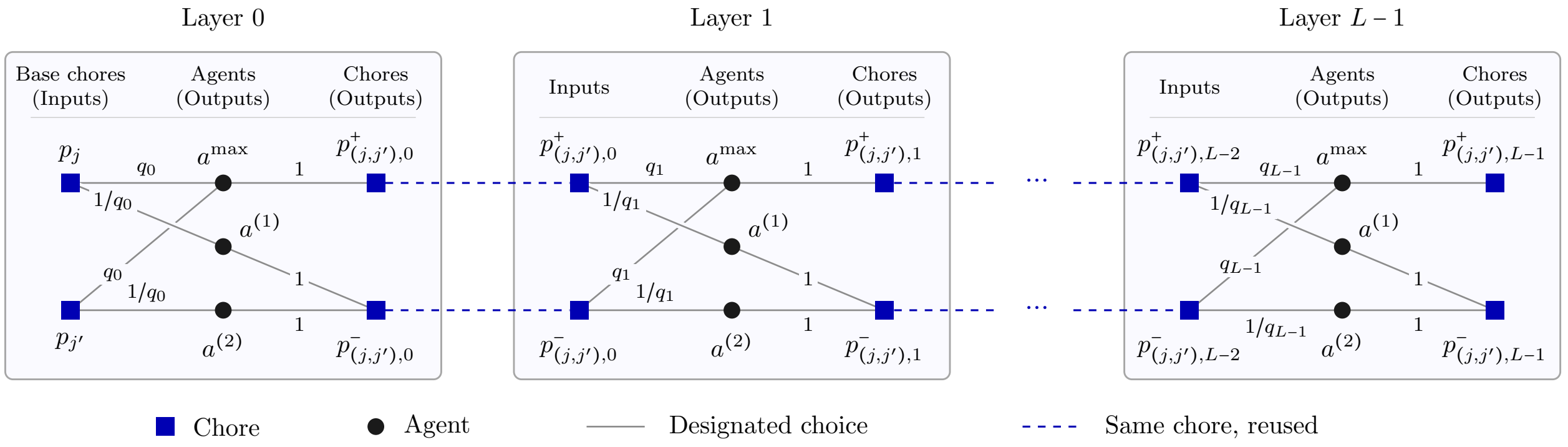

Repeating the construction gives a \emph{chain of gadgets}, as illustrated in \cref{fig:gadget-chain-vector}. We define it formally below. 
By a slight abuse of notation, the output symbols also denote their prices.



\begin{definition}[Chain of Maximum-Minimum gadgets]
\label{def:gadget-chain}
Fix base chores $(j,j')$ with $j<j'$ and an integer $L\geq1$.
An \emph{$L$-layer chain} on $(j,j')$ consists of Maximum-Minimum
gadgets indexed by $k=0,\ldots,L-1$, starting with the gadget on
the base chores. Each layer gadget adds three new agents and two new output chores, connected as follows:
\[
\begin{aligned}
\text{Outputs of layer }k &: \quad
  \bigl(p^+_{(j,j'),k},\,p^-_{(j,j'),k}\bigr),
  &&0\leq k<L,\\[3pt]
\text{Inputs to layer }k &: \quad
  \begin{cases}
    (j,j'), & k=0,\\[2pt]
    \bigl(p^+_{(j,j'),k-1},\,p^-_{(j,j'),k-1}\bigr), & 1\leq k<L.
  \end{cases}
\end{aligned}
\]

At layer $k$, the disutility parameter, common earning requirement of
the three new agents, and common supply of the two output chores
are, respectively,
\[
q_k>1,\qquad B_{(j,j'),k}>0,\qquad S_{(j,j'),k}>0.
\]
The three agents at layer $k$ are named and assigned disutilities
by the following matrix, whose first two columns refer to the
ordered input pair specified above. For $k=0$, the first two column labels are interpreted as $j$ and $j'$, respectively.
\begin{center}
\renewcommand{\arraystretch}{1.25}
\setlength{\tabcolsep}{9pt}
\begin{tabular}{@{}cccccc@{}}
\toprule
\multirow{2}{*}{\small Agent}
 & \small First input & \small Second input
 & \multicolumn{2}{c}{\small Outputs}
 & \multirow{2}{*}{\small All other chores}\\[2pt]
 & $p^+_{(j,j'),k-1}$ & $p^-_{(j,j'),k-1}$
 & $p^+_{(j,j'),k}$ & $p^-_{(j,j'),k}$ & \\
\midrule
 $a^{\max}_{(j,j'),k}$ & $q_k$ & $q_k$ & $1$ & -- & --\\[4pt]
 $a^{(1)}_{(j,j'),k}$ & $1/q_k$ & -- & -- & $1$ & --\\[4pt]
 $a^{(2)}_{(j,j'),k}$ & -- & $1/q_k$ & -- & $1$ & --\\
\bottomrule
\end{tabular}
\end{center}

\end{definition}

\begin{remark}
\label{rem:symbol-dash-gadget-chain}
Similar to \cref{rem:symbol-dash-gadget},
here each dash denotes a finite positive disutility
to be chosen sufficiently large that the agent does not choose that chore
at equilibrium.
\end{remark}

\paragraph{Price Condition for Layer \(k\).}
Now we give a general version of the \eqref{eq:gadget-payment-window} in \cref{subsec:max-min-gadget}.
Fix a chain on $(j,j')$, write $p^+_{(j,j'),-1}:=p_j$ and
$p^-_{(j,j'),-1}:=p_{j'}$. For $0\leq k<L$, let
$N_{\mathrm{out},k}^{\pm}$ denote the agents outside layer $k$ who use
the corresponding output chore $p^{\pm}_{(j,j'),k}$.
The \emph{$k$-th layer price condition} is
\begin{equation}
\tag{\(k\)-th price condition}
\label{eq:gadget-layer-payment-window}
\begin{aligned}
&\sum_{a\in N_{\mathrm{out},k}^{+}} B_a < S_{(j,j'),k}\,p^{+}_{(j,j'),k}< B_{(j,j'),k},\\
&\sum_{a\in N_{\mathrm{out},k}^{-}} B_a < S_{(j,j'),k}\,p^{-}_{(j,j'),k}< B_{(j,j'),k}
\end{aligned}
\end{equation}
\begin{remark}
\label{rem:price-is-x-function}
At an equilibrium satisfying \eqref{eq:gadget-layer-payment-window},
an analysis similar to \cref{lem:max-min-gadget} determines the output prices
$p^+_{(j,j'),k}$ from the input prices $p^+_{(j,j'),k-1}$\footnote{The same discussion holds for \(p^-_{(j,j'),k-1}.\)}
by the same max--min formulas.
These formulas recursively define
all output prices as functions of $x$, starting from $p_j(x),p_{j'}(x)$.
We omit the dependence on $x$ in the notation.
\end{remark}
\begin{remark}
\label{rem:elements-N}
By the disutility matrix in \cref{def:gadget-chain} and the convention
in \cref{rem:symbol-dash-gadget-chain}, only agents in the next layer
can use $p^+_{(j,j'),k}$ from outside layer $k$. Hence,
\[
\begin{aligned}
N^+_{\mathrm{out},k}
&\subseteq
\bigl\{a^{\max}_{(j,j'),k+1},\,a^{(1)}_{(j,j'),k+1}\bigr\},
&&0\leq k<L-1,\\[4pt]
N^+_{\mathrm{out},L-1}&=\varnothing.
\end{aligned}
\]
The same argument gives
\[
\begin{aligned}
N^-_{\mathrm{out},k}
&\subseteq
\bigl\{a^{\max}_{(j,j'),k+1},\,a^{(2)}_{(j,j'),k+1}\bigr\},
&&0\leq k<L-1,\\[4pt]
N^-_{\mathrm{out},L-1}&=\varnothing.
\end{aligned}
\]
\end{remark}
It remains to ensure that \eqref{eq:gadget-layer-payment-window} holds at every equilibrium of our constructed market.
We again claim it here and defer the verification to the proof of \cref{lem:equilibrium-recovery} in \cref{sec:equilibrium-recovery}.
\begin{claim}
\label{clm:final-parameter-choice}
Choose \(L\) such that \(2^{-L}=t\). In our complete constructed market, \eqref{eq:gadget-layer-payment-window} holds at every equilibrium and every layer when the gadget parameters are chosen as
\begin{equation}
\label{eq:chain-approximation-parameters}
\begin{aligned}
q_k&=1+\frac{t}{4}\left(\frac12-4t\right)^k,
&s_k&=2\log q_k,\\
B_{(j,j'),k}&=w_{jj'}t\left(\frac12-4t\right)^k,
&S_{(j,j'),k}&=(1-4t)B_{(j,j'),k}.
\end{aligned}
\end{equation}
\end{claim}

\paragraph{Recovery of the Quadratic Term.}
Now for chores pair \((j, j^{\prime})\), we have added \(L\)-layer chain to it, where each layer contains three agents and two chores.
Similar to the analysis of \cref{lem:max-min-gadget}, the contributions to $\Psi$ at a general layer $k$ is
as follows. 
\begin{align*}
\text{Layer }k:\quad
&-\frac12 B_{(j,j'),k}|\Delta_{(j,j'),k}|\nonumber\\*
&\quad-\frac32 B_{(j,j'),k}
\bigl(\log p^+_{(j,j'),k-1}+\log p^-_{(j,j'),k-1}\bigr)
-B_{(j,j'),k}\log q_k.
\end{align*}
As previously mentioned, $p^+_{(j,j'),-1}=p_j$ and $p^-_{(j,j'),-1}=p_{j'}$. Summing over all $L$ layers gives the total contribution
\begin{equation}
\label{eq:gadget-chain-contribution}
\begin{aligned}
&-\frac32 B_{(j,j'),0}(\log p_j+\log p_{j'})-\frac32\sum_{k=1}^{L-1}B_{(j,j'),k}
\bigl(\log p^+_{(j,j'),k-1}+\log p^-_{(j,j'),k-1}\bigr)\\
&\quad+P_{(j,j')}
-\sum_{k=0}^{L-1}B_{(j,j'),k}\log q_k,
\end{aligned}
\end{equation}
where
\begin{equation*}
P_{(j,j')}:=-\frac12\sum_{k=0}^{L-1}B_{(j,j'),k}|\Delta_{(j,j'),k}|
\end{equation*}
collects the absolute-value terms. We now show how
$P_{(j,j')}$ combines these terms into a negative square.
The remaining terms are treated
in \cref{subsec:extra-terms}.

Recall that in \eqref{eq:gadget-square-identity}, we need to construct the final decomposition
:
\[
-\frac{w_{jj'}}2\Delta_{(j,j'),0}^2
=-w_{jj'}\sum_{k=0}^{L-1}s_k|\Delta_{(j,j'),k}|
+
\left[-\frac{w_{jj'}}2\Delta_{(j,j'),L}^2
+\frac{w_{jj'}}2\sum_{k=0}^{L-1}s_k^2\right]
.
\]
We use $P_{(j,j')}$ to express the absolute-value term,
\begin{align*}
  -w_{jj'}\sum_{k=0}^{L-1}s_k|\Delta_{(j,j'),k}|
  =
  P_{(j,j')}
  +
  \sum_{k=0}^{L-1}\left(\frac{B_{(j,j'),k}}2-w_{jj'}s_k\right)|\Delta_{(j,j'),k}|.
\end{align*}
Substituting this expression yields
\begin{equation*}
\label{eq:gadget-rational-telescoping}
\begin{aligned}
-\frac{w_{jj'}}2\Delta_{(j,j'),0}^2
={}&P_{(j,j')}
\underbrace{-\frac{w_{jj'}}2\Delta_{(j,j'),L}^2
+\frac{w_{jj'}}2\sum_{k=0}^{L-1}s_k^2}
_{\text{remainder in \eqref{eq:gadget-square-identity}}}\\[1.2ex]
&+\underbrace{\sum_{k=0}^{L-1}
\left(\frac{B_{(j,j'),k}}2-w_{jj'}s_k\right)|\Delta_{(j,j'),k}|}
_{\text{coefficient correction}}.
\end{aligned}
\end{equation*}
Define
\[
\tilde{R}_{(j,j^{\prime})}:=
-\frac{w_{jj'}}2\Delta_{(j,j'),L}^2
+\frac{w_{jj'}}2\sum_{k=0}^{L-1}s_k^2
+
\sum_{k=0}^{L-1}
\left(\frac{B_{(j,j'),k}}2-w_{jj'}s_k\right)|\Delta_{(j,j'),k}|
\]
Recall from \cref{rem:price-is-x-function} that the gadget prices, and hence
$\Delta_{(j,j'),k}$, are functions of $x$. Thus $P_{(j,j')}$ and $\tilde{R}_{(j,j')}$ are also
functions of $x$, whose dependence we omit in the notation.

Now we use the gadget chains to approximate \eqref{eq:desired-contribution} with \(P_{(j,j^{\prime})}\).
Recall that in \eqref{eq:gadget-log-square-approximation}, we use the Taylor expansion to get
\[
-\frac{w_{jj'}}{2}t^2(x_j-x_{j'})^2 = -\frac{w_{jj'}}{2}\Delta_{(j,j'),0}^2 + \underbrace{\hat{R}_{(j,j^{\prime})}(x)}_{\text{which is }O(t^3)} 
\]
then we have our final approximation:
\begin{equation*}
  -\frac{w_{jj'}}{2}t^2(x_j-x_{j'})^2 = P_{(j,j')} + R_{(j,j')}, \quad R_{(j,j')} = \tilde{R}_{(j,j')} + \hat{R}_{(j,j')} 
\end{equation*}

\begin{theorem}[Target quadratic approximation]
\label{thm:chain-quadratic-approximation}
For every $x\in[0,1]^n$,
\begin{equation}
\label{eq:chain-quadratic-approximation}
  -\frac{w_{jj'}}{2}t^2(x_j-x_{j'})^2 = P_{(j,j')}(x) + R_{(j,j')}(x), 
\end{equation}
In our complete constructed market, the remainder $R_{(j,j')}$
satisfies
\begin{equation*}
\left|R_{(j,j')}(x+he_\ell)-R_{(j,j')}(x)\right|
\leq12w_{jj'}t^3|h|
\end{equation*}
for every $\ell\in[n]$ and every coordinate segment from $x$
to $x+he_\ell$ contained in the box.
Moreover, for every feasible coordinate direction
$\widetilde v\in\{\pm e_\ell:\ell\in[n]\}$ at $x$, the right
directional derivative exists and satisfies
\[
\left|
R_{(j,j')}'(x;\widetilde v)
\right|
\leq12w_{jj'}t^3.
\]
\end{theorem}

The proof is deferred to \Cref{app:error-chain}, where
we collect the bounds for all error terms in the reduction.

\subsection{Refining the Objective through Earning Requirements}
\label{subsec:extra-terms}
We adjust the basic agents' earning requirements to make two
corrections to the logarithmic objective: cancel the additional
logarithmic terms contributed by the gadget chains
in~\eqref{eq:gadget-chain-contribution} and introduce
the missing linear term $\alpha c^\top x$.
The latter includes the factor $\alpha$ because we approximate
the scaled objective $\alpha Q(x)$ to establish \cref{item:market-error-bound}
of \Cref{prop:market-target}.
Let $N_{\mathrm{base}}$ denote the set of basic agents.
For each basic agent $a$ associated with chore $j$, define the
final earning requirement by
\begin{equation}
\label{eq:adjusted-base-earnings}
\begin{aligned}
B_a^{\mathrm{base}}:=1
&-\underbrace{
\frac32\sum_{k=0}^{L-1}
\left(
\sum_{j'<j}B_{(j',j),k}
+\sum_{j'>j}B_{(j,j'),k}
\right)
}_{\text{cancel additional logarithms}}-\underbrace{\frac{\alpha}{t}c_j}
_{\text{recover }\alpha c^\top x},
\qquad a\in N_{\mathrm{base}},
\end{aligned}
\end{equation}
where the coefficient for the reference chore is
\begin{equation}
\label{eq:linear-reference-coefficient}
c_0:=-\sum_{j=1}^{n}c_j.
\end{equation}
Here $c_0$ is an auxiliary coefficient for the reference chore;
the input vector remains $c=(c_1,\ldots,c_n)^\top\in\mathbb Q^n$.
The gadget agents retain their earning requirements.
We explain the two corrections below, then verify positivity
and bound the additional approximation error along feasible
coordinate changes.

Recall from~\eqref{eq:gadget-chain-contribution} that the contribution
of the chain associated with $(j,j')$ is
$$
\begin{aligned}
&-\frac32 B_{(j,j'),0}(\log p_j+\log p_{j'})
-\frac32\sum_{k=1}^{L-1}B_{(j,j'),k}
\bigl(\log p^+_{(j,j'),k-1}
+\log p^-_{(j,j'),k-1}\bigr)\\
&+P_{(j,j')}(p)
-\sum_{k=0}^{L-1}B_{(j,j'),k}\log q_k.
\end{aligned}
$$
As shown in~\cref{subsec:gadget-telescoping}, $P_{(j,j')}(p)$
provides the desired quadratic correction, with approximation
error controlled along coordinate changes.
The last summation is independent of $x$ and can be absorbed into
an additive constant. We therefore first cancel the two additional
logarithmic sums.

\paragraph{Canceling the Additional Logarithms.}
By~\eqref{eq:gadget-output-prices}, for $k=1,\ldots,L-1$,
the output prices satisfy
\begin{align*}
p^+_{(j,j'),k}p^-_{(j,j'),k}
&=
\left(
\frac{\max\{p^+_{(j,j'),k-1},p^-_{(j,j'),k-1}\}}{q_k}
\right)
\left(
q_k\min\{p^+_{(j,j'),k-1},p^-_{(j,j'),k-1}\}
\right)\\
&=p^+_{(j,j'),k-1}p^-_{(j,j'),k-1}.
\end{align*}
Starting from the basic input prices and applying this identity
successively along a chain gives
$$
p^+_{(j,j'),k}p^-_{(j,j'),k}=p_jp_{j'},
\qquad k=0,\ldots,L-1.
$$
Consequently, the first two terms of
\eqref{eq:gadget-chain-contribution} combine into
\begin{equation}
\label{eq:chain-extra-logarithms}
-\frac32\left(\sum_{k=0}^{L-1}B_{(j,j'),k}\right)
(\log p_j+\log p_{j'}).
\end{equation}

Recall from~\cref{sec:basic-market} that, for the price family
defined there, the basic market objective is
$$
\Psi_0(p)
=-\sum_{a\in N_{\mathrm{base}}}\log\mu_a(p)
=-\sum_{j=0}^{n}\log p_j.
$$
The second equality follows from the disutility choices,
which ensure that the basic agent associated with chore $j$
has optimal ratio $p_j$.

Thus, for each chain $(j,j')$, the additional terms in
\eqref{eq:chain-extra-logarithms} can be canceled by reducing
the earning requirement of each of the two corresponding basic
agents by $\frac32\sum_{k=0}^{L-1}B_{(j,j'),k}$.
Collecting the reductions over all incident chains gives the first
correction in~\eqref{eq:adjusted-base-earnings}. Applied on its own,
it changes the earning requirement of the basic agent associated
with chore $j$ from $1$ to
$$
1-\frac32\sum_{k=0}^{L-1}
\left(
\sum_{j'<j}B_{(j',j),k}
+\sum_{j'>j}B_{(j,j'),k}
\right).
$$
At fixed prices, this change preserves the agent's optimal
ratio $p_j$ and changes only the coefficient of $\log p_j$
in her objective contribution.
Summing over all basic agents, the reduction therefore adds
$$
\frac32\sum_{0\leq j<j'\leq n}
\left(\sum_{k=0}^{L-1}B_{(j,j'),k}\right)
(\log p_j+\log p_{j'})
$$
to the objective, exactly canceling the additional logarithmic
contributions in~\eqref{eq:chain-extra-logarithms}
summed over all chains.

\paragraph{Recovering the Linear Term.}
To establish \cref{item:market-error-bound} of \Cref{prop:market-target}, we approximate
the scaled objective $\alpha Q(x)$, so the remaining correction is
$\alpha c^\top x$.
Recall that the basic price family from~\cref{sec:basic-market} is
$$
p_j(x)=\frac{1+tx_j}{1+t\bar x},
\qquad
\bar x=\frac{1}{n+1}\sum_{j=0}^{n}x_j.
$$
Thus, an individual logarithm has expansion
$$
\begin{aligned}
\log p_j(x)
&=\log(1+tx_j)-\log(1+t\bar x)\\
&=t(x_j-\bar x)
-\frac{t^2}{2}(x_j^2-\bar x^2)+O(t^3).
\end{aligned}
$$
However, summing over $j=0,\ldots,n$ cancels the linear terms:
$$
\sum_{j=0}^{n}\log p_j(x)
=-\frac{t^2}{2}
\left(\sum_{j=0}^{n}x_j^2-(n+1)\bar x^2\right)
+O((n+1)t^3).
$$
The second correction in~\eqref{eq:adjusted-base-earnings}
adjusts the coefficients of these logarithms to recover the
desired linear contribution.
Since $\sum_{j=0}^{n}c_j=0$ by
\eqref{eq:linear-reference-coefficient} and $x_0=0$,
Taylor expansion gives
\begin{equation}
\label{eq:linear-log-expansion}
\begin{aligned}
\sum_{j=0}^{n}\frac{\alpha}{t}c_j\log p_j(x)
&=\sum_{j=0}^{n}\frac{\alpha}{t}c_j
\bigl(\log(1+tx_j)-\log(1+t\bar x)\bigr)\\
&=\sum_{j=1}^{n}\frac{\alpha}{t}c_j\log(1+tx_j)\\
&=\alpha c^\top x-\frac{\alpha t}{2}\sum_{j=1}^{n}c_jx_j^2
+O\left(\alpha t^2\sum_{j=1}^{n}|c_j|\right),
\end{aligned}
\end{equation}
uniformly over $x\in[0,1]^n$.

To introduce this expression into the objective, replace the
current earning requirement $B_a$ of the basic agent associated
with chore $j$ by $B_a-\frac{\alpha}{t}c_j$.
Her contribution then becomes
$$
-\left(B_a-\frac{\alpha}{t}c_j\right)\log p_j(x)
=
-B_a\log p_j(x)
+\frac{\alpha}{t}c_j\log p_j(x).
$$
Applying this adjustment to all basic agents therefore adds
$$
\sum_{j=0}^{n}\frac{\alpha}{t}c_j\log p_j(x)
=\alpha c^\top x+R_{\mathrm{lin}}(x),
$$
where
$$
R_{\mathrm{lin}}(x)
:=
\sum_{j=1}^{n}\frac{\alpha}{t}c_j
\bigl(\log(1+tx_j)-tx_j\bigr).
$$
The factor $\alpha/t$ converts the first-order term $tx_j$
of each logarithm into $\alpha x_j$, while
$R_{\mathrm{lin}}(x)$ collects the higher-order terms.

\paragraph{Validity and Approximation Error.}
Our choice of $\alpha$ in \eqref{eq:choice-of-alpha} ensures that the adjusted earning requirements remain positive\footnote{Moreover, the basic and gadget agents together have total
earning requirement
$
\sum_{a\in N}B_a=n+1.
$ See \cref{lem:adjusted-base-earnings} in \cref{appendix: subsec6.4} for a proof.}.
This choice also gives a uniform third-order bound for the
linear approximation error.

\ignore{
Along the constructed price family $p(x)$, the designated-choice
comparison ensures that adding the gadgets preserves the basic
agents' optimal ratios $p_j(x)$.
The following theorem quantifies the error introduced by the
linear adjustment.}

\begin{theorem}[Linear Approximation]
\label{thm:linear-approximation}
For every $x\in[0,1]^n$,
\begin{equation}
\label{eq:linear-approximation}
\frac{\alpha}{t}\sum_{j=0}^{n}c_j\log p_j(x)
=\alpha c^\top x+R_{\mathrm{lin}}(x).
\end{equation}
For every $\ell\in[n]$ and every coordinate segment from $x$
to $x+he_\ell$ contained in $[0,1]^n$,
\begin{equation}
\label{eq:linear-remainder-lipschitz}
\left|R_{\mathrm{lin}}(x+he_\ell)-R_{\mathrm{lin}}(x)\right|
\leq\frac{t^3}{2(n+1)}|h|.
\end{equation}
Moreover, for every feasible coordinate direction
$\tilde v\in\{\pm e_\ell:\ell\in[n]\}$ at $x$, the right
directional derivative exists and satisfies
$$
\left|R_{\mathrm{lin}}'(x;\tilde v)\right|
\leq\frac{t^3}{2(n+1)}.
$$
\end{theorem}

The proof of linear approximation is deferred to \hyperref[app:error-linear]{Appendix~\ref*{app:error-bounds}}.

\paragraph{Objective Approximation.}
Combining
\cref{thm:basic-quadratic-approximation,thm:chain-quadratic-approximation,thm:linear-approximation}
gives
\begin{equation}
\label{eq:current-market-objective}
\begin{aligned}
\Psi(p(x))
&=-\sum_{a\in N}B_a\log\left(
\max_{j\in M}\frac{p_j(x)}{d_{aj}}\right)\\
&=C+\alpha Q(x)
+R_0(x)-\sum_{0\leq j<j'\leq n}R_{(j,j')}(x)
+R_{\mathrm{lin}}(x).
\end{aligned}
\end{equation}
Here $Q(x)=\frac12x^\top Ax+c^\top x$, $C$ is independent of $x$,
and $N$ and $M$ contain all basic and gadget agents and chores,
respectively. The basic agents have the adjusted earning
requirements in~\eqref{eq:adjusted-base-earnings}.
The same theorems provide third-order bounds on the remainder
terms along feasible coordinate changes.

The earning adjustments above yield the objective approximation
in~\eqref{eq:current-market-objective}.
It remains to extend the normalization, which currently accounts
only for the basic chores, to include the auxiliary chores while
preserving this approximation.

\subsection{Normalization of the Full Price Vector}
\label{subsec:normalization-final-program}

Since our reduction encodes $x$ in prices and recovers it from
an equilibrium, the full price family must satisfy payment
balance (\Cref{def:CE}).
We enforce this condition by adjusting the basic supplies and
normalizing the full price vector, introducing only third-order
error along feasible coordinate changes.

Since the adjusted earning requirements preserve total earnings
at $n+1$ (\cref{lem:adjusted-base-earnings}), payment balance requires
$$
\sum_{j\in M}S_jp_j=\sum_{a\in N}B_a=n+1
\qquad\text{for every }x\in[0,1]^n.
$$

Recall that the gadget in layer $k$ of chain $(j,j')$ introduces
two auxiliary chores, each with supply $S_{(j,j'),k}$.
The basic chores, with prices $p_j(x)$, still have unit supplies.
The auxiliary prices are defined recursively by the gadget
max--min formulas.
All prices are positive, and the prices of the basic chores satisfy
$$
\sum_{j=0}^{n}p_j=n+1.
$$
Once the auxiliary chores are included, total payment becomes
$$
\begin{aligned}
\sum_{j\in M}S_jp_j
&=
\underbrace{\sum_{j=0}^{n}p_j}
_{\text{basic chores: already balanced}}
+
\underbrace{
\sum_{0\leq j<j'\leq n}\sum_{k=0}^{L-1}
S_{(j,j'),k}
\bigl(p^+_{(j,j'),k}+p^-_{(j,j'),k}\bigr)
}_{\text{auxiliary chores: introduce the imbalance}}.\\
&>n+1=\sum_{a\in N}B_a
\end{aligned}
$$
Thus the current price family does not yet satisfy payment balance.
We first adjust the basic supplies to offset the additional
payments, then rescale the full price vector to eliminate
the remaining imbalance.

\paragraph{Adjusting Supplies and Normalizing Prices.}
For each layer $k$ of chain $(j,j')$, we reduce the supply of each
basic endpoint, $j$ and $j'$, by $S_{(j,j'),k}$, while keeping
the auxiliary supplies unchanged.

Collecting these reductions over all incident chains, we define
\begin{equation}
\label{eq:adjusted-basic-supplies}
S_j
:=
1-\sum_{k=0}^{L-1}
\sum_{\substack{j'=0\\j'\neq j}}^{n}
S_{(j,j'),k},
\qquad j=0,\ldots,n,
\end{equation}
using the symmetric convention
$S_{(j,j'),k}=S_{(j',j),k}$.
We defer the verification of the rationality and positivity
of $S_j$ to \Cref{lem:adjusted-basic-supplies} in \Cref{app:normalization}.

With these supplies fixed, let $T(x):=\sum_{j\in M}S_jp_j(x)$ be the total payment at the
current prices.
Since all supplies and prices are positive, $T(x)>0$.
We then rescale the full price vector by setting
$$
\widetilde p_j(x)
:=
\frac{n+1}{T(x)}\,p_j(x),
\qquad j\in M.
$$
For every $x\in[0,1]^n$, the resulting prices satisfy
payment balance:
$$
\sum_{j\in M}S_j\widetilde p_j(x)
=
\frac{n+1}{T(x)}
\sum_{j\in M}S_jp_j(x)
=n+1
=\sum_{a\in N}B_a.
$$

Common scaling preserves the gadget price relations because
both maximum and minimum scale by the same positive factor.
It also scales all of an agent's price-to-disutility ratios
equally, preserving her optimal choices.
Thus, for every agent $a$,
$$
\mu_a(\widetilde p(x))
=
\frac{n+1}{T(x)}\mu_a(p(x)).
$$

Since $\sum_{a\in N}B_a=n+1$, the logarithmic objective becomes
\begin{equation}
\label{eq:normalization-objective-identity}
\begin{aligned}
\Psi(\widetilde p(x))
&=-\sum_{a\in N}B_a\log\mu_a(\widetilde p(x))\\
&=\Psi(p(x))
+\left(\sum_{a\in N}B_a\right)\log\frac{T(x)}{n+1}\\
&=\Psi(p(x))+(n+1)\log\frac{T(x)}{n+1}.
\end{aligned}
\end{equation}
Normalization therefore introduces the additive correction
$
R_{\mathrm{norm}}(x)
:=
(n+1)\log\frac{T(x)}{n+1}.
$

The following theorem shows that this correction equals the
change in the logarithmic objective and satisfies the required
third-order error bounds.

\begin{theorem}[Normalization Error]
\label{thm:normalization-error}
For every $x\in[0,1]^n$,
$$
\Psi(\widetilde p(x))
=
\Psi(p(x))+R_{\mathrm{norm}}(x).
$$

The remainder satisfies
\begin{equation*}
\left|
R_{\mathrm{norm}}(x+he_\ell)-R_{\mathrm{norm}}(x)
\right|
\leq3t^3|h|
\end{equation*}
for every $\ell\in[n]$ and every coordinate segment from $x$
to $x+he_\ell$ contained in $[0,1]^n$.
Moreover, for every feasible coordinate direction
$\widetilde v\in\{\pm e_\ell:\ell\in[n]\}$ at $x$, the right
directional derivative exists and satisfies
$$
\left|
R_{\mathrm{norm}}'(x;\widetilde v)
\right|
\leq3t^3.
$$
\end{theorem}
We defer the proof to \cref{app:error-normalization}.

\begin{remark}
A natural alternative is to retain unit supplies for the basic
chores and normalize the full price vector directly.
Although this enforces payment balance, the resulting correction
to the logarithmic objective can contain an $x$-dependent term
of order $t^2$.
Thus, without the supply adjustment, normalization need not
satisfy the third-order error bounds along feasible coordinate
changes.
The detailed derivation appears in \Cref{Failure of Direct Normalization}.
\end{remark}

\section{Verifying the Market Properties in \texorpdfstring{\cref{prop:market-target}}{Proposition \ref{prop:market-target}}}
\label{sec:completed-market}

In this section, we complete the market construction and verify
that the constructed market satisfies
Properties~\ref{item:market-price-mapping}\ref{item:market-decoding},
\ref{item:market-price-mapping}\ref{item:market-right-derivative},
and~\ref{item:market-error-bound} of \Cref{prop:market-target}.

\paragraph{The Completed Market.}
The completed market consists of the basic market from
\Cref{sec:basic-market} and an $L$-layer gadget chain from
\Cref{def:gadget-chain} for every pair of basic chores
$0\leq j<j'\leq n$.
There are $n+1$ basic agents and $n+1$ basic chores.
Each gadget layer adds three agents and two auxiliary chores.
We complete the disutility matrix by assigning $(1+t)^2$
to every unspecified agent--chore pair, obtaining
$$
\mathcal F=(N,M,\{d_{aj}\},\{B_a\},\{S_j\}).
$$

For the input $\mathcal I=(A,c,\varepsilon)$, choose
\begin{equation*}
t:=\max\left\{
2^{-\ell}:\ell\in\mathbb Z_{\ge8},\
2^{-\ell}\leq
\frac{\varepsilon}{
128(n+1)\left(
1+\sum_{i,j=1}^n|A_{ij}|+\sum_{i=1}^n|c_i|
\right)}
\right\}.
\end{equation*}

The chain length $L$ is determined by $t=2^{-L}$,
as required in \cref{clm:final-parameter-choice}.
The weights and scaling factor are given by
\eqref{eq:basic-weights} and \eqref{eq:choice-of-alpha}.
We use the gadget parameters from
\cref{clm:final-parameter-choice} and the adjusted basic
earning requirements and supplies from
\eqref{eq:adjusted-base-earnings} and
\eqref{eq:adjusted-basic-supplies}, respectively.

All market data are strictly positive and rational and are
fixed independently of $x$, with
$$
\sum_{a\in N}B_a=\sum_{j\in M}S_j=n+1.
$$
The normalization in \eqref{eq:recovery-family} ensures
$\sum_{j\in M}S_j\widetilde p_j(x)=n+1$ for every
$x\in[0,1]^n$.
By \cref{lem:disutility-completion}, completing the
disutility matrix preserves the optimal-ratio formulas
along $\widetilde p(x)$.
We give a detailed specification of the completed market
in \Cref{sec:equilibrium-recovery}.

\subsection{Verification of Decoding Mapping}
\label{sec:equilibrium-recovery}

In this subsection, we verify
property~\ref{item:market-price-mapping}\ref{item:market-decoding}
of \Cref{prop:market-target}.
It suffices to show that every equilibrium price vector of the completed market
lies in the constructed price family and that a corresponding point
in the unit box can be recovered from it in polynomial time (\cref{lem:equilibrium-recovery}).
\paragraph{Market Specification.}
We now specify the completed market
$\mathcal F=(N,M,\allowbreak {}\{d_{aj}\},\{B_a\},\{S_j\})$
constructed from $\mathcal I=(A,c,\varepsilon)$.
Set
$$
\lambda:=
\frac{1}{2(n+1)\left(
1+\sum_{i,j=1}^n|A_{ij}|+\sum_{i=1}^n|c_i|
\right)}.
$$
The choice of $t$ at the beginning of \Cref{sec:completed-market},
together with the resulting chain length and scaling factor,
can then be written as
\begin{equation}
\label{eq:reduction-parameter-choice}
\begin{aligned}
t&:=\max\left\{
2^{-\ell}:\ell\in\mathbb Z_{\ge8},\
2^{-\ell}\leq\frac{\lambda\varepsilon}{64}
\right\},\\
L&:=\log_2(1/t),
\qquad
\alpha:=\lambda t^2.
\end{aligned}
\end{equation}
Thus $L\in\mathbb Z_{\ge8}$, $t=2^{-L}$, and
$0<t\leq\min\{1/256,\lambda\varepsilon/64\}$.
Let $c_0:=-\sum_{i=1}^n c_i$.
As in \eqref{eq:basic-weights}, define
$$
w_{jj'}:=\frac1{n+1}+\lambda A_{jj'}
\quad (1\leq j<j'\leq n),
\qquad
w_{0j}:=\frac1{n+1}-\lambda\sum_{j'=1}^n A_{jj'}
\quad (j\in[n]),
$$
and write $w_{j'j}=w_{jj'}$.

\begin{enumerate}[label=(\roman*)]
    \item \textbf{Agents and Chores.}
    For each basic chore $j\in\{0,\ldots,n\}$, let $a_j$
    denote its associated basic agent.
    For every pair $0\leq j<j'\leq n$, add an $L$-layer
    gadget chain as in \Cref{def:gadget-chain}.
    Thus
    $$
    N=
    \{a_j:0\leq j\leq n\}
    \cup
    \bigcup_{\substack{0\leq j<j'\leq n\\0\leq k<L}}
    \left\{
    a^{\max}_{(j,j'),k},
    a^{(1)}_{(j,j'),k},
    a^{(2)}_{(j,j'),k}
    \right\},
    $$
    and
    $$
    M=
    \{0,\ldots,n\}
    \cup
    \bigcup_{\substack{0\leq j<j'\leq n\\0\leq k<L}}
    \left\{
    p^+_{(j,j'),k},p^-_{(j,j'),k}
    \right\}.
    $$
    Here $p^+_{(j,j'),k}$ and $p^-_{(j,j'),k}$ name the
    plus and minus output chores of the corresponding gadget.
    All auxiliary agents and chores are distinct across
    pairs and layers and from the basic agents and chores.
    The ordered inputs of layer $0$ are $(j,j')$;
    those of layer $k\geq1$ are
    $(p^+_{(j,j'),k-1},p^-_{(j,j'),k-1})$.

    \item \textbf{Gadget Parameters, Earning Requirements,
    and Supplies.}
    For every chain $(j,j')$ and layer $k=0,\ldots,L-1$,
    set, as in \eqref{eq:chain-approximation-parameters},
    $$
    q_k:=1+\frac t4\left(\frac12-4t\right)^k,
    \qquad
    B_{(j,j'),k}:=
    w_{jj'}t\left(\frac12-4t\right)^k,
    \qquad
    S_{(j,j'),k}:=(1-4t)B_{(j,j'),k}.
    $$
    Each of the three agents has earning requirement
    $B_{(j,j'),k}$, and each output chore has supply
    $S_{(j,j'),k}$:
    $$
    B_{a^{\max}_{(j,j'),k}}
    =
    B_{a^{(1)}_{(j,j'),k}}
    =
    B_{a^{(2)}_{(j,j'),k}}
    =
    B_{(j,j'),k},
    \qquad
    S_{p^+_{(j,j'),k}}
    =
    S_{p^-_{(j,j'),k}}
    =
    S_{(j,j'),k}.
    $$
    For use in sums, write
    $B_{(j',j),k}=B_{(j,j'),k}$ and
    $S_{(j',j),k}=S_{(j,j'),k}$.

    \item \textbf{Basic Earning Requirements and Supplies.}
    For every $j=0,\ldots,n$, the adjustments in
    \eqref{eq:adjusted-base-earnings} and
    \eqref{eq:adjusted-basic-supplies} give
    $$
    B_{a_j}:=
    1-\frac32
    \sum_{k=0}^{L-1}
    \sum_{\substack{j'=0\\j'\neq j}}^n
    B_{(j,j'),k}
    -\frac{\alpha}{t}c_j,
    \qquad
    S_j:=
    1-
    \sum_{k=0}^{L-1}
    \sum_{\substack{j'=0\\j'\neq j}}^n
    S_{(j,j'),k}.
    $$

    \item \textbf{Disutilities.}
    For every basic agent $a_j$, set
    $$
    d_{a_j\ell}:=
    \begin{cases}
        1,
        & \ell\in\{0,j\},\\
        1+t,
        & \ell\in\{0,\ldots,n\}\setminus\{0,j\},\\
        (1+t)^2,
        & \ell\in M\setminus\{0,\ldots,n\}.
    \end{cases}
    $$
    For a gadget in layer $k$, we omit its pair and
    layer indices. The disutilities are
    $$
    \begin{array}{c|ccccc}
        & \text{first input}
        & \text{second input}
        & \text{plus output}
        & \text{minus output}
        & \text{all other chores}\\
        \hline
        a^{\max}
        & q_k & q_k & 1 & (1+t)^2 & (1+t)^2\\
        a^{(1)}
        & 1/q_k & (1+t)^2 & (1+t)^2 & 1 & (1+t)^2\\
        a^{(2)}
        & (1+t)^2 & 1/q_k & (1+t)^2 & 1 & (1+t)^2
    \end{array}
    $$
    These are the disutilities from
    \Cref{sec:basic-market,def:gadget-chain}, with every
    unspecified entry set to $(1+t)^2$ as in
    \cref{lem:disutility-completion}.
\end{enumerate}

All market data are strictly positive and rational, are fixed
independently of $x$, and satisfy
$$
\sum_{a\in N}B_a=\sum_{\ell\in M}S_\ell=n+1
$$
by \cref{lem:adjusted-base-earnings,lem:adjusted-basic-supplies}.
The construction has
$$
|N|=n+1+3L\binom{n+1}{2},
\qquad
|M|=n+1+2L\binom{n+1}{2}.
$$
Since $L$ and the encoding lengths of all entries are
polynomially bounded in the encoding length of $\mathcal I$,
the market can be constructed in polynomial time.

\paragraph{Price Mapping.} We then recall the price mapping from \Cref{subsec:normalization-final-program}.
For $x\in[0,1]^n$, let $x_0:=0$ and $\bar x:=\frac1{n+1}\sum_{j=0}^nx_j$. The
price vector $p(x)$ has basic prices $p_j(x)=(1+tx_j)/(1+t\bar x)$ for
$j=0,\ldots,n$~\eqref{eq:naive-price}, and its output prices are defined layer by layer by the
max--min rule~\eqref{eq:gadget-output-prices}: if the first and second inputs of a gadget at layer $k$ have
prices $u$ and $v$, then its plus and minus outputs have prices
$\max\{u,v\}/q_k$ and $q_k\min\{u,v\}$. The price family $\widetilde p$ rescales
$p(x)$ to satisfy payment balance:
\begin{equation}
\label{eq:recovery-family}
\widetilde p(x):=\frac{n+1}{T(x)}\,p(x),
\qquad\text{where}\quad
T(x):=\sum_{\ell\in M}S_\ell\,p_\ell(x).
\end{equation}
Thus $\sum_{\ell\in M}S_\ell\,\widetilde p_\ell(x)=n+1$ and
$\widetilde p_j(x)/\widetilde p_0(x)=1+tx_j$ for every $j\in[n]$. 

Accordingly,
we define the decoder $g$ from \Cref{prop:market-target}: for an instance
$\mathcal{I}$ and an equilibrium $(p,z)$ of the completed market $f(\mathcal{I})$,
\begin{equation}
\label{eq:decoder}
g\bigl(\mathcal{I},(p,z)\bigr):=\frac1t\left(\frac{p_1}{p_0}-1,\ \ldots,\
\frac{p_n}{p_0}-1\right).
\end{equation}
The decoder $g$ first computes $t$ from the instance $\mathcal{I}$.
It then recovers $x$ using only $t$ and the equilibrium prices,
according to \eqref{eq:decoder}.

\begin{lemma}[Equilibrium Recovery]
\label{lem:equilibrium-recovery}
Let $(p^\star,z^\star)$ be an equilibrium of the completed market $f(\mathcal{I})$, and
let $x^\star:=g\bigl(\mathcal{I},(p^\star,z^\star)\bigr)$. Then $x^\star\in[0,1]^n$ and
$p^\star=\widetilde p(x^\star)$.
\end{lemma}

\paragraph{Proof Plan.}
Each agent has \emph{designated choices}
(\Cref{def:max-min-gadget}):
the basic chores for a basic agent, and its inputs and its designated output
for a gadget agent. Their disutilities, specified
in~\Cref{sec:basic-market} and \Cref{def:gadget-chain}, namely
in $[(1+t)^{-1},1+t]$, are low. All other agent--chore pairs are \emph{undesignated
choices} and have the high disutility $(1+t)^2$
(\cref{lem:disutility-completion}). The main step is to show that, at
every equilibrium, every agent uses only its designated choices
(\cref{lem:recovery-fine}). Together with
bounds on the payments of the output chores
(\cref{lem:recovery-coarse}, \cref{item:coarse-window}),
this shows that every gadget
satisfies the assumption of \cref{lem:max-min-gadget}, so its output prices
follow the max--min rule, and recovering $x^\star$ is straightforward.
The bounds on basic prices in \cref{item:coarse-basic} ensure that $x^\star\in[0,1]^n$.
These price relations and payment balance then give
$p^\star=\widetilde p(x^\star)$, proving
\cref{lem:equilibrium-recovery}.

\paragraph{Setup.}
Fix an equilibrium $(p^\star,z^\star)$ of the completed market.
For simplicity, write $(p,z)$ for $(p^\star,z^\star)$ throughout
the remainder of \cref{sec:equilibrium-recovery}. We say that agent $a$ uses chore $\ell$ if
$z_{a\ell}>0$. By \Cref{def:CE}, all prices are positive, each agent
earns exactly $B_a$, and she uses only chores maximizing
$p_\ell/d_{a\ell}$. Payment balance \eqref{eq:market_clearing} gives
$\sum_{\ell\in M}S_\ell p_\ell=n+1=\sum_{\ell\in M}S_\ell$.

\paragraph{Necessary Condition for Using Undesignated Choices.}In this part, we establish a necessary condition for an agent
to use an undesignated choice
(\cref{lem:recovery-undesignatedd}).
We later combine this condition with bounds on equilibrium prices
to prove that every agent uses only designated choices at equilibrium
(\cref{lem:recovery-fine}).

\begin{lemma}
\label{lem:recovery-undesignatedd}
If an agent $a$ uses an undesignated choice $\ell$, then
$p_\ell=p_{\max}\geq(1+t)^2p_{\min}$, where $p_{\max}$ and
$p_{\min}$ denote the largest and smallest prices among all chores,
respectively. Consequently, a chore with price less than $p_{\max}$
is used only by agents for which it is a designated choice, and
if $p_{\max}<(1+t)^2p_{\min}$, then no agent uses an undesignated choice.
\end{lemma}

\begin{proof}
Suppose agent $a$ uses an undesignated chore $\ell$, so
$d_{a\ell}=(1+t)^2$.
Take a chore $r$ with price $p_r=p_{\max}$.
Since $d_{ar}\leq(1+t)^2$, individual optimality~(\Cref{def:CE})
 gives
$$
\frac{p_\ell}{(1+t)^2}
=\mu_a(p)
\geq\frac{p_{\max}}{d_{ar}}
\geq\frac{p_{\max}}{(1+t)^2}.
$$
Hence $p_\ell\geq p_{\max}$, and therefore $p_\ell=p_{\max}$.

Moreover, agent $a$ has a designated chore $s$ with
$d_{as}=1$. Thus
$$
\frac{p_{\max}}{(1+t)^2}
=\mu_a(p)
\geq\frac{p_s}{d_{as}}
=p_s
\geq p_{\min},
$$
which implies $p_{\max}\geq(1+t)^2p_{\min}$.

\end{proof}

By \cref{lem:recovery-undesignatedd}, it suffices to prove
$p_{\max}<(1+t)^2p_{\min}$ to rule out undesignated choices at equilibrium.

\paragraph{Ruling Out Undesignated Choices.}
\label{subsec:recovery-fine}
We now show that, at every equilibrium of the completed market, each agent
uses only the chores designated for it in the preceding construction.
We begin by introducing next-layer agents and recalling \eqref{eq:gadget-layer-payment-window}.
We then establish bounds on output payments and basic prices
(\cref{lem:recovery-coarse}). These bounds force certain agents to use
chores within their gadgets, allowing us to control the highest and lowest
prices and exclude undesignated choices (\cref{lem:recovery-fine}).

\emph{Next-layer Agents.}
Let \emph{out} denote an output chore of a gadget in layer $k$.
It is a designated choice of $a^{\max}$ if it is the plus output,
and of $a^{(1)}$ and $a^{(2)}$ if it is the minus output. If $k<L-1$, then \emph{out} is also an input of layer
$k+1$, and it is a designated choice of two agents there. We call them the
\emph{next-layer agents} of \emph{out}. By~\eqref{eq:chain-approximation-parameters},
every agent of the gadget at layer $k$ of chain $(j,j')$ has earning
requirement $B_{(j,j'),k}=w_{jj'}t(\frac12-4t)^k$. So the next-layer agents of
an output of layer $k$ of chain $(j,j')$ have total earning requirement
\begin{equation}
\label{eq:next-layer-requirement}
2B_{(j,j'),k+1}=(1-8t)B_{(j,j'),k}
\qquad\text{if }k<L-1.
\end{equation}
Outputs of the last layer have no next-layer agents.
 
\emph{Price Condition.}
Each output of layer $k$ of chain $(j,j')$ has supply
$S_{(j,j'),k}=(1-4t)B_{(j,j'),k}$ by
\eqref{eq:chain-approximation-parameters}.
Recall
\eqref{eq:gadget-layer-payment-window}: for each sign,
$$
\sum_{a\in N_{\mathrm{out},k}^{\pm}}B_a
<S_{(j,j'),k}\,p^\pm_{(j,j'),k}
<B_{(j,j'),k},
$$
where $N_{\mathrm{out},k}^{\pm}$ consists of agents outside
the gadget who actually use the corresponding output.
In \cref{lem:recovery-coarse}, \cref{item:coarse-window},
we first establish the upper bound for every output and
the lower bound for outputs priced below $p_{\max}$.
After \cref{lem:recovery-fine} rules out undesignated choices,
we extend the lower bound to every output in the final
equilibrium-recovery proof (\cref{lem:equilibrium-recovery}).

\begin{lemma}[Price Bounds]
\label{lem:recovery-coarse}
The equilibrium price vector $p$ satisfies the following properties.
\begin{lemmaitems}
\item\label{item:coarse-range}
$(1+t)^{-3}\leq p_\ell\leq(1+t)^3$ for every chore $\ell$.
\item\label{item:coarse-window}
For every chain $(j,j')$, layer $k$, and sign
$\sigma\in\{+,-\}$,
\begin{equation}
\label{eq:recovery-windows}
\begin{gathered}
S_{(j,j'),k}\,p^\sigma_{(j,j'),k}<B_{(j,j'),k},\\
p^\sigma_{(j,j'),k}<p_{\max}
\quad\Longrightarrow\quad
\sum_{a\in N_{\mathrm{out},k}^{\sigma}}B_a
<S_{(j,j'),k}\,p^\sigma_{(j,j'),k}.
\end{gathered}
\end{equation}
\item\label{item:coarse-basic}
$p_0\leq p_j\leq(1+t)p_0$ for every $j\in[n]$.
\end{lemmaitems}
\end{lemma}
 
\begin{proof}
\textbf{\cref{item:coarse-range}.}
Let $\ell_{\min}$ and $\ell_{\max}$ be chores with prices
$p_{\min}$ and $p_{\max}$, respectively.
By market clearing, some agent $a$ uses $\ell_{\min}$.
Optimality in \cref{def:CE} gives
$$
\frac{p_{\min}}{d_{a\ell_{\min}}}
=\mu_a(p)
\geq
\frac{p_{\max}}{d_{a\ell_{\max}}}.
$$
Since all disutility coefficients lie in
$[(1+t)^{-1},(1+t)^2]$, it follows that
$$
\frac{p_{\max}}{p_{\min}}
\leq
\frac{d_{a\ell_{\max}}}{d_{a\ell_{\min}}}
\leq
\frac{(1+t)^2}{(1+t)^{-1}}
=(1+t)^3.
$$
Moreover, payment balance \eqref{eq:market_clearing} implies
that the supply-weighted average price is $1$:
$$
\frac{\sum_{\ell\in M}S_\ell p_\ell}
     {\sum_{\ell\in M}S_\ell}
=1.
$$
Hence $p_{\min}\leq1\leq p_{\max}$.
Combining this with the ratio bound yields
$$
(1+t)^{-3}\leq p_{\min}\leq p_\ell
\leq p_{\max}\leq(1+t)^3
\qquad\text{for every }\ell\in M,
$$
which proves \cref{item:coarse-range}.

\textbf{\cref{item:coarse-window}.}
Fix a sign $\sigma\in\{+,-\}$ and the corresponding output
\emph{out} of layer $k$ of chain $(j,j')$,
and write $B=B_{(j,j'),k}$.
Since $t\leq1/256$, we have
$(1+t)^3<1+4t$ and $(1+t)^{-3}\geq1-3t$.
Thus \cref{item:coarse-range} and
$S_{(j,j'),k}=(1-4t)B$ give
$$
S_{(j,j'),k}\,p_{\text{\emph{out}}}
\leq(1-4t)(1+4t)B<B.
$$

Now suppose $p_{\text{\emph{out}}}<p_{\max}$.
By \cref{lem:recovery-undesignatedd}, every agent using
\emph{out} has it as a designated choice.
Hence the outside users of \emph{out} are among its
next-layer agents.
If $k<L-1$, \eqref{eq:next-layer-requirement} gives
$$
\begin{aligned}
\sum_{a\in N_{\mathrm{out},k}^{\sigma}}B_a
&\leq2B_{(j,j'),k+1}
=(1-8t)B\\
&<(1-4t)(1-3t)B
\leq S_{(j,j'),k}\,p_{\text{\emph{out}}},
\end{aligned}
$$
where the last inequality uses \cref{item:coarse-range}.
If $k=L-1$, there are no next-layer agents, so
$N_{\mathrm{out},k}^{\sigma}=\varnothing$ and the lower
bound follows from $S_{(j,j'),k}\,p_{\text{\emph{out}}}>0$.
This proves \cref{item:coarse-window}.

\textbf{\cref{item:coarse-basic}.}
We first show that every basic chore $\ell$ with
$p_\ell<p_{\max}$ is used by some basic agent.
By \cref{lem:recovery-undesignatedd}, every agent using
$\ell$ must have $\ell$ as a designated choice.
Besides the basic agents, these are the two agents in layer $0$
of each chain containing $\ell$.
By \eqref{eq:chain-approximation-parameters}, their total
earning requirement is
$$
2t\sum_{j'\neq\ell}w_{\ell j'}<2t,
$$
where the inequality follows from
$\sum_{j'\neq\ell}w_{\ell j'}<1$
(\Cref{app:normalization}).
Hence these gadget agents earn less than $2t$ from $\ell$.

On the other hand, $S_\ell>1-2t$ by
\cref{lem:adjusted-basic-supplies}, and
$p_\ell\geq(1+t)^{-3}\geq1-3t$ by
\cref{item:coarse-range}.
By market clearing, the total earnings from $\ell$ are therefore
$$
S_\ell p_\ell>(1-2t)(1-3t)>2t.
$$
Consequently, some basic agent must use $\ell$.

Now fix $j\in[n]$.
By \Cref{sec:basic-market}, every basic agent has disutility
$1$ for chore $0$ and disutility in $[1,1+t]$ for every
basic chore.
To prove $p_j\geq p_0$, first suppose $p_j<p_{\max}$.
Then some basic agent $a$ uses $j$, and
\cref{def:CE} gives
$$
p_j
\geq\frac{p_j}{d_{aj}}
=\mu_a(p)
\geq\frac{p_0}{d_{a0}}
=p_0.
$$
Otherwise, $p_j=p_{\max}\geq p_0$.

To prove $p_j\leq(1+t)p_0$, first suppose $p_0<p_{\max}$.
Then some basic agent $b$ uses chore $0$, and optimality gives
$$
p_0
=\frac{p_0}{d_{b0}}
=\mu_b(p)
\geq\frac{p_j}{d_{bj}}
\geq\frac{p_j}{1+t}.
$$
Otherwise, $p_j\leq p_0=p_{\max}$, which also implies
$p_j\leq(1+t)p_0$.
Thus $p_0\leq p_j\leq(1+t)p_0$, proving
\cref{item:coarse-basic}.
\end{proof}

The preliminary price-ratio bound
$p_{\max}/p_{\min}\leq(1+t)^3$ does not rule out undesignated choices.
Increasing the disutility of undesignated choices does not help: this
estimate is always $1+t$ times this disutility, and a larger disutility weakens
the resulting price bounds used to establish \cref{item:coarse-window}.
Instead, we use the payment bounds to show that
all prices lie in $[p_0/q_0,(1+t)q_0p_0]$, so
$p_{\max}/p_{\min}\leq(1+t)q_0^2<(1+t)^2$. We cannot use the max--min rule
here, since it requires that agents use only designated choices. We only
use two consequences of \cref{item:coarse-window}: every agent of a gadget also uses
a chore other than its designated output, and an output with price less than
$p_{\max}$ is used by an agent of its gadget. We apply them to a most expensive
chore and to a cheapest chore, and use that $q_k$ is decreasing in $k$.

\begin{lemma}[Excluding Undesignated Choices]
\label{lem:recovery-fine}
The equilibrium price vector $p$ satisfies
\begin{equation}
\label{eq:recovery-fine-bounds}
p_{\max}\leq(1+t)q_0\,p_0
\qquad\text{and}\qquad
p_{\min}\geq\frac{p_0}{q_0}.
\end{equation}
Consequently, no agent uses an undesignated choice.
\end{lemma}
 
\begin{proof}
Recall from~\eqref{eq:chain-approximation-parameters} that
$q_k=1+\frac t4(\frac12-4t)^k$. Thus
$q_0=1+\frac t4>q_1>\cdots>q_{L-1}>1$, and
$q_0^2<1+t$. So the consequence follows from
\cref{lem:recovery-undesignatedd}, since~\eqref{eq:recovery-fine-bounds}
gives $p_{\max}/p_{\min}\leq(1+t)q_0^2<(1+t)^2$.

To prove~\eqref{eq:recovery-fine-bounds}, recall from
\cref{lem:recovery-coarse} that $p$ satisfies
\eqref{eq:recovery-windows} and that
$p_0\leq p_j\leq(1+t)p_0$ for every $j\in[n]$.
We use two consequences of the bounds in
\eqref{eq:recovery-windows}. Let \emph{out} be an output
of layer $k$ of chain $(j,j')$.
\begin{enumerate}[label=(O\arabic*),ref=(O\arabic*)]
    \item\label{item:O-other}
    \emph{If \emph{out} is the designated output of an agent $a$,
    then $a$ uses some chore $\ell\neq\text{\emph{out}}$, and
    $p_{\text{\emph{out}}}\leq p_\ell/d_{a\ell}$.}
    Indeed, \emph{out} pays
    $S_{(j,j'),k}\,p_{\text{\emph{out}}}<B_a$ in total
    by \cref{item:coarse-window}. Since $a$ earns exactly $B_a$,
    she uses some chore $\ell\neq\text{\emph{out}}$, and
    $p_{\text{\emph{out}}}
    =p_{\text{\emph{out}}}/d_{a,\text{\emph{out}}}
    \leq\mu_a(p)=p_\ell/d_{a\ell}$
    by~\cref{def:CE}.

    \item\label{item:O-own}
    \emph{If $p_{\text{\emph{out}}}<p_{\max}$, then \emph{out}
    is used by an agent whose designated output is \emph{out}.}
    Indeed, by \cref{lem:recovery-undesignatedd}, only agents
    for which \emph{out} is a designated choice use \emph{out}.
    Write $\sigma$ for the sign of \emph{out}.
    Since $p_{\text{\emph{out}}}<p_{\max}$,
    \cref{item:coarse-window} gives
    $$
    \sum_{a\in N_{\mathrm{out},k}^{\sigma}}B_a
    <S_{(j,j'),k}\,p_{\text{\emph{out}}}.
    $$
    The outside agents earn at most the left-hand side
    from \emph{out}. Since the agents earn
    $S_{(j,j'),k}\,p_{\text{\emph{out}}}$ from \emph{out}
    in total, some agent of its gadget uses \emph{out}.
    As \emph{out} is a designated choice of this agent,
    it is her designated output.
\end{enumerate}
When the chain $(j,j')$ is clear, we write
$p^\pm_k=p^\pm_{(j,j'),k}$.
The choices of $a^{\max}_{(j,j'),k}$ other than $p^+_k$
have disutility $q_k$ (its inputs) or $(1+t)^2$,
so~\cref{item:O-other} gives
\begin{equation}
\label{eq:recovery-plus-upper}
p^+_k\leq\frac{p_{\max}}{q_k}<p_{\max}
\qquad\text{for every chain $(j,j')$ and every layer $k$.}
\end{equation}

\emph{Upper Bound.}
If some basic chore has price $p_{\max}$, then
$p_{\max}\leq(1+t)p_0$~(\cref{item:coarse-basic}).
Otherwise, let \emph{out} be an output of layer $k$ of some
chain $(j,j')$ with $p_{\text{\emph{out}}}=p_{\max}$.
Then $\text{\emph{out}}=p^-_k$
by~\eqref{eq:recovery-plus-upper}.
We apply~\cref{item:O-other} to $a^{(1)}_{(j,j'),k}$.
Its choices other than $p^-_k$ are its first input, with
disutility $1/q_k$, and undesignated choices, whose ratio
is at most $p_{\max}/(1+t)^2<p_{\max}$.
Hence $p_{\max}\leq q_ku$, where $u$ is the price
of the first input of layer $k$.
If $k\geq1$, then
$u=p^+_{k-1}\leq p_{\max}/q_{k-1}$
by~\eqref{eq:recovery-plus-upper}, which gives
$p_{\max}\leq(q_k/q_{k-1})\,p_{\max}<p_{\max}$,
a contradiction.
So $k=0$ and $u=p_j$, and thus
$p_{\max}\leq q_0p_j\leq(1+t)q_0p_0$.

\emph{Lower Bound.}
Assume for the sake of contradiction that $p_{\min}<p_0/q_0$.
Since basic prices are at least $p_0$~(\cref{item:coarse-basic}), a chore attaining
$p_{\min}$ must be an output.
Let \emph{out} be such an output of layer $k$ of some
chain $(j,j')$. Then
$p_{\text{\emph{out}}}=p_{\min}<p_0\leq p_{\max}$.
By~\cref{item:O-own}, \emph{out} is used by an agent $a$
whose designated output is \emph{out}.
So $\mu_a(p)=p_{\min}$ by~\cref{def:CE}, and the ratio
of $a$ on each of its inputs is at most $p_{\min}$.
Thus $a$ is neither $a^{(1)}_{(j,j'),k}$
nor $a^{(2)}_{(j,j'),k}$, since each of them has an input
with ratio at least $q_kp_{\min}>p_{\min}$.
Hence $a=a^{\max}_{(j,j'),k}$, and both inputs of layer $k$
have price at most $q_kp_{\min}<p_0$.
So they are not basic chores, that is, $k\geq1$, and
$p^-_{k-1}\leq q_kp_{\min}<p_{\max}$.
By~\cref{item:O-own}, some agent
$a'\in\{a^{(1)}_{(j,j'),k-1},a^{(2)}_{(j,j'),k-1}\}$
uses $p^-_{k-1}$.
Comparing with her input, which has price at least $p_{\min}$
and disutility $1/q_{k-1}$, we have
$p^-_{k-1}=\mu_{a'}(p)\geq q_{k-1}p_{\min}>q_kp_{\min}$,
a contradiction.
\end{proof}

Having established the bounds in \eqref{eq:recovery-windows} and ruled out undesignated choices
(\cref{lem:recovery-coarse,lem:recovery-fine}), we now apply
\cref{lem:max-min-gadget} to complete the proof of
\cref{lem:equilibrium-recovery}.

\paragraph{Proof of \cref{lem:equilibrium-recovery}.}

Fix an equilibrium $(p^\star,z^\star)$ of the completed market,
and write $(p,z)=(p^\star,z^\star)$.
Let $x:=g\bigl(\mathcal{I},(p,z)\bigr)$ and $x_0:=0$.

\begin{enumerate}[label=\textbf{Step~\arabic*.},ref=\arabic*,leftmargin=*]
    \item \label{step:recovery-decoding}
    By \cref{lem:recovery-coarse},
    \cref{item:coarse-range,item:coarse-basic}, we have $p_0>0$ and
    $$
    p_0\leq p_j\leq(1+t)p_0
    \qquad\text{for every }j\in[n].
    $$
    Since $t>0$, the decoder \eqref{eq:decoder} gives
    $$
    0\leq x_j
    =\frac1t\left(\frac{p_j}{p_0}-1\right)
    \leq1
    \qquad\text{for every }j\in[n].
    $$
    Hence $x\in[0,1]^n$.

    \item \label{step:recovery-basic-scaling}
    Let
    $
    \bar x:=\frac1{n+1}\sum_{j=0}^n x_j,
    \kappa:=p_0(1+t\bar x)>0.
    $
    By the decoder~\eqref{eq:decoder} and the choice $x_0=0$,
    $1+tx_j=p_j/p_0$ for every $j=0,\ldots,n$.
    Thus, by~\eqref{eq:naive-price},
    $$
    p_j(x)
    =\frac{1+tx_j}{1+t\bar x}
    =\frac{p_j}{p_0(1+t\bar x)}
    =\frac{p_j}{\kappa}.
    $$
    Therefore $p_j=\kappa p_j(x)$ for every basic chore $j$.

    \item \label{step:recovery-price-condition}
    Fix a gadget in layer $k$ of chain $(j,j')$.
    Let $N_{\mathrm{out},k}^{+}$ and $N_{\mathrm{out},k}^{-}$
    be the sets of agents outside this gadget who actually use
    its plus and minus outputs, respectively.
    By \cref{lem:recovery-fine}, every agent uses only designated
    choices. Hence, for $k<L-1$, \Cref{def:gadget-chain} gives
    $$
    \begin{aligned}
    N_{\mathrm{out},k}^{+}
    &\subseteq
    \{a^{\max}_{(j,j'),k+1},a^{(1)}_{(j,j'),k+1}\},\\
    N_{\mathrm{out},k}^{-}
    &\subseteq
    \{a^{\max}_{(j,j'),k+1},a^{(2)}_{(j,j'),k+1}\}.
    \end{aligned}
    $$
    For $k<L-1$, each next-layer agent has earning requirement
    $B_{(j,j'),k+1}$. By \eqref{eq:next-layer-requirement},
    \cref{item:coarse-range}, and
    $S_{(j,j'),k}=(1-4t)B_{(j,j'),k}$, we obtain
    $$
    \begin{aligned}
    \sum_{a\in N_{\mathrm{out},k}^{\pm}}B_a
    &\leq2B_{(j,j'),k+1}
    =(1-8t)B_{(j,j'),k}\\
    &<(1-4t)(1-3t)B_{(j,j'),k}
    \leq S_{(j,j'),k}\,p^\pm_{(j,j'),k},
    \end{aligned}
    $$
    where the last inequality uses
    $p^\pm_{(j,j'),k}\geq(1+t)^{-3}\geq1-3t$.
    For $k=L-1$, both outside-user sets are empty, so the
    strict lower bound follows from
    $S_{(j,j'),k}\,p^\pm_{(j,j'),k}>0$.

    Combining this with the upper bound from
    \cref{lem:recovery-coarse}, \cref{item:coarse-window}, yields
    $$
    \sum_{a\in N_{\mathrm{out},k}^{\pm}}B_a
    <S_{(j,j'),k}\,p^\pm_{(j,j'),k}
    <B_{(j,j'),k}
    $$
    for each sign separately.
    Thus \eqref{eq:gadget-layer-payment-window} holds at every layer.
    By \cref{lem:max-min-gadget}, applied with parameters
    $q_k$, $B_{(j,j'),k}$, and $S_{(j,j'),k}$, the equilibrium
    output prices follow the max--min rule
    \eqref{eq:gadget-output-prices}, which also defines the
    constructed output prices.

    \item \label{step:recovery-output-scaling}
    At layer $0$, the input chores are basic, so their equilibrium
    prices are $\kappa$ times their constructed prices by
    Step~\ref{step:recovery-basic-scaling}.
    Suppose a gadget at layer $k$ has constructed input prices
    $u,v$ and equilibrium input prices $\kappa u,\kappa v$.
    By Step~\ref{step:recovery-price-condition} and $\kappa>0$,
    its output prices satisfy
    $$
    \begin{aligned}
    p^+_{(j,j'),k}
    &=\frac{\max\{\kappa u,\kappa v\}}{q_k}
    =\kappa\frac{\max\{u,v\}}{q_k}
    =\kappa p^+_{(j,j'),k}(x),\\
    p^-_{(j,j'),k}
    &=q_k\min\{\kappa u,\kappa v\}
    =\kappa q_k\min\{u,v\}
    =\kappa p^-_{(j,j'),k}(x).
    \end{aligned}
    $$
    These outputs are the inputs to the next layer.
    Induction over the layers of every chain therefore gives
    $
    p=\kappa p(x).
    $

    \item \label{step:recovery-normalization}
    By \eqref{eq:market_clearing} and
    Step~\ref{step:recovery-output-scaling},
    $
    n+1
    =\sum_{\ell\in M}S_\ell p_\ell
    =\kappa\sum_{\ell\in M}S_\ell p_\ell(x)
    =\kappa T(x).
    $
    Since $T(x)>0$, we obtain $\kappa=(n+1)/T(x)$.
    By \eqref{eq:recovery-family},
    $$
    p=\kappa p(x)
    =\frac{n+1}{T(x)}p(x)
    =\widetilde p(x).
    $$
    Returning to the original notation, we have
    $x^\star\in[0,1]^n$ and
    $p^\star=\widetilde p(x^\star)$, as required.
\end{enumerate}

\paragraph{Completing the Verification.}
For any exact rational equilibrium $(p^\star,z^\star)$, the decoder
\eqref{eq:decoder} returns a rational point $x^\star$, since $t$
and the prices are rational and $p_0^\star>0$.
The parameter choice above~\eqref{eq:reduction-parameter-choice} makes $t$ computable in polynomial time
with polynomial encoding length. The decoder uses $O(n)$ rational
arithmetic operations and therefore runs in time polynomial in the
encoding lengths of $\mathcal I$ and the supplied equilibrium.
Together with \cref{lem:equilibrium-recovery}, this verifies
property~\ref{item:market-price-mapping}\ref{item:market-decoding}
of \Cref{prop:market-target}.

\subsection{Verification of the Right Derivative Property}
\label{subsec:market-right-derivative}

In this subsection, we verify
property~\ref{item:market-price-mapping}\ref{item:market-right-derivative}
of \Cref{prop:market-target} for the normalized price family
$\widetilde p$ defined in \eqref{eq:recovery-family}.

\begin{lemma}[Right Derivatives of the Price Family]
\label{lem:price-family-right-derivative}
For every $x\in[0,1]^n$ and every feasible coordinate
direction $\widetilde v$ at $x$, the curve
$h\mapsto\widetilde p(x+h\widetilde v)$ has a finite
right derivative at $h=0$. That is, the following limit exists:
$$
\lim_{h\downarrow0}
\frac{\widetilde p(x+h\widetilde v)-\widetilde p(x)}{h}
\in\mathbb R^{M}.
$$
\end{lemma}

\begin{proof}
Fix $x\in[0,1]^n$ and a feasible coordinate direction
$\widetilde v\in\{\pm e_i:i\in[n]\}$.
Recall that the basic prices in \eqref{eq:naive-price} are
$$
p_j(x)=\frac{1+tx_j}{1+t\bar x},
\qquad
x_0=0,
\qquad
\bar x=\frac1{n+1}\sum_{j=0}^n x_j,
$$
and the auxiliary prices are defined recursively by the
max--min rule \eqref{eq:gadget-output-prices}.
The normalized price family introduced in
\Cref{subsec:normalization-final-program} is
$$
\widetilde p(x)=\frac{n+1}{T(x)}p(x),
\qquad
T(x):=\sum_{\ell\in M}S_\ell p_\ell(x).
$$

We first prove that the basic prices have finite right derivatives
along $x+h\widetilde v$.
We then apply the max--min rule layer by layer to establish
the same property for the auxiliary prices.
Finally, we show that normalization preserves this property.

For these steps, we use the following facts.
For a real-valued function $u$ defined for sufficiently small
$h\geq0$, write
$$
u'(0+):=\lim_{h\downarrow0}\frac{u(h)-u(0)}{h}
$$
whenever this limit exists.
We use the following facts about right derivatives.
Facts~\ref{fact:price-affine-quotient}
and~\ref{fact:price-quotient-rule} follow from
\cite[Definition~5.1 and Theorem~5.3]{Rudin1976},
which include right derivatives at left endpoints.
The statement about right derivatives of maxima and minima
in Fact~\ref{fact:price-max-min} is the one-variable case of
\cite[Remark~3.2]{BaierFarkhiRoshchina2014}.

\begin{enumerate}[label=(F\arabic*),ref=F\arabic*]
    \item \label{fact:price-affine-quotient}
    If $u$ and $v$ are affine functions and $v(0)\neq0$,
    then $u/v$ has a finite right derivative at $0$.

    \item \label{fact:price-max-min}
    If $u$ and $v$ have finite right derivatives at $0$,
    then so do $\max\{u,v\}$ and $\min\{u,v\}$.
    If $u(0)=v(0)$, their right derivatives are
    $$
    \begin{aligned}
    \bigl(\max\{u,v\}\bigr)'(0+)
    &=\max\{u'(0+),v'(0+)\},\\
    \bigl(\min\{u,v\}\bigr)'(0+)
    &=\min\{u'(0+),v'(0+)\}.
    \end{aligned}
    $$
    If $u(0)\neq v(0)$, the derivatives are those of the
    larger and smaller functions at $0$, respectively.

    \item \label{fact:price-quotient-rule}
    Finite linear combinations preserve finite right derivatives.
    Moreover, if $u$ and $v$ have finite right derivatives
    at $0$ and $v(0)\neq0$, then
    $$
    \left(\frac{u}{v}\right)'(0+)
    =
    \frac{u'(0+)v(0)-u(0)v'(0+)}{v(0)^2}.
    $$
\end{enumerate}

For each basic chore $j$, both the numerator and denominator
of $p_j(x+h\widetilde v)$ are affine functions of $h$,
and the denominator is positive.
Thus Fact~\ref{fact:price-affine-quotient} gives a finite
right derivative for every basic price.

Next, fix a gadget in layer $k$ with input chores
$\ell_1$ and $\ell_2$, and define
$$
u(h):=p_{\ell_1}(x+h\widetilde v),
\qquad
v(h):=p_{\ell_2}(x+h\widetilde v).
$$
Suppose that both functions have finite right derivatives
at $h=0$.
By \eqref{eq:gadget-output-prices}, the gadget's output
prices along the same coordinate segment are
$$
\frac{\max\{u(h),v(h)\}}{q_k}
\qquad\text{and}\qquad
q_k\min\{u(h),v(h)\},
$$
where $q_k>0$ is fixed independently of $h$.
Facts~\ref{fact:price-max-min}
and~\ref{fact:price-quotient-rule} therefore give finite
right derivatives for both output prices, even when the two input prices are equal.
Starting from the basic prices and proceeding layer by layer,
we conclude that every coordinate of $p(x+h\widetilde v)$
has a finite right derivative at $h=0$.

Finally,
$$
T(x+h\widetilde v)
=\sum_{\ell\in M}S_\ell p_\ell(x+h\widetilde v)
$$
has a finite right derivative by
Fact~\ref{fact:price-quotient-rule}, since the supplies are fixed.
Moreover, $T(x)>0$ because all supplies and prices are positive.
Applying the quotient rule in Fact~\ref{fact:price-quotient-rule} to
$$
\widetilde p_\ell(x+h\widetilde v)
=
\frac{(n+1)p_\ell(x+h\widetilde v)}
{T(x+h\widetilde v)}
\qquad\text{for every }\ell\in M
$$
shows that every normalized price has a finite right derivative
at $h=0$.
Hence
$$
\lim_{h\downarrow0}
\frac{\widetilde p(x+h\widetilde v)-\widetilde p(x)}{h}
\in\mathbb R^{M},
$$
as required.
\end{proof}

\subsection{Verification of the Error Bound}
\label{subsec:market-error-bound}

The following lemma verifies
property~\ref{item:market-error-bound} of
\Cref{prop:market-target}.

\begin{lemma}[Total Error Bound]
\label{lem:market-error-bound}
Let $\alpha>0$ be
given by \eqref{eq:choice-of-alpha} and $t$ chosen as in
\eqref{eq:reduction-parameter-choice}.
There exist a constant $C\in\mathbb R$, independent of $x$,
and a function $R:[0,1]^n\to\mathbb R$ such that
\begin{equation}
\label{eq:normalized-market-objective}
\Psi(\widetilde p(x))=C+\alpha Q(x)+R(x)
\qquad\text{for every }x\in[0,1]^n.
\end{equation}
Moreover, $\beta:=20t^3$ satisfies
$0\leq\beta\leq\alpha\varepsilon$, and
$$
|R(x+he_i)-R(x)|\leq\beta|h|
$$
for every $i\in[n]$ and every coordinate segment from
$x$ to $x+he_i$ contained in $[0,1]^n$.
\end{lemma}

\begin{proof}
Combining \eqref{eq:current-market-objective} with
\cref{thm:normalization-error} gives
\eqref{eq:normalized-market-objective}, where $C$
is independent of $x$ and
$$
R(x):=
R_0(x)
-\sum_{0\leq j<j'\leq n}R_{(j,j')}(x)
+R_{\mathrm{lin}}(x)
+R_{\mathrm{norm}}(x).
$$

Fix $i\in[n]$ and a coordinate segment from $x$ to
$x+he_i$ contained in $[0,1]^n$.
The chain remainder $R_{(j,j')}$ in
\eqref{eq:chain-quadratic-approximation} depends only on
$x_j$ and $x_{j'}$.
Thus changing $x_i$ affects only chains incident to $i$.
By \eqref{eq:basic-weights} and \eqref{eq:choice-of-alpha},
their weights satisfy
$$
W_i:=
\sum_{\substack{j=0\\j\neq i}}^n w_{ij}
=
\frac{n}{n+1}-\frac{\alpha}{t^2}A_{ii}
<1.
$$
Applying the bounds from
\cref{thm:basic-quadratic-approximation,thm:chain-quadratic-approximation,thm:linear-approximation,thm:normalization-error}
and the triangle inequality yields
\begin{equation*}
\begin{aligned}
|R(x+he_i)-R(x)|
&\leq
\left(
2+12W_i+\frac{1}{2(n+1)}+3
\right)t^3|h|\leq20t^3|h|
=\beta|h|.
\end{aligned}
\end{equation*}

Finally, by \eqref{eq:choice-of-alpha} and
\eqref{eq:reduction-parameter-choice},
\begin{equation*}
\begin{aligned}
\frac{\beta}{\alpha}
&=
40(n+1)\left(
1+\sum_{i,j=1}^n|A_{ij}|+\sum_{i=1}^n|c_i|
\right)t\leq\frac{5}{16}\varepsilon
\leq\varepsilon.
\end{aligned}
\end{equation*}
Hence $0\leq\beta\leq\alpha\varepsilon$, completing the proof.
\end{proof}

\section*{AI Methodology}
We used OpenAI's GPT-6 Astra
to assist with mathematical analysis across all sections of
the paper, as well as language editing, notation consistency
checks, and \LaTeX{} formatting. The authors reviewed and
verified all mathematical statements, proofs, calculations,
and citations, including any content developed with AI
assistance, and take full responsibility for the accuracy,
originality, and integrity of the paper.

\bibliography{refer}

\appendix

\section{Omitted proof for Section 2}\label{proof-first-order-character}
\begin{proof}[Proof of \Cref{thm:first-order-character}]
\citet{Chaudhury24competitive} introduce the
\emph{Chores Dual Redundant} program, which in our notation is
\begin{equation}
\begin{aligned}
\max_{p\geq0,\,\widehat\mu>0}\quad
&-\sum_{a=1}^n B_a\log\widehat\mu_a\\
\text{s.t.}\quad
&p_j\leq d_{aj}\widehat\mu_a\qquad(a\in[n],\ j\in[m]),\\
&\sum_{j=1}^m S_jp_j=\sum_{a=1}^n B_a.
\end{aligned}
\label{eq:first-order-lifted-program}
\end{equation}
Here $\widehat\mu_a$ is an auxiliary upper bound on $\mu_a(p)$,
and we drop the constant term $\sum_jS_jp_j=\sum_a B_a$ from their objective.
Their unit-supply formulation covers arbitrary $S_j>0$ by the
change of units $\widehat p_j=S_jp_j$, $\widehat d_{aj}=S_jd_{aj}$,
and $\widehat z_{aj}=z_{aj}/S_j$.

Fix $p^*\in\mathcal P$, set $\widehat\mu_a^*:=\mu_a(p^*)$, and let
$J_a:=\{j:p_j^*=d_{aj}\mu_a(p^*)\}$.
The feasible directions $(v,w)$ of \eqref{eq:first-order-lifted-program}
at $(p^*,\widehat\mu^*)$ are exactly those satisfying
\[
\sum_{j=1}^m S_jv_j=0,
\qquad
w_a\geq\max_{j\in J_a}\frac{v_j}{d_{aj}}
\quad\text{for every }a\in[n].
\]
For the second part, substituting $(p^*+hv,\widehat\mu^*+hw)$ into
$p_j\leq d_{aj}\widehat\mu_a$ gives
$h(v_j-d_{aj}w_a)\leq d_{aj}\mu_a(p^*)-p_j^*.$
For $j\in J_a$, the right-hand side is zero, so
$w_a\geq v_j/d_{aj}$, yielding the stated lower bound.
For $j\notin J_a$, the right-hand side is strictly positive,
so the inequality holds for sufficiently small $h>0$.

Since the objective is strictly decreasing in each $\widehat\mu_a$, every KKT point
satisfies $\widehat\mu_a=\mu_a(p)$.
\cite[Theorem~3]{Chaudhury24competitive} gives the first equivalence:
\begin{equation}
\label{eq:equivalence-chaudhury}
\begin{aligned}
&p^*\text{ is an equilibrium price vector}\\
\Longleftrightarrow\quad&
(p^*,\widehat\mu^*)\text{ is a KKT point of \eqref{eq:first-order-lifted-program}}\\
\Longleftrightarrow\quad&
-\sum_a\frac{B_aw_a}{\mu_a(p^*)}\leq0
\quad\text{for every feasible direction }(v,w).
\end{aligned}
\end{equation}

For any feasible direction $(v,w)$,
\begin{align}\label{eq:first-order-derivative-equivalence}
-\sum_a\frac{B_aw_a}{\mu_a(p^*)}
\leq-\sum_a\frac{B_a}{\mu_a(p^*)}
\max_{j\in J_a}\frac{v_j}{d_{aj}}=\Psi'(p^*;v),
\end{align}
with equality when $w_a=\max_{j\in J_a}v_j/d_{aj}$ for every $a$.

To verify the last equality, fix $v$. For sufficiently small $h>0$,
indices outside $J_a$ remain below the maximum, so
$\mu_a(p^*+hv) =\mu_a(p^*)+h\max_{j\in J_a}\frac{v_j}{d_{aj}}.$
Consequently,
\[
\begin{aligned}
\Psi'(p^*;v)
&=\lim_{h\downarrow0}
\frac{\Psi(p^*+hv)-\Psi(p^*)}{h}\\
&=-\sum_a B_a\lim_{h\downarrow0}
\frac{\log\mu_a(p^*+hv)-\log\mu_a(p^*)}{h}\\
&=-\sum_a B_a\lim_{h\downarrow0}
\frac{\log\left(\mu_a(p^*)+h\max_{j\in J_a}\frac{v_j}{d_{aj}}\right)
-\log\mu_a(p^*)}{h}\\
&=-\sum_a\frac{B_a}{\mu_a(p^*)}
\max_{j\in J_a}\frac{v_j}{d_{aj}}.
\end{aligned}
\]
The last step uses $\mu_a(p^*)>0$ and $(\log t)'=1/t$.

If $p^*$ is an equilibrium price vector, for each feasible direction $v$ in~\eqref{eq:first-order-condition}, choose appropriate $w$ to achieve equality in \eqref{eq:first-order-derivative-equivalence}.
Then $\Psi'(p^*;v)=-\sum_a B_aw_a/\mu_a(p^*)\leq0$.
Conversely, if $\Psi'(p^*;v)\leq0$ for every feasible $v$,
\eqref{eq:first-order-derivative-equivalence} gives
$-\sum_a B_aw_a/\mu_a(p^*)\leq0$ for every feasible $(v,w)$.
Together with \eqref{eq:equivalence-chaudhury}, this proves the claim.
\end{proof}

\section{Omitted proof for Section \ref{sec:basic-market}}
\begin{proof}[\normalfont\bfseries Proof of \Cref{thm:basic-quadratic-approximation}]
\phantomsection\label{app:error-basic}
Let $0<t\leq1$, as in \Cref{sec:basic-market}.
The logarithmic objective of the basic market is
$\Psi_0(p(x)) =(n+1)\log(1+t\bar x)-\sum_{j=0}^{n}\log(1+tx_j).$
The identity
$$
\sum_{0\leq j<j'\leq n}(x_j-x_{j'})^2
=(n+1)\sum_{j=0}^{n}x_j^2-(n+1)^2\bar x^2
$$
shows that the remainder in
\eqref{eq:basic-quadratic-approximation} is exactly
$R_0(x) =\Psi_0(p(x))-\frac{t^2}{2} \left(\sum_{j=0}^{n}x_j^2-(n+1)\bar x^2\right).$
This function is smooth on a neighborhood of the box. For $i\in[n]$,
direct differentiation gives
$$
\begin{aligned}
\partial_i R_0(x)
&=\frac{t}{1+t\bar x}-\frac{t}{1+tx_i}-t^2(x_i-\bar x)=t^3\left(
\frac{\bar x^2}{1+t\bar x}-\frac{x_i^2}{1+tx_i}
\right).
\end{aligned}
$$
Since $0\leq\bar x,x_i\leq1$, we obtain
$$
|\partial_i R_0(x)|
\leq t^3\left(
\frac{\bar x^2}{1+t\bar x}+\frac{x_i^2}{1+tx_i}
\right)
\leq2t^3.
$$
The mean value theorem along a coordinate segment yields
$|R_0(x+he_i)-R_0(x)|\leq2t^3|h|.$
For a feasible direction $\widetilde v=\pm e_i$, smoothness gives
$R_0'(x;\widetilde v)=\pm\partial_iR_0(x)$, so the stated
right-derivative bound follows as well.
\end{proof}

\section{Omitted proof for Section \ref{sec:gadgets}}
\label{app:error-bounds}

\subsection{Supplementary Proofs for \texorpdfstring{\Cref{subsec:gadget-telescoping}}{Section \ref{subsec:gadget-telescoping}}}
\begin{proof}[\normalfont\bfseries Proof of \Cref{thm:chain-quadratic-approximation}]
\phantomsection\label{app:error-chain}
Fix a chain $(j,j')$. For the completed market,
\eqref{eq:reduction-parameter-choice} gives $t=2^{-L}\leq1/256$.
We bound the two main-text remainders in
$R_{(j,j')}=\tilde R_{(j,j')}+\hat R_{(j,j')}$ from
\eqref{eq:chain-quadratic-approximation}.
Recall that
$$
\Delta_{(j,j'),0}(x)=\log\frac{1+tx_j}{1+tx_{j'}},
\qquad
\Delta_{(j,j'),k+1}(x)=|\Delta_{(j,j'),k}(x)|-s_k.
$$
In particular, $|\Delta_{(j,j'),0}(x)|\leq t$, and every
$\Delta_{(j,j'),k}$ depends only on the two endpoint coordinates.

\emph{The chain remainder $\tilde R_{(j,j')}$.}
By \eqref{eq:chain-approximation-parameters},
$B_{(j,j'),k}/2-w_{jj'}s_k
=2w_{jj'}(q_k-1-\log q_k)\geq0$.
The bound $y-\log(1+y)\leq y^2/2$ for $y\geq0$ and the geometric
series therefore give
$$
\sum_{k=0}^{L-1}\left(\frac{B_{(j,j'),k}}2-w_{jj'}s_k\right)
\leq w_{jj'}\sum_{k=0}^{L-1}(q_k-1)^2
\leq\frac{w_{jj'}t^2}{12}.
$$
For comparison only, initialize
$\overline\Delta_{(j,j'),0}(x)=\Delta_{(j,j'),0}(x)$ and set
$$
\overline\Delta_{(j,j'),k+1}(x)
=|\overline\Delta_{(j,j'),k}(x)|-t2^{-(k+1)}.
$$
Induction gives $|\overline\Delta_{(j,j'),k}(x)|\leq t2^{-k}$,
hence $|\overline\Delta_{(j,j'),L}(x)|\leq t^2$.
Moreover, $s_k\leq B_{(j,j'),k}/(2w_{jj'})\leq t2^{-(k+1)}$.
The preceding coefficient bound and geometric sums imply
$$
\sum_{k=0}^{L-1}\bigl(t2^{-(k+1)}-s_k\bigr)
\leq\frac{8t^2}{1+8t}+\frac{t^2}{12}.
$$
Since absolute value is $1$-Lipschitz, comparison of the two
recurrences yields, uniformly over $x$,
$$
|\Delta_{(j,j'),L}(x)|
\leq|\overline\Delta_{(j,j'),L}(x)|
+\sum_{k=0}^{L-1}\bigl(t2^{-(k+1)}-s_k\bigr)
\leq\left(9+\frac1{12}\right)t^2.
$$

Fix a coordinate segment in an endpoint $i\in\{j,j'\}\cap[n]$.
The bound $|\partial_i\Delta_{(j,j'),0}|\leq t$ and the same
$1$-Lipschitz recurrence give
$$
|\Delta_{(j,j'),k}(x+he_i)-\Delta_{(j,j'),k}(x)|
\leq t|h|,\qquad 0\leq k\leq L.
$$
In the main-text formula for $\tilde R_{(j,j')}$, the constant
term cancels. The terminal squared term contributes at most
$(9+1/12)w_{jj'}t^3|h|$, and the coefficient corrections contribute
at most $w_{jj'}t^3|h|/12$. Thus, including segments crossing ties,
$$
|\tilde R_{(j,j')}(x+he_i)-\tilde R_{(j,j')}(x)|
\leq10w_{jj'}t^3|h|.
$$

\emph{The logarithmic approximation remainder $\hat R_{(j,j')}$.}
The main-text definition is
$$
\hat R_{(j,j')}(x)
=\frac{w_{jj'}}2\left(\Delta_{(j,j'),0}(x)^2-t^2(x_j-x_{j'})^2\right).
$$
The logarithmic expression for $\Delta_{(j,j'),0}$ gives, when $j>0$,
$$
|\partial_j\hat R_{(j,j')}(x)|
=w_{jj'}t^2\left|
\int_{x_{j'}}^{x_j}
\left(\frac1{(1+tx_j)(1+tz)}-1\right)\,dz
\right|
\leq2w_{jj'}t^3.
$$
Here the integrand has absolute value at most $2t$ and
$|x_j-x_{j'}|\leq1$. The same bound holds for the other endpoint.
The mean value theorem and the bound for $\tilde R_{(j,j')}$ yield
$$
|R_{(j,j')}(x+he_i)-R_{(j,j')}(x)|
\leq12w_{jj'}t^3|h|.
$$
For every nonendpoint coordinate, the difference is zero.

\emph{Right derivatives, including at ties.}
Along a feasible coordinate direction, $\Delta_{(j,j'),0}$ has a
finite right derivative. If $f$ has a finite right derivative at
zero, then $|f|$ does too: it equals
$\operatorname{sgn}(f(0))f'(0+)$ when $f(0)\neq0$, and $|f'(0+)|$
when $f(0)=0$.
Induction through the fold recurrence therefore gives right
derivatives of every $\Delta_{(j,j'),k}$, including at ties.
The formulas for $\tilde R_{(j,j')}$ and $\hat R_{(j,j')}$ then
give a finite right derivative of $R_{(j,j')}$.
Dividing the finite-difference bound by a positive step and
taking the right limit proves
$|R_{(j,j')}'(x;\widetilde v)|\leq12w_{jj'}t^3$
for every feasible coordinate direction $\widetilde v$.
\end{proof}

\subsection{Supplementary Proofs for \texorpdfstring{\Cref{subsec:extra-terms}}{Section \ref{subsec:extra-terms}}}
\label{appendix: subsec6.4}
In this section, we prove the adjusted earning requirements remain positive and the total earning requirement is $n+1$.

\begin{lemma}[Validity of Earning Requirements]
\label{lem:adjusted-base-earnings}
Let $L\geq1$ be an integer, and let $t$ be rational with
$0<t<1/8$.
Suppose the gadget earning requirements are given by
\eqref{eq:chain-approximation-parameters}, with the weights
in~\eqref{eq:basic-weights}.
Then the adjusted basic earning requirements
in~\eqref{eq:adjusted-base-earnings} are rational and satisfy
$$
B_a^{\mathrm{base}}>0
\qquad\text{for every }a\in N_{\mathrm{base}}.
$$
Moreover, the basic and gadget agents together have total
earning requirement
$$
\sum_{a\in N}B_a=n+1.
$$
\end{lemma}

\begin{proof}
The parameters $t$, $\alpha$, $c_j$ for $j=0,\ldots,n$, and
$B_{(j,j'),k}$ in \eqref{eq:adjusted-base-earnings} are rational
by the input assumptions and
\eqref{eq:choice-of-alpha}, \eqref{eq:linear-reference-coefficient},
\eqref{eq:basic-weights}, and \eqref{eq:chain-approximation-parameters}.
Since $B_a^{\mathrm{base}}$ is obtained from these rational quantities
by finitely many arithmetic operations, it is rational.

By \eqref{eq:basic-weights} and the choice of $\alpha$ in
\eqref{eq:choice-of-alpha},
\[
0<w_{jj'}<\frac{3}{2(n+1)},
\qquad
\frac{\alpha}{t}|c_j|<\frac{t}{2(n+1)}
\]
for every pair $0\leq j<j'\leq n$ and every $j=0,\ldots,n$,
respectively. The second bound also holds for the reference
coefficient because $|c_0|\leq\sum_{i=1}^n|c_i|$.
Set $\rho:=\frac12-4t\in(0,\frac12)$.
The geometric sum in \eqref{eq:chain-approximation-parameters} gives
\[
\sum_{k=0}^{L-1}B_{(j,j'),k}
=w_{jj'}t\sum_{k=0}^{L-1}\rho^k
<2w_{jj'}t
<\frac{3t}{n+1}.
\]
For each basic agent $a$ associated with a basic chore $j$, we have
\[
\begin{aligned}
B_a^{\mathrm{base}}&=
1-\frac32\sum_{k=0}^{L-1}
\left(
\sum_{j'<j}B_{(j',j),k}
+\sum_{j'>j}B_{(j,j'),k}
\right)
-\frac{\alpha}{t}c_j\\
&>1-\frac32\,n\,\frac{3t}{n+1}-\frac{t}{2(n+1)}>0,
\end{aligned}
\]
where we use $t<1/8$.

Finally, each layer $k$ of a chain $(j,j')$ contributes
$3B_{(j,j'),k}$ to the total earning requirement of its three gadget
agents. We can compute
\[
\begin{aligned}
\sum_{a\in N}B_a
&=\sum_{a\in N_{\mathrm{base}}}B_a^{\mathrm{base}}
+3\sum_{0\leq j<j'\leq n}\sum_{k=0}^{L-1}B_{(j,j'),k}\\
&=n+1-\frac{3}{2}\sum_{j=0}^n\sum_{k=0}^{L-1}
\left(
\sum_{j'<j}B_{(j',j),k}
+\sum_{j'>j}B_{(j,j'),k}
\right)
-\frac{\alpha}{t}\sum_{j=0}^n c_j+3\sum_{0\leq j<j'\leq n}\sum_{k=0}^{L-1}B_{(j,j'),k}\\
&=(n+1)-\frac{\alpha}{t}\sum_{j=0}^n c_j
=n+1,
\end{aligned}
\]
where the last equality uses \eqref{eq:linear-reference-coefficient}.
\end{proof}

\begin{proof}[\normalfont\bfseries Proof of \Cref{thm:linear-approximation}]
\phantomsection\label{app:error-linear}
Let $0<t<1/8$.
Since $\sum_{j=0}^n c_j=0$ and $x_0=0$, the cancellation in
\eqref{eq:linear-log-expansion} gives
\[
\frac{\alpha}{t}\sum_{j=0}^n c_j\log p_j(x)-\alpha c^\top x
=\frac{\alpha}{t}\sum_{j=1}^n c_j\bigl(\log(1+tx_j)-tx_j\bigr)
=R_{\mathrm{lin}}(x).
\]
This proves \eqref{eq:linear-approximation}. The remainder is smooth
on an open neighborhood of $[0,1]^n$.
For every $\ell\in[n]$, by our definition of $\alpha$ in~\eqref{eq:choice-of-alpha}, we have
\[
\begin{aligned}
\left|\partial_\ell R_{\mathrm{lin}}(x)\right|
&=\left|\frac{\alpha}{t}c_\ell
\left(\frac{t}{1+tx_\ell}-t\right)\right|=\left|\frac{\alpha t c_\ell x_\ell}{1+tx_\ell}\right|\leq\alpha t|c_\ell|\\
&=\frac{t^3|c_\ell|}{2(n+1)\left(
1+\sum_{i,j=1}^{n}|A_{ij}|+\sum_{i=1}^{n}|c_i|
\right)}\leq\frac{t^3}{2(n+1)},
\end{aligned}
\]
where the first inequality uses $0\leq x_\ell\leq 1$. For any coordinate segment from $x$ to $x+he_\ell$
contained in $[0,1]^n$, it follows that
\[
\begin{aligned}
\left|R_{\mathrm{lin}}(x+he_\ell)-R_{\mathrm{lin}}(x)\right|
&=\left|h\int_0^1
\partial_\ell R_{\mathrm{lin}}(x+she_\ell)\,ds\right|\leq |h|\int_0^1\frac{t^3}{2(n+1)}\,ds
=\frac{t^3}{2(n+1)}|h|,
\end{aligned}
\]
proving~\eqref{eq:linear-remainder-lipschitz}.
Finally, smoothness implies that, for every feasible
$\tilde v=\pm e_\ell$, the right directional derivative exists
and satisfies
\[
\left|R_{\mathrm{lin}}'(x;\tilde v)\right|
=\left|\partial_\ell R_{\mathrm{lin}}(x)\right|
\leq\frac{t^3}{2(n+1)}.
\]
\end{proof}

\subsection{Supplementary Proofs for \texorpdfstring{\Cref{subsec:normalization-final-program}}{Section \ref{subsec:normalization-final-program}}}
\label{app:normalization}

We use the parameter choices of \cref{lem:adjusted-base-earnings},
with an integer $L\geq1$, rational $0<t<1/8$, and $r=1/2-4t$. Recall that
$$
\alpha=\frac{t^2}{2(n+1)\bigl(1+\sum_{i,j=1}^{n}|A_{ij}|+\sum_{j=1}^{n}|c_j|\bigr)},
\qquad q_k=1+\frac t4r^k,\qquad S_{(j,j'),k}=(1-4t)w_{jj'}tr^k.
$$
We use the symmetric conventions $w_{jj'}=w_{j'j}$ and
$S_{(j,j'),k}=S_{(j',j),k}$, and write
$W_j=\sum_{\substack{j'=0\\j'\neq j}}^n w_{jj'}$.
The definition of $\alpha$ gives
$\frac{\alpha}{t^2}\sum_{i,j=1}^n|A_{ij}|<\frac1{2(n+1)}$,
so the correction weights satisfy $\frac1{2(n+1)}<w_{jj'}<\frac3{2(n+1)}<1$ and
$$
W_j=\frac n{n+1}-\frac{\alpha}{t^2}A_{jj}\quad(j\in[n]),
\qquad W_0=\frac n{n+1}-\frac{\alpha}{t^2}\sum_{i,j=1}^n A_{ij}.
$$
In particular, $W_j<1-\frac1{2(n+1)}<1$ for every $j=0,\ldots,n$.

\begin{lemma}[Auxiliary supply bounds]
\label{lem:chain-supply-bound}
For every chain $(j,j')$, $\sum_{k=0}^{L-1}S_{(j,j'),k}<2w_{jj'}t$.
Moreover, $\sum_{0\leq j<j'\leq n}\sum_{k=0}^{L-1}S_{(j,j'),k}<(n+1)t$ and
$\sum_{\substack{j'=0\\j'\neq j}}^n\sum_{k=0}^{L-1}S_{(j,j'),k}<2t$
for every $j=0,\ldots,n$.
\end{lemma}

\begin{proof}
The geometric series gives
$$
\sum_{k=0}^{L-1}S_{(j,j'),k}\leq\frac{1-4t}{1-r}w_{jj'}t<2w_{jj'}t.
$$
Summing over chains incident to $j$ gives the bound $2tW_j<2t$;
summing over all chains gives $t\sum_{j=0}^n W_j<(n+1)t$.
\end{proof}

\begin{lemma}[Rationality and positivity of the adjusted basic supplies]
\label{lem:adjusted-basic-supplies}
Each adjusted basic supply $S_j$ is rational and satisfies
$S_j>1-2t>0$ for $j=0,\ldots,n$.
\end{lemma}

\begin{proof}
The correction weights and auxiliary supplies are rational, so
\eqref{eq:adjusted-basic-supplies} makes $S_j$ rational.
By \cref{lem:chain-supply-bound}, $1-S_j<2t$, proving positivity.
\end{proof}

Recall that $x_0=0$, $\bar x=\frac1{n+1}\sum_{j=0}^n x_j$, and
$p_j(x)=\frac{1+tx_j}{1+t\bar x}$ for $j=0,\ldots,n$.
The auxiliary prices follow the gadget max--min formulas.

\begin{lemma}[Uniform price bounds and auxiliary-price sums]
\label{lem:auxiliary-price-expansion}
For every $x\in[0,1]^n$,
$|p_j(x)-1|\leq t$ and $|p^\pm_{(j,j'),k}(x)-1|\leq2t$. Moreover,
$$
p^+_{(j,j'),k}(x)+p^-_{(j,j'),k}(x)=p_j(x)+p_{j'}(x)+O(t^2)=2+t(x_j+x_{j'}-2\bar x)+O(t^2).
$$
All estimates are uniform over $x$, the chains, and the layers,
with absolute constants.
\end{lemma}

\begin{proof}
The basic-price formula gives $|p_j(x)-1|\leq t$ and
$(1+t)^{-1}\leq p_j(x)\leq1+t$.
Fix a chain $(j,j')$ and layer $k$.
Since $\sum_{\ell=0}^k\log q_\ell\leq t/[4(1-r)]<t/2$,
iteration along the chain gives
$e^{-t/2}/(1+t)\leq p^\pm_{(j,j'),k}(x)\leq(1+t)e^{t/2}$,
hence $|p^\pm_{(j,j'),k}(x)-1|\leq2t$.
Every gadget preserves the product of its input prices, so
$p^+_{(j,j'),k}p^-_{(j,j'),k}=p_jp_{j'}$ and
$$
\begin{aligned}
|p^+_{(j,j'),k}+p^-_{(j,j'),k}-p_j-p_{j'}|
&=|(p_j-1)(p_{j'}-1)-(p^+_{(j,j'),k}-1)(p^-_{(j,j'),k}-1)|\leq5t^2.
\end{aligned}
$$
Finally, $p_j(x)=1+t(x_j-\bar x)+O(t^2)$ with error at most $t^2$,
proving the second expansion with error at most $7t^2$.
\end{proof}

\paragraph{Failure of Direct Normalization and the Supply Adjustment.}
\label{Failure of Direct Normalization}

Retain unit supplies for the basic chores, leaving the auxiliary
supplies and price family unchanged. In this subsection, sums over
$(j,j',k)$ range over $0\leq j<j'\leq n$ and $0\leq k<L$.
The total payment is
$$
T_{\mathrm{dir}}(x)=n+1+\sum_{j<j',k}S_{(j,j'),k}\bigl(p^+_{(j,j'),k}(x)+p^-_{(j,j'),k}(x)\bigr).
$$
Scaling prices by $(n+1)/T_{\mathrm{dir}}(x)$ enforces payment balance
and, by homogeneity and total earnings $n+1$, adds
$(n+1)\log(T_{\mathrm{dir}}(x)/(n+1))$ to the logarithmic objective.

\begin{lemma}[Effect of direct normalization]
\label{lem:direct-normalization}
Uniformly over $x\in[0,1]^n$,
$$
\begin{aligned}
(n+1)\log\frac{T_{\mathrm{dir}}(x)}{n+1}
={}&2\sum_{j<j',k}S_{(j,j'),k}
-\frac2{n+1}\left(\sum_{j<j',k}S_{(j,j'),k}\right)^2\\
&+t\sum_{j<j',k}S_{(j,j'),k}(x_j+x_{j'}-2\bar x)+O((n+1)t^3),
\end{aligned}
$$
where the implicit constant is absolute.
\end{lemma}

\begin{proof}
By \cref{lem:chain-supply-bound,lem:auxiliary-price-expansion},
$$
T_{\mathrm{dir}}(x)=n+1+2\sum_{j<j',k}S_{(j,j'),k}
+t\sum_{j<j',k}S_{(j,j'),k}(x_j+x_{j'}-2\bar x)+O((n+1)t^3).
$$
Since $\sum_{j<j',k}S_{(j,j'),k}<(n+1)t$ and the $x$-dependent term is $O((n+1)t^2)$,
the claim follows from $\log(1+u)=u-u^2/2+O(u^3)$.
\end{proof}

The first two terms are independent of $x$, but the third can vary
at order $t^2$: the incident supply sums need not be equal, so its
contributions need not cancel. Direct normalization therefore need
not preserve the required third-order error bounds. The supply
adjustment in \eqref{eq:adjusted-basic-supplies} cancels this
second-order contribution before normalization.

\begin{proof}[\normalfont\bfseries Proof of \Cref{thm:normalization-error}]
\phantomsection\label{app:error-normalization}
Let $L\geq1$ and $0<t<1/8$, as in \Cref{lem:adjusted-base-earnings}.
As in \eqref{eq:normalization-objective-identity}, homogeneity
$\mu_a(cp)=c\mu_a(p)$ and total earnings $\sum_a B_a=n+1$ give
\[
\Psi(\widetilde p(x))-\Psi(p(x))
=(n+1)\log\frac{T(x)}{n+1}=R_{\mathrm{norm}}(x).
\]
It remains to bound the coordinate changes of $R_{\mathrm{norm}}$.

Fix a coordinate $i\in[n]$ and a coordinate segment in the box.
The max--min rules commute with common positive scaling, so
$(1+t\bar x)p_j(x)=1+tx_j$ and the scaled auxiliary prices follow
the same rules from these affine inputs. They are continuous and
piecewise affine with finitely many pieces. On each open piece,
all max--min choices can therefore be fixed, including any
consistent ordering of inputs that agree throughout the piece.

For a fixed chain $(j,j')$ and layer $k$, the scaled output pair
then consists, in some order, of
$\theta(1+tx_j),\theta^{-1}(1+tx_{j'})$,
where $\theta>0$ is constant on the piece and is a product of
$q_\ell$ or $q_\ell^{-1}$ for $0\leq\ell\leq k$. Thus
$$
|\log\theta|\leq\sum_{\ell=0}^k\log q_\ell
\leq\frac t4\sum_{\ell=0}^{L-1}\left(\frac12-4t\right)^\ell
<\frac{t}{2+16t}<\frac t2.
$$
Elementary exponential bounds give
$|\theta-1|,|\theta^{-1}-1|<t$ and
$0\leq\theta+\theta^{-1}-2<t^2$.
The difference between this scaled output sum and
$2+t(x_j+x_{j'})$ is
$$
(\theta-1)t(x_j-x_{j'})
+(\theta+\theta^{-1}-2)(1+tx_{j'}),
$$
whose absolute value is less than $(2+t)t^2<3t^2$.
Its derivative in an endpoint coordinate is
$t(\theta-1)$ or $t(\theta^{-1}-1)$, hence has absolute value
less than $t^2$; in every other coordinate it is zero.

By the supply adjustment \eqref{eq:adjusted-basic-supplies},
$$
T(x)-(n+1)=
\sum_{0\leq j<j'\leq n}\sum_{k=0}^{L-1}S_{(j,j'),k}
\bigl(p^+_{(j,j'),k}(x)+p^-_{(j,j'),k}(x)-p_j(x)-p_{j'}(x)\bigr).
$$
The total and incident supply bounds in \Cref{lem:chain-supply-bound}
therefore yield, on every piece,
$$
\begin{aligned}
\bigl|(1+t\bar x)(T(x)-(n+1))\bigr|&<3(n+1)t^3,\\
\left|\partial_i\bigl[(1+t\bar x)(T(x)-(n+1))\bigr]\right|&<2t^3.
\end{aligned}
$$
By continuity, the first bound also holds with $\leq$ at partition points.

Product preservation gives
$(1+t\bar x)^2p^+_{(j,j'),k}(x)p^-_{(j,j'),k}(x)
=(1+tx_j)(1+tx_{j'})\geq1$,
so every scaled auxiliary pair has sum at least $2$.
Since the basic supplies are positive by
\Cref{lem:adjusted-basic-supplies} and the scaled basic prices are
at least $1$, we obtain
$$
(1+t\bar x)T(x)\geq
\sum_{j=0}^n S_j+
2\sum_{0\leq j<j'\leq n}\sum_{k=0}^{L-1}S_{(j,j'),k}=n+1.
$$
Using $\partial_i(1+t\bar x)=t/(n+1)$ and the preceding bounds,
differentiation on each piece gives
$$
\begin{aligned}
|\partial_i R_{\mathrm{norm}}(x)|
&\leq
\frac{(n+1)\left|\partial_i\bigl[(1+t\bar x)(T(x)-(n+1))\bigr]\right|
+t|T(x)-(n+1)|}{(1+t\bar x)T(x)}\\
&<2t^3+3t^4<3t^3.
\end{aligned}
$$
The second numerator term includes the effect of common normalization
on all chains, including those not incident to $i$.

Continuity and summation over the finite partition now give
$|R_{\mathrm{norm}}(x+he_i)-R_{\mathrm{norm}}(x)|\leq3t^3|h|$
for every coordinate segment in the box.
On each piece, $(1+t\bar x)T(x)$ and $1+t\bar x$ are positive affine
functions, so the logarithmic expression for $R_{\mathrm{norm}}$
has finite one-sided derivatives at the endpoints.
The finite partition therefore gives a right derivative along
every feasible coordinate direction, including at ties and box
boundaries; its absolute value is at most $3t^3$ by the
finite-difference bound.
\end{proof}

\section{Omitted proof for Section \ref{sec:completed-market}}
\label{app:completed-market-parameters}
\label{app:disutility-completion}

We now verify that assigning $(1+t)^2$ to previously unspecified
agent--chore pairs preserves the optimal-ratio formulas throughout both
price families. We compare choices before normalization and then use
invariance under common scaling.

\begin{lemma}[Completing the Disutility Matrix]
\label{lem:disutility-completion}
Choose the gadget parameters as in
\eqref{eq:chain-approximation-parameters}, with
$t=2^{-L}\leq1/256$, and assign
$d_{aj}:=(1+t)^2$
to every previously unspecified agent--chore pair.
Then the completed disutility matrix is strictly positive
and rational.

For every $x\in[0,1]^n$, at both $p(x)$ and
$\widetilde p(x)$, every newly specified choice is strictly
inferior to a designated choice of the same agent.
Consequently, each basic agent's own chore remains optimal,
and the gadget agents retain the optimal-ratio formulas
in~\eqref{eq:gadget-optimal-ratios}.
The objective identities and approximation bounds computed
from these designated choices therefore remain valid,
including the normalization correction at $\widetilde p(x)$.
\end{lemma}

\begin{proof}
Since common scaling preserves comparisons between an
agent's price-to-disutility ratios, we first work before
the basic-price normalization.
Set $x_0:=0$ and write
$u_j:=1+tx_j,  j=0,\ldots,n,$
and let $u^\pm_{(j,j'),k}$ denote the auxiliary prices
generated recursively from these basic inputs.

We first show that every resulting price lies in
$\left[\frac1{q_0},(1+t)q_0\right],  q_0=1+\frac t4.$
Fix a chain $(j,j')$ and let
$\Delta_0=\log(u_j/u_{j'})$.
The fold recurrence gives
$\Delta_{k+1}=|\Delta_k|-2\log q_k.$
Since $1<q_k\leq q_0$, induction yields
$$
|\Delta_{k+1}|
\leq\max\{|\Delta_k|,2\log q_k\}
\leq\max\{|\Delta_0|,2\log q_0\}.
$$
Product preservation also gives
$u^+_{(j,j'),k}u^-_{(j,j'),k}=u_ju_{j'}.$
Combining the product and logarithmic-gap bounds, we obtain
$$
\begin{aligned}
u^\pm_{(j,j'),k}
&\geq
\min\left\{
u_j,u_{j'},\frac{\sqrt{u_ju_{j'}}}{q_0}
\right\}
\geq\frac1{q_0},\\
u^\pm_{(j,j'),k}
&\leq
\max\left\{
u_j,u_{j'},q_0\sqrt{u_ju_{j'}}
\right\}
\leq(1+t)q_0,
\end{aligned}
$$
where we used $1\leq u_j,u_{j'}\leq1+t$.
The basic prices also lie in this interval.

A newly specified choice, with disutility $(1+t)^2$,
therefore offers ratio at most
$\frac{(1+t)q_0}{(1+t)^2} = \frac{q_0}{1+t} < \frac1{q_0}.$
The strict inequality follows from
$q_0^2 = \left(1+\frac t4\right)^2 = 1+\frac t2+\frac{t^2}{16} <1+t.$
Every basic agent has her own chore as a designated choice
with disutility $1$, and every gadget agent has a designated
output chore with disutility $1$.
Each such choice offers ratio at least $1/q_0$.
Hence no newly specified choice can be optimal.

For a basic agent associated with chore $j$, her own chore
offers ratio $u_j\geq1$.
The reference chore offers ratio $u_0=1$, while every other
basic chore $j'$ offers ratio
$\frac{u_{j'}}{1+t}\leq1.$
Thus her own chore remains optimal.

For a gadget in layer $k$ with input prices $u,v$,
the output prices are
$u^+=\frac{\max\{u,v\}}{q_k},  u^-=q_k\min\{u,v\}.$
Maximizing over the designated choices gives
$$
\begin{aligned}
\mu_{a^{\max}}
&=\max\left\{\frac{u}{q_k},\frac{v}{q_k},u^+\right\}
=\frac{\max\{u,v\}}{q_k},\\
\mu_{a^{(1)}}
&=\max\{q_ku,u^-\}=q_ku,\\
\mu_{a^{(2)}}
&=\max\{q_kv,u^-\}=q_kv.
\end{aligned}
$$
These are precisely the gadget optimal-ratio formulas.

Both the basic normalization and the final normalization
multiply all prices by a common positive factor.
They preserve the strict comparisons above and scale every
optimal ratio by that factor.
The conclusions therefore hold at both $p(x)$ and
$\widetilde p(x)$.
At either fixed price vector, completing the matrix leaves
the optimal ratios equal to those computed from designated
choices. Hence it introduces no additional term into the
objective or its approximation error.

Finally, the previously specified disutilities are positive
and rational, and the new value $(1+t)^2$ has the same
properties because $t$ is positive and rational.
\end{proof}

\end{document}